\documentclass{aa}  
\usepackage{graphicx}
\usepackage{txfonts}
\usepackage{natbib}
\bibpunct{(}{)}{;}{a}{}{,} 

\newcommand{\nh}{$N_{\rm H}$}
\newcommand{\xmm}{{\it XMM-Newton}}
\newcommand{\cxo}{{\it Chandra}}
\newcommand{\rosat}{{\it ROSAT}}

\newcommand{\bona}{{\sl bona fide}}
\newcommand{\coup}{{\it Chandra Orion Ultradeep Project}}
\newcommand{\xest}{{\it XMM-Newton Extended Survey of the TMC}}
\begin{document}
   \title{X-ray emission from the young brown dwarfs\\
of the Taurus Molecular Cloud}


   \author{N.\ Grosso\inst{1}
\and K.~R.\ Briggs\inst{2}
\and M.\ G{\"u}del\inst{2}
\and S.\ Guieu\inst{1}
\and E.\ Franciosini\inst{3}
\and F.\ Palla\inst{4}
\and\\ C.\ Dougados\inst{1}
\and J.-L.\ Monin\inst{1,5}
\and F.\ M{\'e}nard\inst{1}
\and J.\ Bouvier\inst{1}
\and M.\ Audard\inst{6}
\and A.\ Telleschi\inst{2}
          }

   \offprints{N.\ Grosso}

   \institute{Laboratoire d'Astrophysique de Grenoble,
              Universit{\'e} Joseph-Fourier,
              F-38041 Grenoble cedex 9, France\\
              \email{\tt Nicolas.Grosso@obs.ujf-grenoble.fr}
         \and Paul Scherrer Institut, 5232 Villigen und W{\"u}renlingen,
                Switzerland
         \and INAF - Osservatorio Astronomico di Palermo, Piazza del
              Parlamento 1, I-90134 Palermo, Italy
         \and INAF - Osservatorio Astrofisico di Arcetri, Largo Enrico
              Fermi 5, I-50125 Firenze, Italy
         \and Institut Universitaire de France
         \and Columbia Astrophysics Laboratory, Columbia University,
              550 West 120th Street, New York, NY 10027, USA
             }

   \date{Received 9 May 2006; accepted 5 August 2006}

 
  \abstract
   {}
   {We report the X-ray properties of young ($\sim$3\,Myr) \bona~brown dwarfs of the
Taurus Molecular Cloud (TMC).}
   {The \xest~(XEST) is a large program designed to systematically investigate the
X-ray properties of young stellar/substellar objects in the TMC. In particular, the area
surveyed by 15 \xmm~pointings (of which three are archival observations),
supplemented with one archival \cxo~observation, allows us to study 17
brown dwarfs with M spectral types.
}
   {Half of this sample (9 out of 17 brown dwarfs) is
  detected; 7 brown dwarfs are detected here for the first
  time in X-rays. We observed a flare from
one brown dwarf. We confirm several previous findings on brown dwarf
X-ray activity: a
log-log relation between X-ray and bolometric luminosity for
stars (with $L_*$$\le$10\,L$_\odot$) and brown dwarfs detected in
X-rays, which is consistent with a mean X-ray fractional luminosity
$<\!\log(L_{\rm X}/L_*)\!>\,=-3.5 \pm 0.4$; for the XEST brown dwarfs, the
median of $\log(L_{\rm X}/L_*)$ (including upper limits) is $-4.0$; a
shallow log-log relation between X-ray fractional luminosity and mass; 
a log-log relation between X-ray fractional luminosity and effective temperature;
a log-log relation between X-ray surface flux and effective
temperature. We find no significant log-log correlation between the
X-ray fractional luminosity and $EW({\rm H}\alpha)$. Accreting and
nonaccreting brown dwarfs have a similar X-ray fractional luminosity. 
The median X-ray fractional luminosity of nonaccreting brown dwarfs is
about 4 times lower than the mean saturation value for rapidly rotating low-mass
field stars. Our TMC brown dwarfs have higher X-ray
fractional luminosity than brown dwarfs in the \coup.
}
   {The X-ray fractional luminosity declines from low-mass stars
to M-type brown dwarfs, and as a sample, the brown dwarfs are less efficient X-ray
emitters than low-mass stars. We thus conclude that while the brown dwarf
atmospheres observed here are mostly warm enough to sustain coronal
activity, a trend is seen that may indicate its gradual decline due to
the drop in photospheric ionization degree.}

   \keywords{Stars: low-mass, brown dwarfs -- X-rays: stars -- ISM:
individual objects: the Taurus Molecular Cloud
               }

   \maketitle

\section{Introduction}

%
\begin{figure*}[!t]
\centering
\includegraphics[width=1.8\columnwidth]{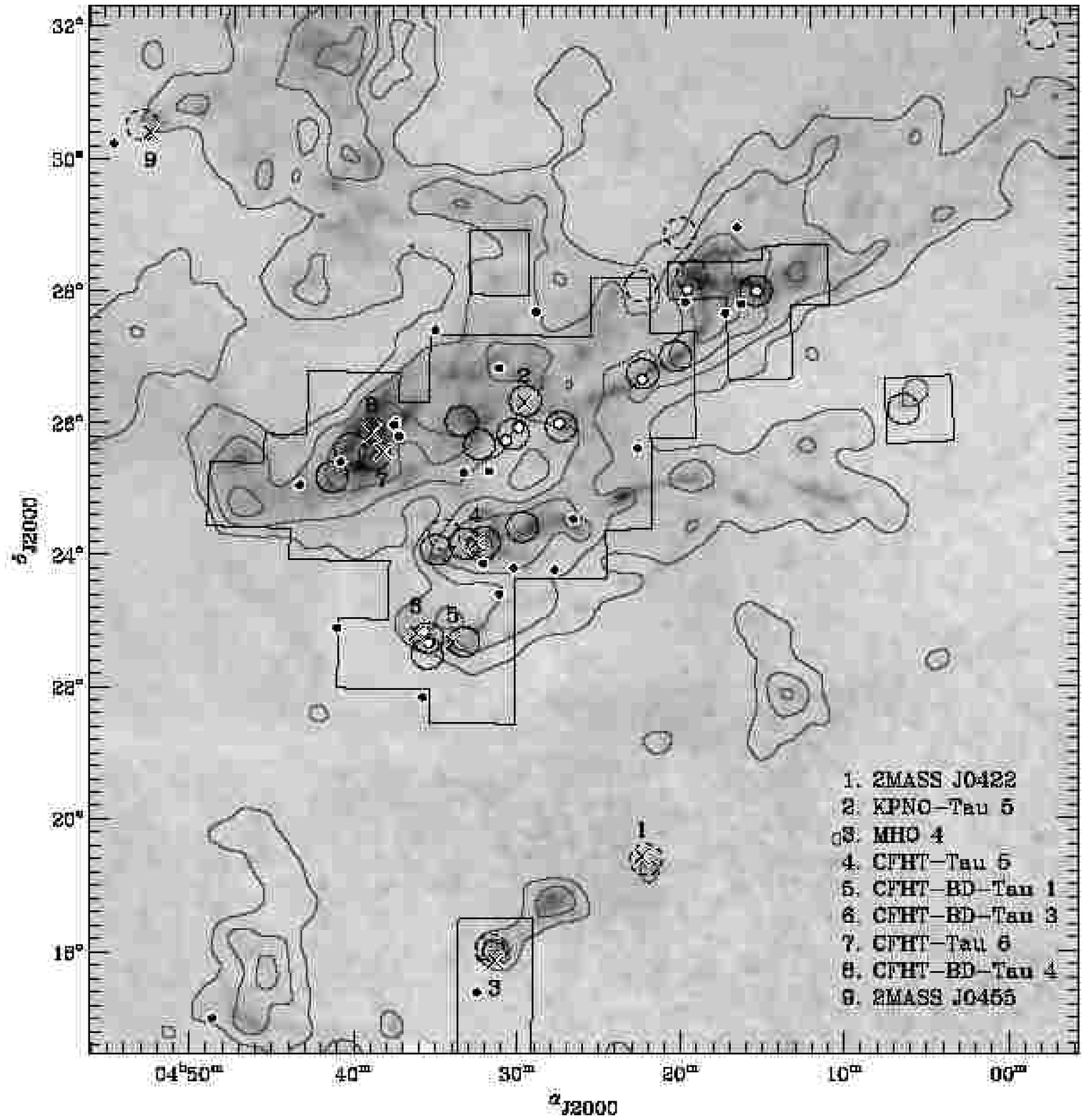}
\caption{The {\it XMM-Newton Extended Survey of the Taurus Molecular
Cloud}~(XEST). Contours show the $^{12}$CO emission \citep{dame87} of
the Taurus Molecular Cloud (TMC), overlaid on a visual
extinction map \citep[linear colour scale, with black colour
corresponding to $A_{\rm V}\sim6$\,mag;
based on DSS~I, {6\arcmin}-resolution map,][]{dobashi05}. 
The 27 \xmm~fields of view of the XEST are plotted with
continuous/dashed circles (of which seven are archival observations
and one is a separate program on T\,Tau; dashed circles). Note the outlying
\xmm~fields of view around SU\,Aur (NE corner) and L1551 (S). The two
squares near labels 3 and 8 show the archival \cxo~fields of view used
in this work (see Table~\ref{table:log}). The continuous line shows the survey of
brown dwarfs (BDs) performed with the Canada-France-Hawaii telescope \citep{guieu06}.
This region hosts 42 young BDs of the TMC 
(\citealt{briceno98,briceno02}; \citealt{luhman00,luhman04,luhman06};
\citealt{martin01}; \citealt{guieu06}). Black dots indicate the
25 BDs of this region which were not surveyed in X-rays with the
XEST. White dots show the 8 BDs not detected in X-rays. Crosses show
the 9 BDs detected in X-rays (Table~\ref{table:detections}) with
labeled numbers referring to BD names. Only 2 BDs of TMC were
previously detected in X-rays by \rosat: MHO\,4
\citep{carkner96,neuhaeuser99} and CFHT-BD-Tau\,4
\citep{briceno99,mokler02}. 
}
\label{fig:map}
\end{figure*}

Pre-main-sequence low-mass stars, i.e.\ T~Tauri stars, show a high
level of X-ray emission, which is generally attributed to an active
corona, an enhanced version of the magnetic corona on the contemporary
Sun \citep[see for a review][]{feigelson99}. The dynamo mechanism
producing the magnetic field in these fully convective stars is still
discussed \citep{preibisch05b}. X-ray observations of brown
dwarfs (BDs) allow investigation of the magnetic activity in the
substellar regime, where masses are lower than about 0.075\,M$_\odot$.

The first X-ray detection of a BD was reported by
\citet{neuhaeuser98}, who identified in the Cha\,I dark cloud an X-ray
source detected by \rosat~-- in early 1991, before the
near-infrared observations of the first BDs were reported 
\citep{nakajima95,rebolo95} -- as the counterpart of a young
BD. \rosat~detected only a few BDs or very low-mass stars, all of them
young and located in star-forming regions
\citep{neuhaeuser99,comeron00,mokler02}.

The new generation of X-ray satellites, \xmm~and \cxo, allows now
detection of more BDs thanks to their increase of sensitivity. The
number of young or intermediate-age BDs detected in X-rays
\citep[e.g.,][]{imanishi01,preibisch01,preibisch02,tsuboi03,briggs04,stelzer04b,ozawa05,stelzer06}
exceeds largely the one of (older) field BDs, which have been detected
so far mainly during X-ray flares
\citep{rutledge00,stelzer04}. Recently, the X-ray properties of 33
young BDs \citep{slesnick04} in the Orion Nebula Cluster (ONC) located
at $\sim$450\,pc were studied by \citet{preibisch05} in the {\it
Chandra Orion Ultradeep Project} (COUP).

\begin{table*}[!ht]
\caption{The sample of the 17 young BDs surveyed in the 
XEST. Col.~(2) gives the reference of the discovery paper. Col.~(3)
give the 2MASS counterparts. Col.~(7) gives the references for the
spectral type (Col.~4), the optical extinction \citep[when not
available derived from $A_{\rm J}$ using][ Col.~5]{rieke85}, and the
reference of the equivalent width of H$\alpha$ ($EW({\rm H}\alpha)$,
given in Col.~6) if it is different. Negligible optical extinction in
Col.~(5) are indicated by 0. The effective temperature in Col.~(8) has
been computed from the spectral type using the temperature scale
$T_{\rm eff}=3841.94-141.17 \times {SpTyp}$, which is valid for young
stars with M spectral type \citep{guieu06}. The visual extinction in
Col.~(5) is taken when available from the literature; for 2MASS\,J0421
and 2MASS\,J0422, we averaged the visual extinction computed from
$J-H$, $H-K_{\rm S}$, and the spectral type, using a dwarf sequence
\citep[compiled from the literature; e.g.,][]{leggett98}.
The bolometric luminosity of the substellar photosphere in
Col.~(9) has been computed from $I$, $J$-band magnitudes and $A_{\rm
V}$ \citep[see][]{guieu06}; for 2MASS\,J0414 and MHO\,4 (without
$I$-band magnitude available) the reference for the luminosity is
\citet{luhman04} and \citet{briceno02}, respectively. Col.~(10)
indicates accreting sources based mainly on $EW({\rm H}\alpha)$
(see \S\ref{accretion}). The last column indicates detection in X-rays
(this work; see Table~\ref{table:detections}). References:
B98=\citet{briceno98}; B02=\citet{briceno02}; G06=\citet{guieu06};
L04=\citet{luhman04}; L06=\citet{luhman06}; M01=\citet{martin01};
M05=\citet{muzerolle05b}.}
\label{table:bd}
\centering
\begin{tabular}{lcclcrccccc}
\hline\hline
\multicolumn{1}{c}{Name} & Ref.\ & 2MASS &
\multicolumn{1}{c}{SpTyp} & $A_{\rm V}$ & \multicolumn{1}{c}{$EW({\rm H}\alpha)$} & Ref.\ & $T_{\rm eff}$ & $L_*$ &  Acc. & X\\
         &                      &   &                         & mag &  \multicolumn{1}{c}{$\AA$}  & &
K & L$_\odot$   \\
\multicolumn{1}{c}{(1)} & (2) & (3) & \multicolumn{1}{c}{(4)} & (5) & \multicolumn{1}{c}{(6)} & (7) & (8) & (9) & (10) & (11)\\
\hline
\object{2MASS\,J0414}   & L04 & J04141188+2811535 &  M6.25 & 1.1 & 250.0    &  M05 & 2960 & 0.015 & y & n\\ 
\object{KPNO-Tau\,2}    & B02 & J04185115+2814332 &  M6.75 & 0.4 & 8.4	   &  G06 & 2889 & 0.007 & n & n\\ 
\object{2MASS\,J0421}   & L06 & J04215450+2652315 &  M8.5  & 3.0 & \dotfill &  L06 & 2642 & 0.003 &\dotfill & n\\
\object{2MASS\,J0422}   & L06 & J04221332+1934392 &  M8    & 1.0 & \dotfill &  L06 & 2713 & 0.017 &\dotfill & y\\
\object{KPNO-Tau\,4}    & B02 & J04272799+2612052 &  M9.5  & 2.5 & 158.1    &  G06 & 2501 & 0.004 & y & n\\ 
\object{KPNO-Tau\,5}    & B02 & J04294568+2630468 &  M7.5  & 0	& 30.0     &  B02 & 2783 & 0.023 & n & y\\ 
\object{KPNO-Tau\,6}    & B02 & J04300724+2608207 &  M9    & 0.9 & 207.9    &  G06 & 2571 & 0.003 & y & n\\ 
\object{KPNO-Tau\,7}    & B02 & J04305718+2556394 &  M8.25 & 0	& 300.0    &  B02 & 2677 & 0.004 & y & n\\ 
\object{MHO\,4}         & B98 & J04312405+1800215 &  M7    & 0.5 & 42.0     &  B02 & 2854 & 0.048 & n & y\\ 
\object{CFHT-Tau\,5}    & G06 & J04325026+2422115 &  M7.5  & 9.2 & 29.8     &  G06 & 2783 & 0.075 & n & y\\ 
\object{CFHT-BD-Tau\,1} & M01 & J04341527+2250309 &  M7    & 3.1 & 19.0     &  M01 & 2854 & 0.017 & n & y\\ 
\object{KPNO-Tau\,9}    & B02 & J04355143+2249119 &  M8.5  & 0	& 20.0     &  B02 & 2642 & 0.001 & n & n\\ 
\object{CFHT-BD-Tau\,2} & M01 & J04361038+2259560 &  M7.5  & 0	& 13.0     &  B02 & 2783 & 0.007 & n & n\\ 
\object{CFHT-BD-Tau\,3} & M01 & J04363893+2258119 &  M7.75 & 0	& 55.0     &  B02 & 2747 & 0.007 & n & y\\ 
\object{CFHT-Tau\,6}    & G06 & J04390396+2544264 &  M7.25 & 0.4 & 63.7     &  G06 & 2818 & 0.024 & y & y\\ 
\object{CFHT-BD-Tau\,4} & M01 & J04394748+2601407 &  M7    & 2.6 & 340.0&  L04,M01 & 2854 & 0.062 & y & y\\ 
\object{2MASS\,J0455}   & L04 & J04552333+3027366 &  M6.25 & 0	& \dotfill &  L04 & 2960 & 0.015 &\dotfill & y\\ 
\hline
\end{tabular}
\end{table*}

We investigate in this paper the X-ray properties of a sample of 17
BDs in the nearby (140\,pc) Taurus Molecular Cloud (TMC), surveyed in
the {\it XMM-Newton Extended Survey of the TMC} \citep[XEST;][]{guedel06b}. 
First, we define in \S\ref{TMC_BD} the TMC BD sample, review previous
X-ray results on it, and report the BDs detected in the XEST. Analysis
of the X-ray variability of the TMC BDs is presented in
\S\ref{lc_analysis}, and X-ray spectral properties are determined in
\S\ref{spectral_properties}. In \S\ref{halpha}, we compare the X-ray
luminosities of the BDs and low-mass stars observed in the
XEST. Comparison of the X-ray fractional luminosities of BDs in the
XEST and COUP is made in \S\ref{XEST_COUP}. We discuss the origin of
the BD X-ray emission in the broader context of the X-ray emission of
cool stars in \S\ref{discussion}. Finally, we summarize our results in
\S\ref{summary}.

\section{The TMC brown dwarf sample}
\label{TMC_BD}

For a typical TMC member, with an age of 3\,Myr, the
stellar/substellar boundary at 0.075\,M$_\odot$ is between M6V and
M6.5V according to pre-main-sequence tracks from \citet{baraffe98}. Therefore, we 
compiled from the literature objects with spectral type later than
M6V to build our \bona~BD sample.\footnote{In XEST, there are only five objects
with a M6V spectral type (namely, V410 X-ray 3, MHO 5, MHO 8, KPNO-Tau 14,
and CFHT-Tau 12), which have luminosities higher than
0.03\,L$_\odot$, placing them above the stellar/substellar boundary
\citep{baraffe98}.}

The first optical imaging and spectroscopic surveys of BDs in the TMC
concentrated on high stellar density regions, while the majority of
the volume occupied by the molecular clouds was left unexplored
\citep{briceno98,luhman00,briceno02,luhman04}. Recently, \citet{guieu06}
performed a large scale optical imaging survey of TMC with the
Canada-France-Hawaii telescope with the CFH12k and MEGACAM large-scale
optical cameras, covering a total area of $\approx$28\,deg$^2$, and
encompassing the densest parts of the cloud as well as their
surroundings. By employing all-sky catalogs,
\citet{luhman06} considered a $15^\circ \times 15^\circ$ area, large
enough to encompass all of the TMC. These optical surveys of the TMC,
conducted since nearly 10 years, yielded the identification of
42~BDs.

\begin{table*}[!t]
\caption{Detection list of young BDs in the XEST. Col.~(1)
numbers correspond to labels in Fig.~\ref{fig:map}. Col.~(2), (3) and (4)
give the BD, satellite and target names, respectively. The naming
of \xmm~sources in Col.~(5) follows the convention of \citet{guedel06b},
where the two and three digits code for the field and the source
number in this field, respectively. Cols.~(6)--(9) give
X-ray source positions, total positional uncertainties, and distance
to the 2MASS position, respectively. References:
[FGMSD03] = \citet{favata03}, [BFR03] = \citet{bally03}. 
There is only pn data for X-ray source
\#1 as it falls in the gap of the MOS CCDs in window mode in this observation.
X-ray source \#4b is affected by pn gap. For X-ray
source \#9 only MOS data are available.}
\label{table:detections}
\centering
\begin{tabular}{clclcccccc}
\hline\hline
 & \multicolumn{1}{c}{BD name}         & Satellite  & \multicolumn{1}{c}{Target}        & X-ray source name
 & $\alpha_{\rm J2000}$ & $\delta_{\rm J2000}$ & err.\ & dist.\\
\#    &                 &       &                &
 &                      &                      & \arcsec & \arcsec \\
(1) & \multicolumn{1}{c}{(2)} & (3) & \multicolumn{1}{c}{(4)} & (5) &
(6) & (7) & (8) & (9) \\
\hline
1~ & 2MASS\,J0422    &  XMM  &  T\,Tau    & XEST-01-062             &  04$^{\rm h}$22$^{\rm m}$13\fs2 & 19\degr34\arcmin40\farcs2 & 1.7 & 1.5 \\
2~ & KPNO-Tau\,5     &  XMM  &  DI\,Tau   & XEST-15-044             &  04$^{\rm h}$29$^{\rm m}$45\fs8 & 26\degr30\arcmin47\farcs9 & 1.8 & 1.9\\
3a & MHO\,4          &  XMM  &V955\,Tau & XEST-22-021=[FGMSD03]\,17 &  04$^{\rm h}$31$^{\rm m}$24\fs2 & 18\degr00\arcmin21\farcs6 & 1.6 & 2.1\\
3b & MHO\,4          &  CXO  &  L1551     & [BFR03]\,18             &  04$^{\rm h}$31$^{\rm m}$24\fs1 & 18\degr00\arcmin21\farcs4 & 0.7 & 0.7\\
4a & CFHT-Tau\,5     &  XMM  &  GK\,Tau   & XEST-04-003             &  04$^{\rm h}$32$^{\rm m}$50\fs3 & 24\degr22\arcmin11\farcs1 & 1.7 & 0.7\\
4b & CFHT-Tau\,5     &  XMM  &  V928\,Tau & XEST-03-031             &  04$^{\rm h}$32$^{\rm m}$50\fs3 & 24\degr22\arcmin11\farcs4 & 1.7 & 0.6\\
5~ & CFHT-BD-Tau\,1  &  XMM  &  CI\,Tau   & XEST-17-068             &  04$^{\rm h}$34$^{\rm m}$15\fs3 & 22\degr50\arcmin33\farcs1 & 1.4 & 2.5\\
6~ & CFHT-BD-Tau\,3  &  XMM  &  HP\,Tau   & XEST-08-080             &  04$^{\rm h}$36$^{\rm m}$38\fs9 & 22\degr58\arcmin13\farcs2 & 1.9 & 1.2\\
7~ & CFHT-Tau\,6     &  XMM  &  TMC\,1A   & XEST-05-005             &  04$^{\rm h}$39$^{\rm m}$04\fs1 & 25\degr44\arcmin26\farcs4 & 1.9 & 1.9\\
8~ & CFHT-BD-Tau\,4  &  CXO  &  L1527     & CXOU\,J043947.5+260140  &  04$^{\rm h}$39$^{\rm m}$47\fs5 & 26\degr01\arcmin40\farcs8 & 0.3 & 0.0\\
9~ & 2MASS\,J0455    &  XMM  &  SU\,Aur   & XEST-26-012             &  04$^{\rm h}$55$^{\rm m}$23\fs1 & 30\degr27\arcmin38\farcs2 & 2.0 & 3.1\\
\hline
\end{tabular}
\end{table*}

        \subsection{Previous X-ray results on the TMC brown dwarfs}

Two BDs of this sample were previously detected in X-rays by the 
{\it ROSAT/Position Sensitive Proportional Counter} (PSPC) -- which was
sensitive only to soft X-rays from 0.1 to 2.4\,keV -- but without
being identified as substellar objects.

The BD MHO\,4 \citep{briceno98,luhman00} was first
detected during a 4.0\,ks {\it ROSAT/PSPC}
pointed observation of V826\,Tau in August 1992 with an X-ray count rate of
4\,counts\,ks$^{-1}$ and proposed as a new weak-line T~Tauri star of
Taurus-Auriga \citep[source RXJ0431.3+1800 of][]{wichmann96}. MHO\,4
was also detected during a 7.7\,ks {\it ROSAT/PSPC} pointed observation of the dark cloud
Lynds~1551 in February 1993 with an X-ray count rate of 3.1\,counts\,ks$^{-1}$ and
proposed as a new member of the TMC \citep[source L1551X~15 of][
associated with a ``faint, very red
star'']{carkner96}. \citet{neuhaeuser99} reported these two X-ray
detections of MHO\,4, considered at that time as a BD
candidate\footnote{V410\,Anon 13, V410\,X-ray\,3, and MHO\,5, which
were considered as BD {\sl candidates} detected in X-rays by
\rosat~\citep{neuhaeuser99}, have now more reliable spectral types
\citep[M5.75V, M6V, and M6V, respectively;][]{briceno02}, which
combined with their luminosities place them above the
stellar/substellar boundary. Therefore these very low-mass stars were
not included in our sample of BDs. However, they were detected in the
XEST and will hence be used in \S\ref{BD_vs_stars} as TMC low-mass
stars detected in X-rays.}.
 
CFHT-BD-Tau\,4, one of the BDs of \citet{martin01}, was
detected during a 10.6\,ks {\it ROSAT/PSPC} pointed observation of the
Heiles dark cloud~2~NW in March 1993 with an X-ray count rate of
4.5\,counts\,ks$^{-1}$ corresponding to a luminosity of
$\sim$$10^{29}$\,erg\,s$^{-1}$ \citep[source HCL~2~NW-7a of][ with no
optical counterpart but a NIR counterpart]{briceno99}. \citet{mokler02}
reported this X-ray source as the counterpart of this BD.

The higher sensitivity of the new generation of X-ray satellites,
\xmm~\citep{jansen01} and \cxo~\citep{weisskopf02}, allows detection 
of BDs which are from $\sim$10 to $\sim$100 times fainter in
X-rays.

        \subsection{X-ray detected brown dwarfs in the XEST}

The {\it XMM-Newton Extended Survey of the TMC} (XEST) is a large
program designed to systematically investigate the high-energy
properties of young stellar and substellar objects in this nearest star-forming region. 
It is composed of 19 dedicated \xmm~pointings plus 8 archival
\xmm~observations. The total \xmm~exposure time is about 300\,h
shared among these 27 pointings \citep{guedel06b}. We will focus here on 
the X-ray detection of BDs with the three X-ray telescopes of
\xmm~-- two {\it European Photon Imaging Cameras}
(EPIC) equipped with MOS CCD arrays \citep{turner01}, and the third
carrying a pn CCD camera \citep{strueder01} --
simultaneous $U$-band observations obtained with the \xmm~optical/UV monitor
are discussed in \citet{grosso06b}.

Fig.~\ref{fig:map} shows the XEST area and the locations of the 42
TMC BDs. Fifteen \xmm~pointings (of which three are archival observations)
surveyed 16 BDs. We looked also for serendipitous \cxo~observations at
the {\it Chandra X-ray Center}'s archives (see online Appendix~\ref{log},
Table~\ref{table:log}). We supplemented the XEST with the
ACIS-I~observation of L1527 which allow
us to survey and detect one more BD (CFHT-BD-Tau\,4); and with the
ACIS-I~observation of L1551 which provides a second epoch observation
of MHO\,4. Therefore, we surveyed in X-rays 17 out of the 42 TMC~BDs,
which allows us to survey the X-ray emission of 40\% of the
known BDs in the TMC.

\begin{figure}[!ht]
\centering
\includegraphics[width=\columnwidth]{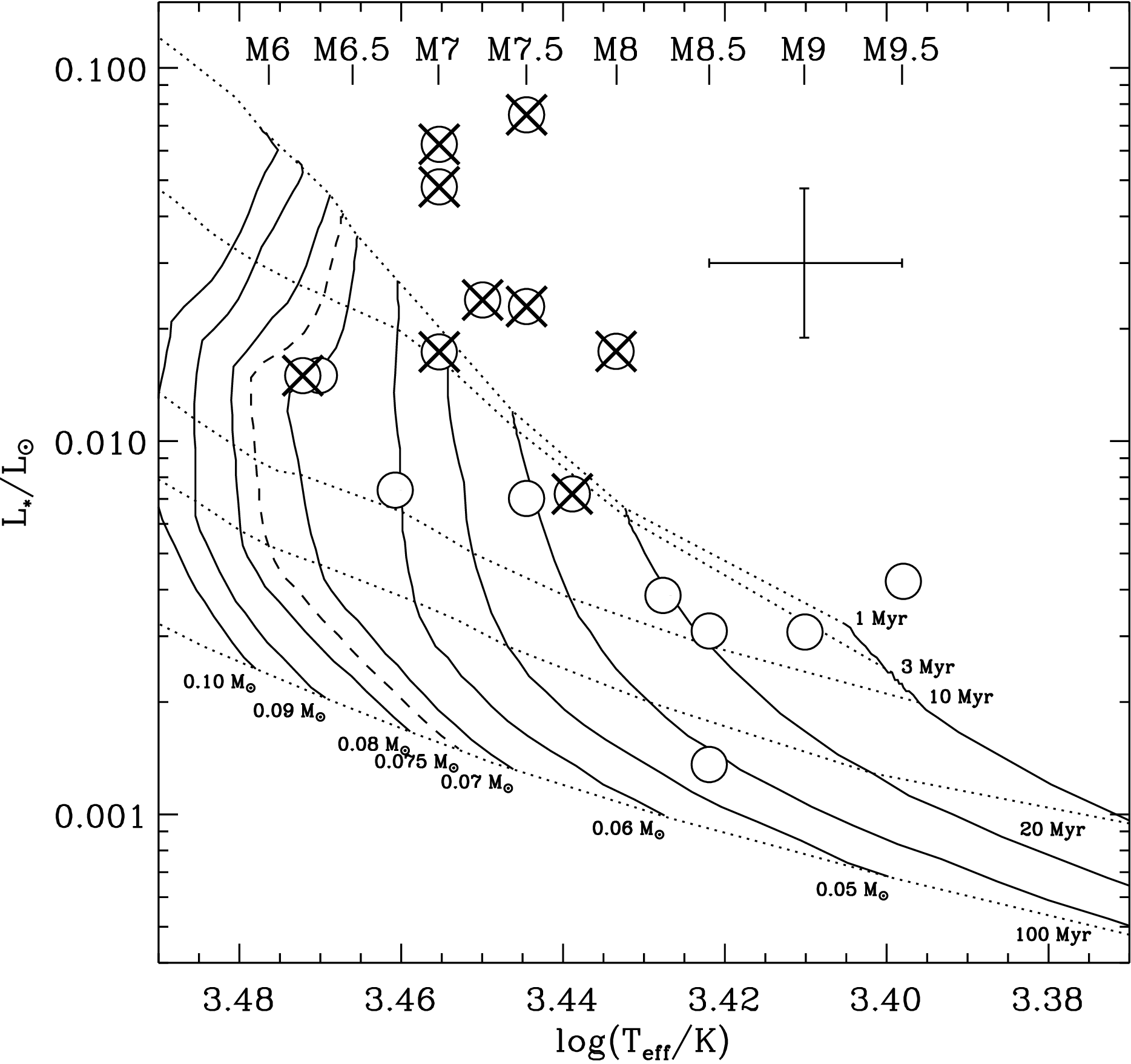}
\caption{HR diagram of the 17 BDs of the TMC surveyed in the XEST. The
references for the computation of the effective temperatures and the
luminosities are given in Table~\ref{table:bd}. The 
pre-main-sequence tracks from
\citet{baraffe98} are shown for comparison. Continuous lines
show mass tracks from 0.1 down to 0.02\,M$_\odot$ in steps of
0.01\,M$_\odot$; the dashed line indicates the stellar/substellar
boundary at 0.075\,M$_\odot$, which is equivalent to
spectral type later than M6V for a
typical TMC member age of 3\,Myr. A cross indicates typical
uncertainties. Two circles have been slightly moved in spectral type
to avoid overlaps. The 9 BDs detected in X-rays are marked with `X'. 
}
\label{fig:hr}
\end{figure}

Table~\ref{table:bd} lists the properties of these 17 BDs.
Fig.~\ref{fig:hr} gives their location in
the HR diagram. The coolest object of the surveyed BD sample has
a spectral type M9.5V, corresponding to about 0.02\,M$_\odot$
\citep{baraffe02}. Their luminosities range from
0.075\,L$_\odot$ to 0.001\,L$_\odot$. The optical extinctions of
this sample range from a negligible value to about 9\,mag with a mean
and a median of 1.5 and 0.5\,mag, respectively. 
The average low extinction of this sample is mainly due to the initial
selection process for TMC candidate BDs based on optical photometry,
and therefore biased towards low extinction BDs.

The list of X-ray sources in the XEST \citep{guedel05} was completed
with X-ray sources of the two archival \cxo~observations (see online
Appendix~\ref{log} for details on the data reduction of \cxo~data). 
We cross-correlated the BD positions with the positions of
all X-ray sources. We detected 9 BDs in X-rays (see
Table~\ref{table:detections}). The two BDs previously detected with
{\it ROSAT} were redetected with \xmm~and \cxo. The detection of
MHO\,4 in L1551 was previously published by \citet{favata03} with
\xmm, and \citet{bally03} with \cxo. Seven BDs are detected here for
the first time in X-rays. The detection rate in X-rays of these 17 BDs
is therefore 53\%. 

We detected mainly the BDs with luminosities greater than
$\sim$0.01\,L$_\odot$, i.e.\ the BDs with spectral type
earlier or equal than M8V (Fig.~\ref{fig:hr}).

\section{X-ray variability analysis of the TMC brown dwarfs}
\label{lc_analysis}

\begin{figure}[!t]
\centering
\includegraphics[height=\columnwidth,angle=90]{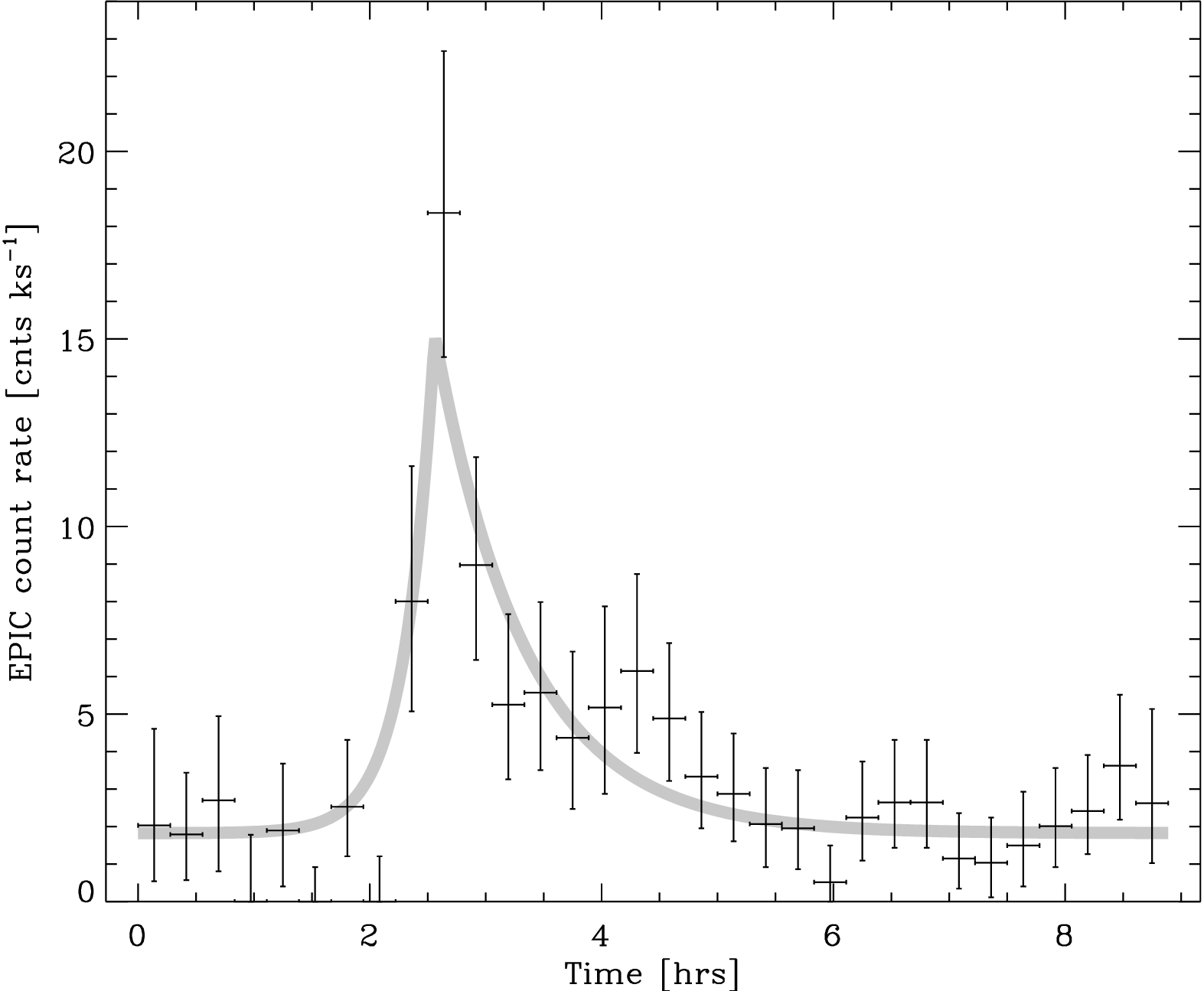}
\caption{Light curve of the X-ray flare from CFHT-BD-Tau\,1. The bin
size is 1000\,s. The grey thick line shows the fit of the non-zero
count rates using a quiescent level plus an exponential rise and decay.
}
\label{fig:flare}
\end{figure}

Obvious variability was observed only from CFHT-BD-Tau\,1, a
nonaccreting M7V BD, which displayed an X-ray (coronal) flare during
$\sim$30\% of the observation period (Fig.~\ref{fig:flare}). The shape
of this light curve, a fast rise followed by exponential decay, is
typical of X-ray flares from
young stellar objects \citep[e.g.,][]{imanishi03,favata05}. We fit the
non-zero count rates with a quiescent level plus an exponential rise and decay
(Fig.~\ref{fig:flare}). We estimate from this light curve fit
that the quiescent level contributes to 54\% of the total counts
(quiescent+flare) detected from CFHT-BD-Tau\,1. 
The quiescent emission level is
$1.8\pm0.5$\,counts\,ks$^{-1}$, the flare amplitude is
$13.3\pm2.9$\,counts\,ks$^{-1}$, i.e.\ about 8 times the quiescent
level. The time scales of the rise and decay phase are $\tau_{\rm
rise}=15\pm5$\,min and $\tau_{\rm decay}=47\pm11$\,min, respectively. Both
values are in the lower range of time scales observed in
the systematic study of X-ray flares from low-mass young stellar objects
in the $\rho$ Ophiuchi star-forming region with
\cxo~\citep{imanishi03}. In this latter star-forming region,
\citet{ozawa05} observed with \xmm~from GY310, a young M8.5V BD
\citep{wilking99}, an X-ray flare with a similar amplitude (peaking
$~6$ times above the quiescent level) and decay time scale, but with a
rise phase about 5 times longer. 
Assuming that the count rate scales linearly with the X-ray luminosity
\citep[e.g.,][]{mitra05}, we compute, from the average X-ray
luminosity of CFHT-BD-Tau\,1 (see Table~\ref{table:spectral_fit}),
that the total energy released by the CFHT-BD-Tau\,1 flare is about
$2\times10^{33}$\,ergs.

The X-ray light curves of the other BDs show no large variability
(Fig.~\ref{fig:lc}). Two BDs, observed at two differents epochs, show
variations in X-ray luminosity lower than a factor of 2
(Fig.~\ref{fig:lc2}; see below Tables~\ref{table:spectral_fit} and
\ref{table:quantiles}). We looked for a possible modulation of the
EPIC light curve of MHO\,4 (see Fig.~\ref{fig:lc2}) using a
Lomb-Scargle periodogram \citep[e.g.,][]{flaccomio05}, but we found no
statistically significant periodic signal. The BD count rates observed
in the XEST are mainly representative of a ``quiescent'' emission,
however (micro)flaring activity could be present even though
unresolved in the light curves as observed in active stars \citep{guedel03}.

\begin{figure*}[!ht]
\centering
\begin{tabular}{@{}c@{}c@{}c@{}}
\includegraphics[height=0.75\columnwidth,angle=-90]{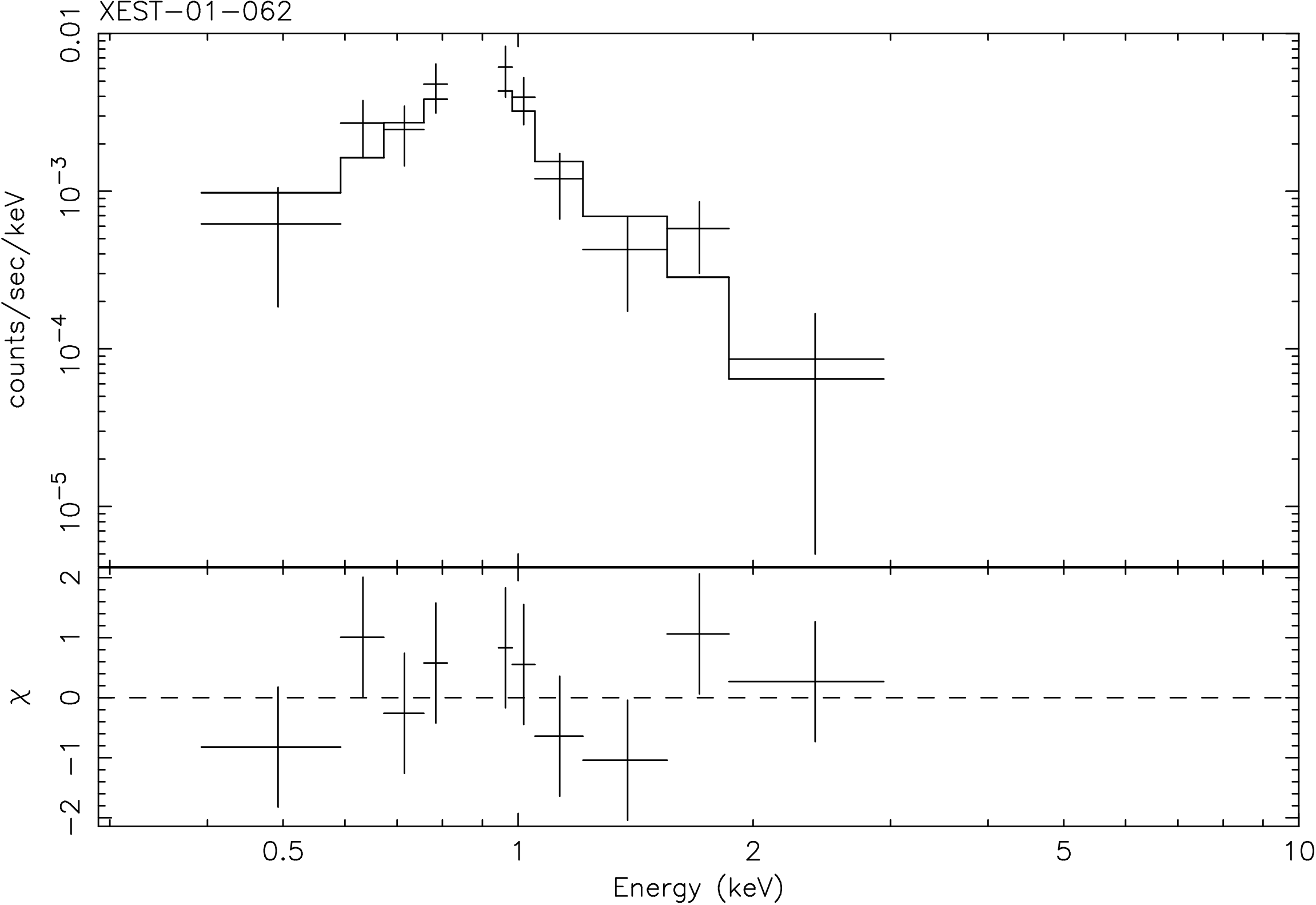}
& \includegraphics[height=0.75\columnwidth,angle=-90]{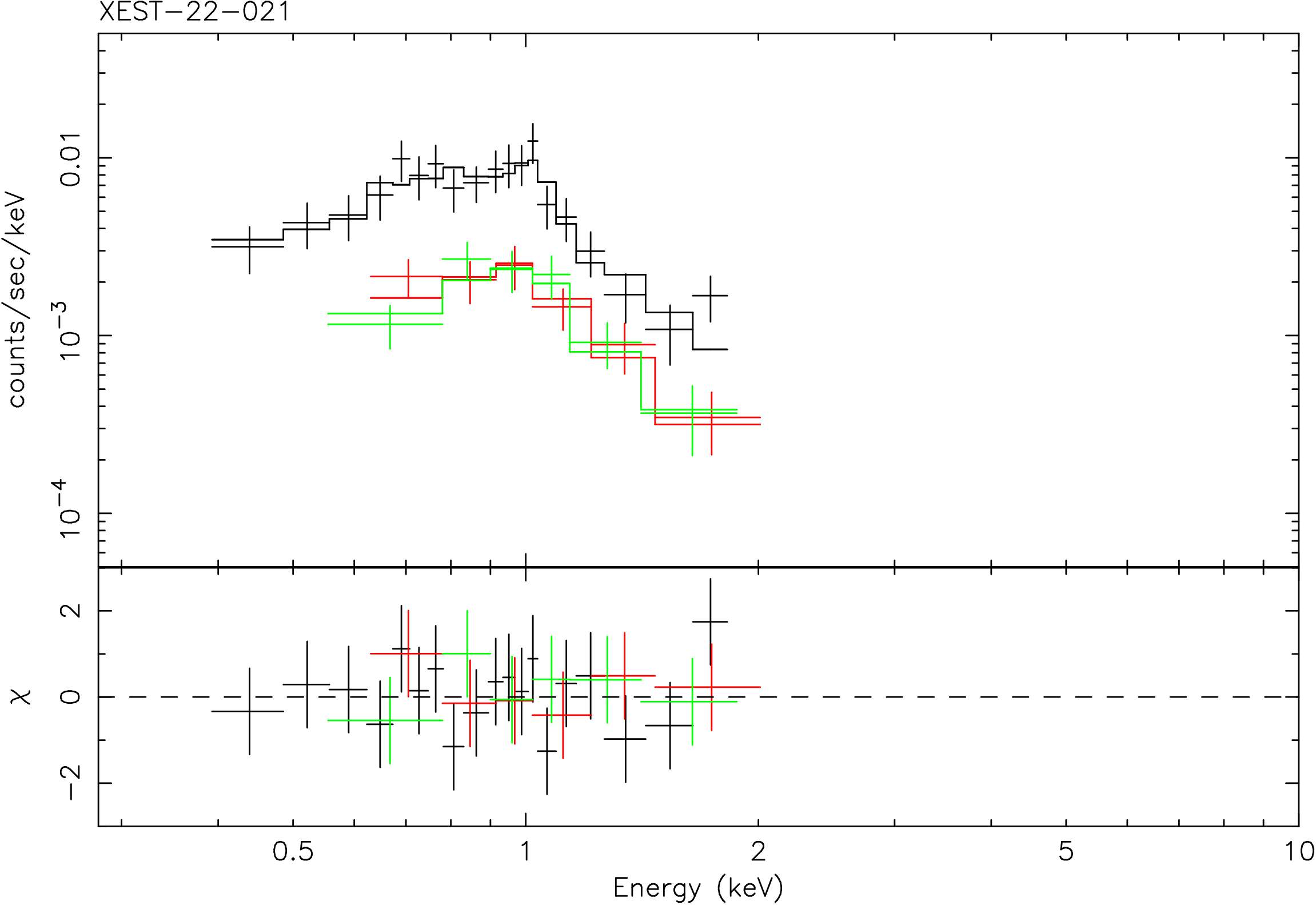} &
\includegraphics[height=0.5\columnwidth,angle=-90]{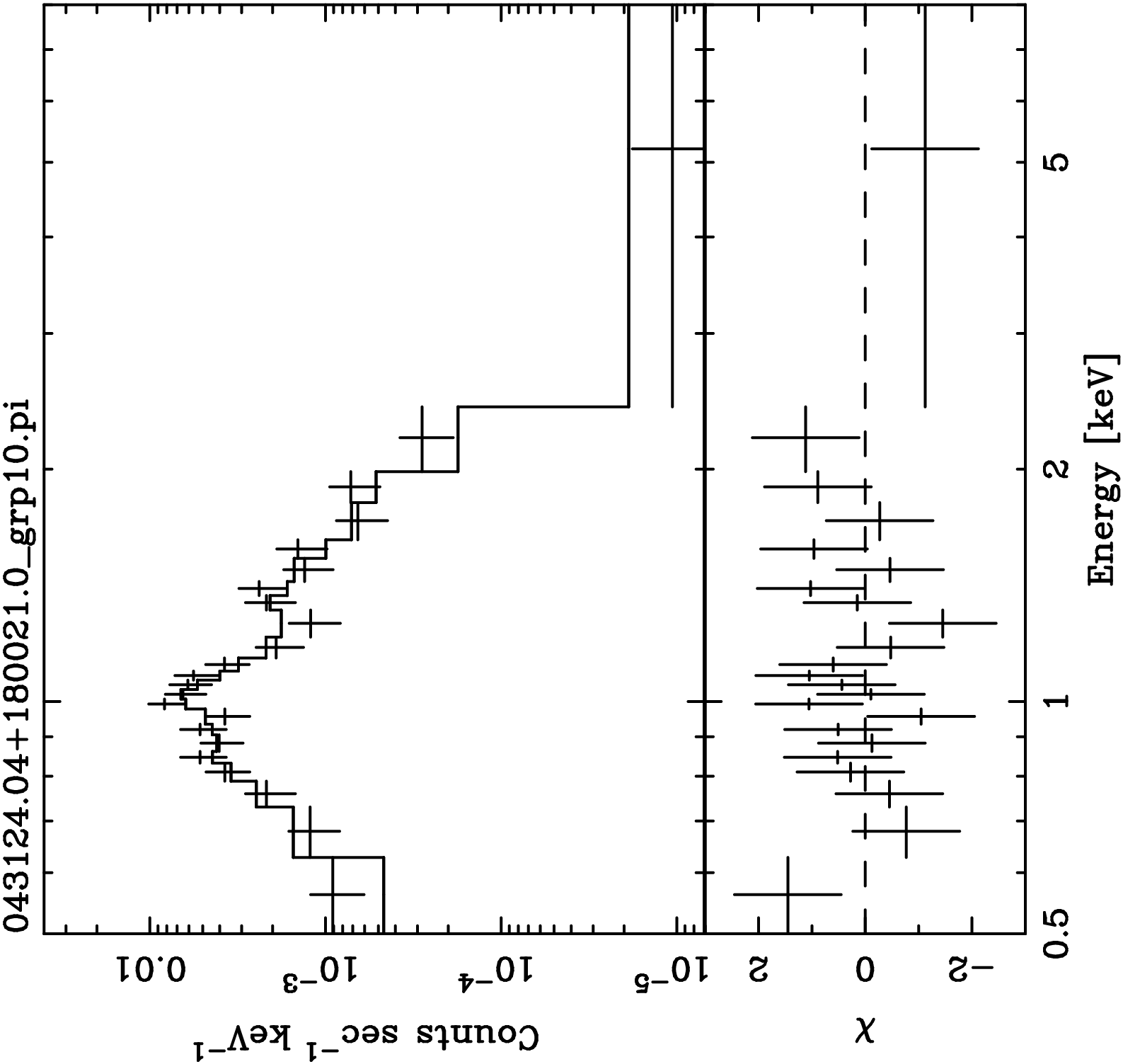}\\
\includegraphics[height=0.75\columnwidth,angle=-90]{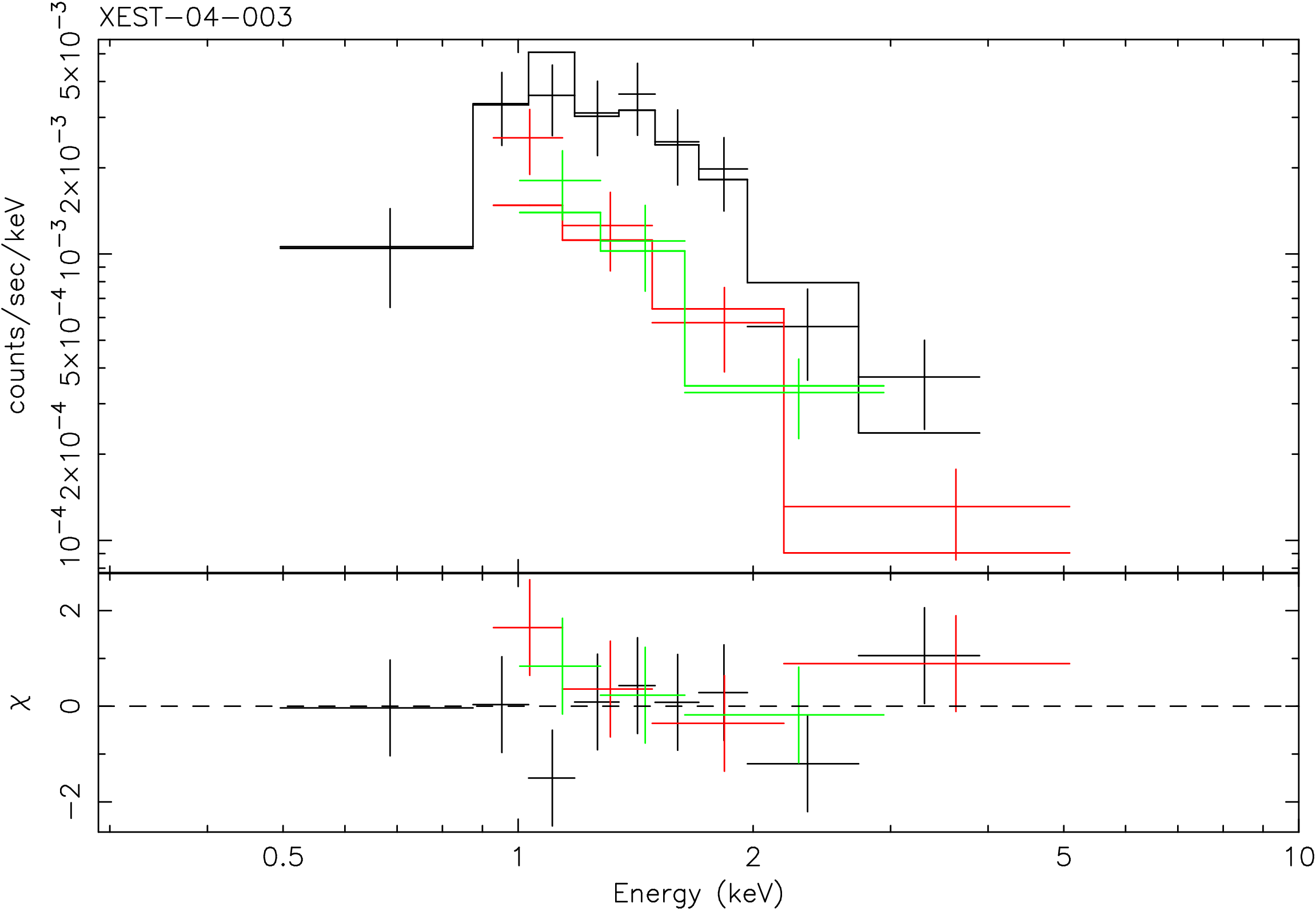}
& \includegraphics[height=0.75\columnwidth,angle=-90]{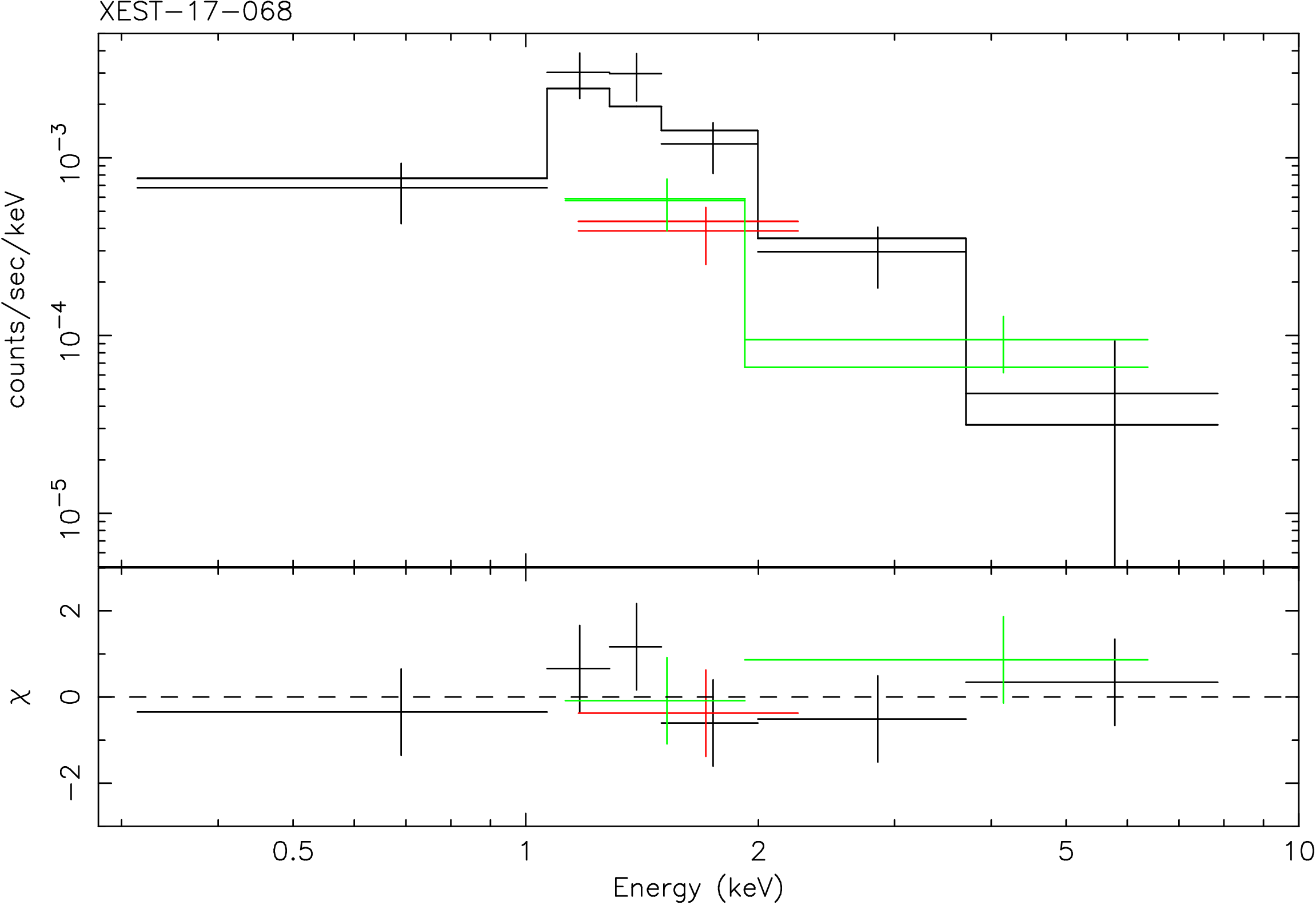}
& \includegraphics[height=0.5\columnwidth,angle=-90]{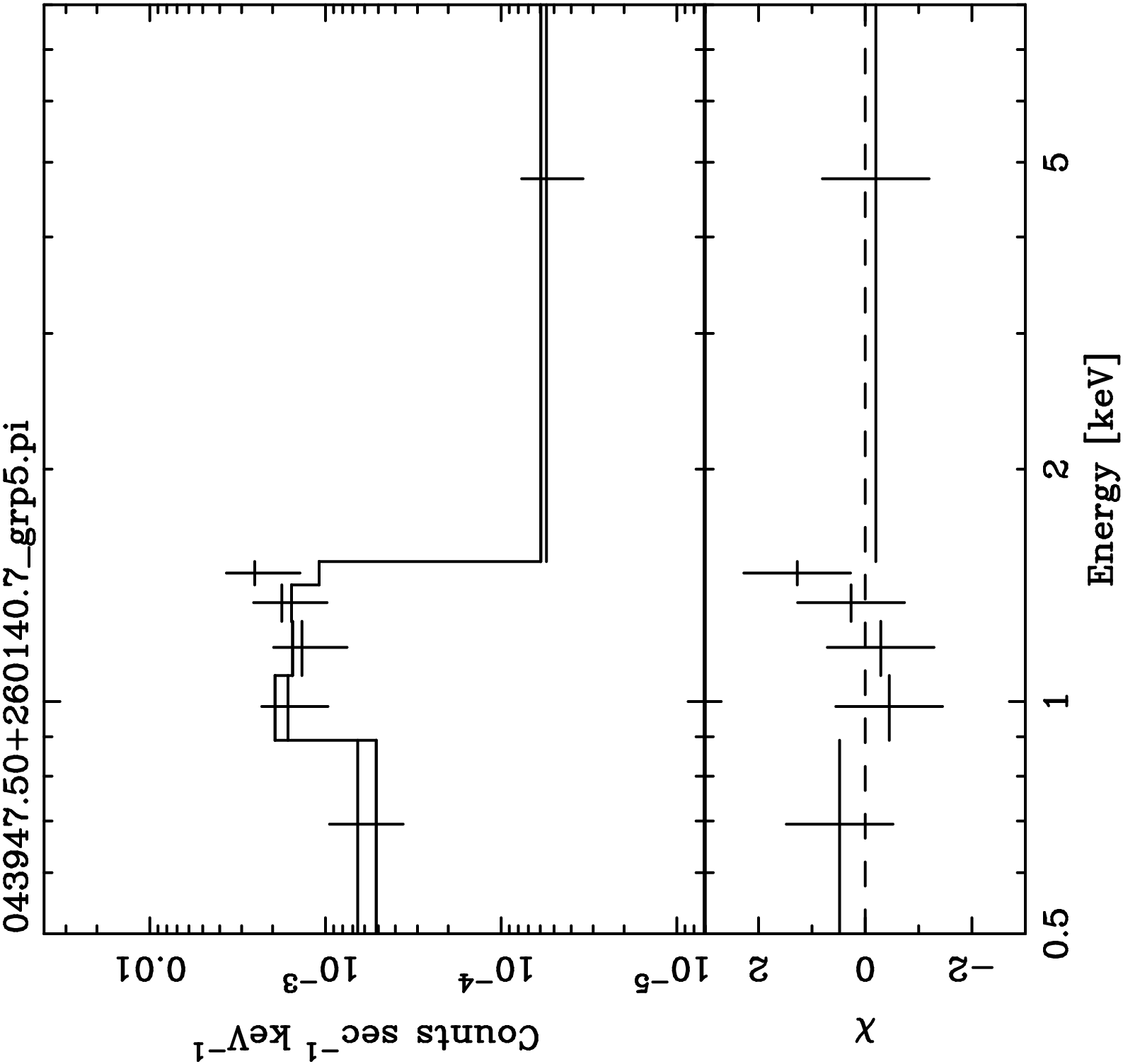}
\end{tabular}
 \caption{X-ray spectra of TMC BDs. From top to bottom and left to right:
2MASS\,J0422 (pn spectrum), MHO\,4 (\xmm~and \cxo~spectra), CFHT-Tau\,5 (\xmm~spectra),
CFHT-BD-Tau\,1 (\xmm~spectra), CFHT-BD-Tau\,4 (\cxo~spectra). Black,
red, and green code for \xmm/EPIC pn, MOS1, and MOS2 spectra,
respectively. The continuous lines show our best fits obtained with an
absorbed single-temperature or two-temperature (MHO\,4 spectra) plasma
model (see Table~\ref{table:spectral_fit}).  
}
\label{fig:spectra}
\end{figure*}

\begin{table*}[!ht]
\caption{Spectral properties of BDs obtained from spectral fitting. To
fit the spectra (Fig.~\ref{fig:spectra}), we used a {\tt WABS}
absorption model \citep{morrison83} combined with one or two {\tt
MEKAL} optically thin thermal plasma model \citep{mewe95} with 0.3 times
the solar elemental abundances. {\tt MEKAL} plasma models were computed
rather than
interpolated from a pre-calculated table. We used $\chi^2$ statistics with
standard weighting. Col.~(4) gives the net source counts collected by
the instrument given in Col.~(3). Confidence ranges at the 68\% level
($\Delta\chi^2=1$; corresponding to $\sigma=1$ for Gaussian
statistics) are given in parentheses. The value of reduced $\chi^2$ and $\nu$,
the degrees of freedom, are indicated in Col.~(10). The emission
measures in Cols.~(8) and (9) and the X-ray luminosity in the
0.5--8\,keV energy range corrected for absorption in Col.~(11) were
computed assuming a distance of 140\,pc for the TMC. 
The X-ray fractional luminosity, $\eta=\log (L_{\rm X}/L_*)$, is
given in the last column. For CFHT-BD-Tau\,1, the second line gives an
estimate of the quiescent X-ray luminosity derived for the light curve
fit (see \S\ref{lc_analysis}).}
\label{table:spectral_fit}
\centering
\begin{tabular}{@{}llcclccccccc@{}}
\hline\hline
& \multicolumn{1}{c}{BD name} & Instr. & $N$ & \multicolumn{1}{c}{$N_{\rm H}$}          & $kT_1$ &
$kT_2$ & $EM_1$ & $EM_2$ & $\chi^2_\nu$ ($\nu$) & $L_{\rm X}$ & 
$\eta$ \\
\#& &                             &      & \multicolumn{1}{c}{$10^{21}$\,cm$^{-2}$} & \multicolumn{2}{c}{keV} &
\multicolumn{2}{c}{$10^{51}$\,cm$^{-3}$}  & & $10^{28}$\,erg\,s$^{-1}$\\
\multicolumn{1}{c}{(1)} & \multicolumn{1}{c}{(2)} & (3) & (4) &
\multicolumn{1}{c}{(5)} & (6) & (7) & (8) & (9) & (10) & (11) & (12) \\
\hline
1   & 2MASS\,J0422    & pn    & 102  &  1.7 (0.9--2.8)  &   0.7 (0.6--0.8) & \dotfill        & ~2 & \dotfill  & 0.83 (07) &  02.0 & -3.5 \\
3a  & MHO\,4          & EPIC  & 571  &  1.2 (0.7--1.7)  &   0.4 (0.3--0.5)  & 1.1 (1.0--1.2)  &  ~5  &  6     & 0.56 (26) &  08.3  & -3.3\\
3b  & MHO\,4          & ACIS-I& 271  &  ~~0  (0--0.2)   &   0.5 (0.4--0.6)  & 1.1 (1.1--1.3)  &  ~4  &  6     & 0.85 (18) &  08.0  & -3.4\\
4a  & CFHT-Tau\,5     & EPIC  & 314  &  4.9 (3.7--7.7)  &   1.5 (1.2--1.9)  & \dotfill        & 14  & \dotfill& 0.82 (13) &  11.8  & -3.4\\
5~  & CFHT-BD-Tau\,1  & EPIC  & 206  &  6.0 (3.9--10.2) &   1.7 (1.1--2.6)  & \dotfill        & 17  & \dotfill& 0.59 (06) &  14.3  & -2.7\\
    &  &  &  &  &   &  &  &  &  &  07.7  & -2.9\\
8~  & CFHT-BD-Tau\,4  & ACIS-I&  33  &  8.6 (5.6--13.9) &   0.5 (0.2--0.9)  & \dotfill        & 27  & \dotfill& 0.74 (3)  &  24.3  & -3.0\\
\hline
\end{tabular}
\end{table*}

\section{X-ray spectral properties of the TMC brown dwarfs}
\label{spectral_properties}


        \subsection{X-ray spectra of the brightest X-ray sources}
	\label{spectral_fitting}

In total, we have 11 X-ray observations of the 9 BDs detected in the XEST
(Table~\ref{table:detections}). We performed spectral fits for
the two \cxo~sources, and for these \xmm~sources with more than $\sim$100
counts (Fig.~\ref{fig:spectra}). The computation of \xmm~source
spectra with associated
redistribution matrix files (RMFs) and auxiliary response files (ARFs)
is detailed in \citet{guedel06b}, for \cxo~observations see online
Appendix~\ref{log}. Spectral fitting was performed with {\tt XSPEC}
\citep[version 11.3;][]{arnaud96} using one or two optically thin
thermal plasma models \citep[{\tt MEKAL};][]{mewe95} with 0.3 times
the solar elemental abundances, combined with an X-ray absorption
model \citep[{\tt WABS};][]{morrison83}, i.e.\ the same model as used
in COUP to allow us to directly compare with the result from
\citet{preibisch05}. For \xmm~spectra, we fitted simultaneously
the EPIC~pn, MOS1, and MOS2 spectra.

Table~\ref{table:spectral_fit} gives our best fit parameters obtained
for 4 BDs in 3 \xmm~and 2~\cxo~observations. The \xmm~and \cxo~spectra
of the brightest source, MHO\,4, are better described with a
two-temperature plasma. The two plasma temperatures found separately
from \xmm~and \cxo~spectra are consistent with each other. For CFHT-Tau\,5,
the hydrogen column density, $N_{\rm H}$, determined by one
temperature spectral fitting is 3 (with an acceptable range of 2--4)
times lower than the value
predicted from $A_{\rm V}=9.2\pm0.8$\,mag \citep{guieu06} using the
$N_{\rm H}/A_{\rm J}$ ratio of \citet{vuong03} combined with
the $A_{\rm V}/A_{\rm J}$ ratio of \citet{cardelli89} for $R_{\rm
V}=3.1$, which leads to $N_{\rm H} =  1.6 \times 10^{21} A_{\rm
V}$\,cm$^{-2}$\,mag$^{-1}$. This may point out a non-canonical
dust-to-gas ratio on the line of sight towards this object.

        \subsection{Quantile analysis of the faintest X-ray sources}

For the faintest X-ray sources, we used a quantile analysis, the new
spectral classification technique for X-ray sources proposed by
\citet{hong04}. This technique uses a quantile diagram based on the
X-ray colours: $x\equiv\log (Q_{50}/(1-Q_{50}))$, and $y\equiv 3
\times Q_{75}/Q_{25}$, where $Q_{\rm x} \equiv (E_{\rm x\%}-E_{\rm
i})/(E_{\rm f}-E_{\rm i})$, with $E_{\rm x\%}$ the energy where the
net counts are x\% of the source net counts in the $E_{\rm
i}$--$E_{\rm f}$ energy range. We take here $E_{\rm i}=0.5$\,keV and
$E_{\rm f}=7.3$\,keV.

\begin{figure*}[!ht]
\centering
\begin{tabular}{@{}cc@{}}
\includegraphics[width=\columnwidth]{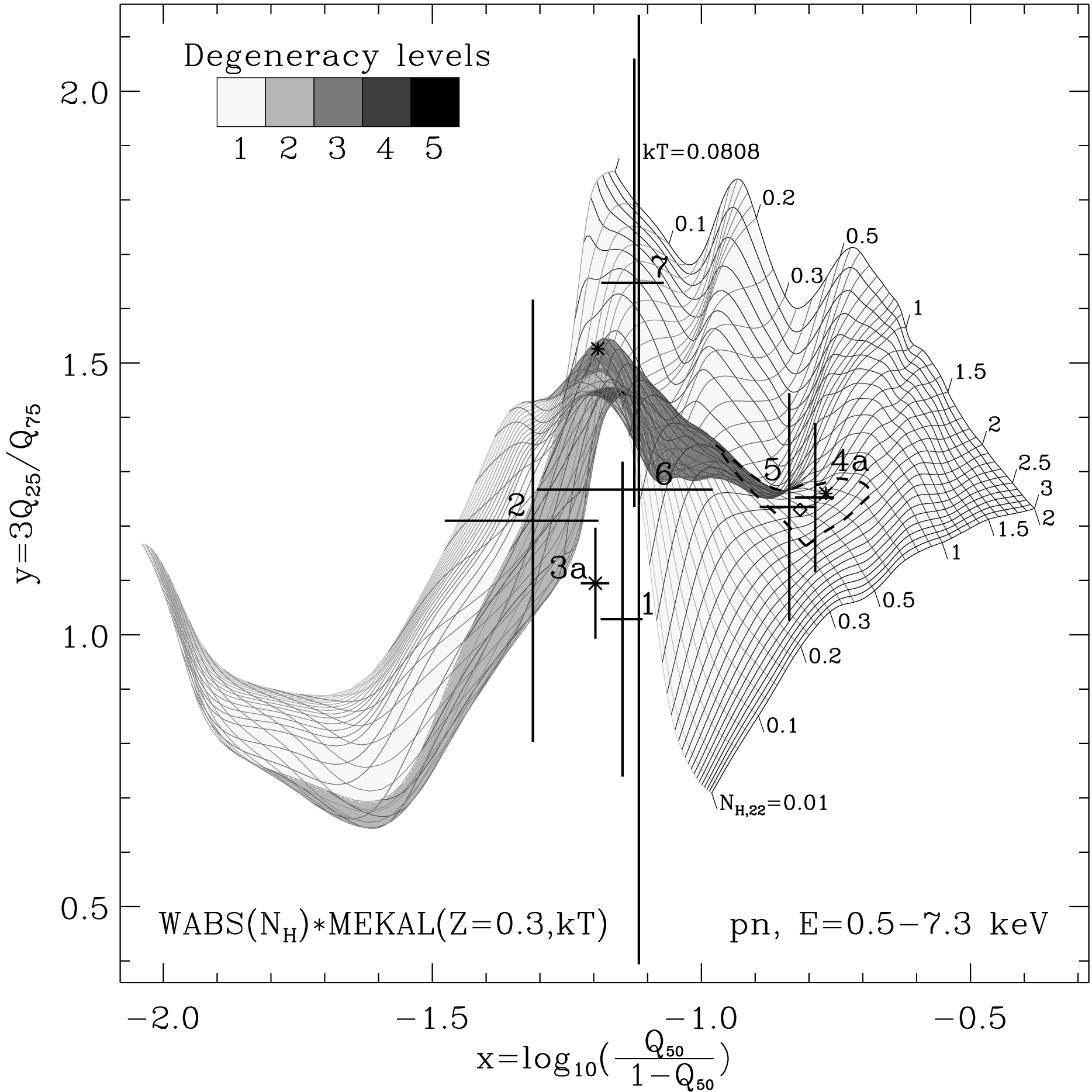} &
\includegraphics[width=\columnwidth]{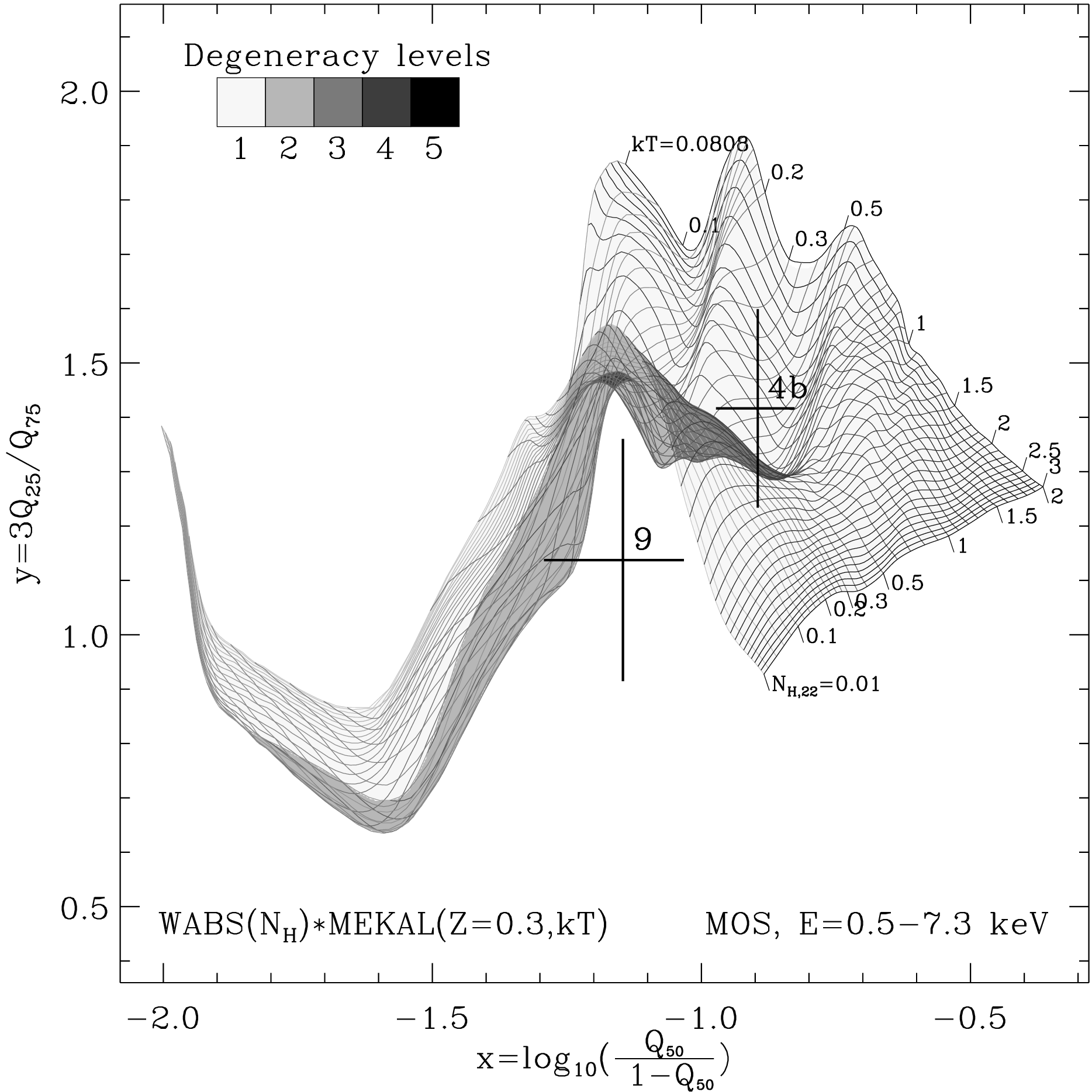}
\end{tabular}
\caption{Quantile diagram for pn (left) and MOS1+MOS2 (right). The
grid of hydrogen column density and plasma temperature were computed
following the prescriptions of \citet{hong04} using in {\tt SHERPA} an
absorption model \citep[{\tt WABS};][]{morrison83} multiplied with a
single optically thin thermal plasma model \citep[{\tt
MEKAL};][]{mewe95} with 0.3 times the solar
elemental abundances. The background map in grey levels indicates the degeneracy
level of grid parameters (see online Appendix~\ref{degeneracy}). The MOS1+MOS2
quantile diagram is used only for X-ray sources without available pn
data. In the PN quantile diagram, asterisks and diamond mark the plasma parameters
obtained from spectral fitting; the dashed line indicates the X-ray colour locus
corresponding to the parameter uncertainties of source \#4a
(diamond; Table~\ref{table:spectral_fit}). The spectrum of \#3 requires a two-temperature plasma model (see
Table~\ref{table:spectral_fit}).
}
\label{fig:QDx}
\end{figure*}

With the spectral model of \S\ref{spectral_fitting}, we calculated for
pn and MOS1+MOS2, from the Perl script provided by
\citet{hong04}\footnote{The version 1.7 of the IDL
and Perl quantile analysis softwares of \citet{hong04} is available at {\tt
http://hea-www.harvard.edu/ChaMPlane/quantile:}\,.}, a grid of hydrogen
column densities and plasma temperatures (\nh, $kT$) in {\tt
SHERPA}\footnote{{\tt SHERPA} is a part of the {\tt CIAO} package.}
and the corresponding X-ray colours. We adopted on-axis RMF and ARF;
and for MOS1+MOS2, the MOS1 and MOS2 RMFs were averaged and the ARFs
added. Fig.~\ref{fig:QDx} shows the resulting quantile diagrams. Grid
lines indicate loci of iso-column densities and iso-temperatures.

\begin{table*}[!ht]
\caption{ 
Spectral properties of BDs obtained from quantile analysis. Col.~(3)
gives the instrument name (MOS stands for MOS1+MOS2). Col.~(4) and (5)
indicate the exposure and the
net source counts collected by this instrument (i.e.\ for MOS the
exposure average and the sum of net source counts). Col.~(6) gives the
energies below which the net counts are 25\%, 50\%, and 75\% of the
source net counts in the 0.5--7.3\,keV energy range. The resulting
position in the quantile diagram (Fig.~\ref{fig:QDx}) is given in
Col.~(7). The observed optical extinctions (see Table~\ref{table:bd}) were used to
disentangle temperature double solutions in the quantile diagram (\#2 on pn), and
to estimate the hydrogen column density (Col.~8) for sources with low
constraint in the quantile diagram (\#2, \#6 and \#7 on pn, and \#9
on MOS), using the relation $N_{\rm H} =  1.6 \times
10^{21} A_{\rm V}$~cm$^{-2}$\,mag$^{-1}$ \citep{vuong03,cardelli89};
these values are between brackets in Col.~8. Where negligible optical
extinction was measured, we adopted $N_{\rm H} = 0.1 \times
10^{21}$~cm$^{-2}$. When no constraint on the plasma temperature was
obtained, we adopted 1\,keV (value between brackets in Col.~9). The
X-ray luminosity in the 0.5--8\,keV energy range corrected for
absorption in Col.~(10) was computed assuming a distance of 140\,pc
for the TMC. The X-ray fractional luminosity, $\eta=\log (L_{\rm
X}/L_*)$, is given in the last column.
}
\label{table:quantiles}
\centering
\begin{tabular}{rlccccccccc}
\hline\hline
  & \multicolumn{1}{c}{BD name} & Instr. & Exp. & $N$ & $E_{\rm
25\%}/E_{\rm 50\%}/E_{\rm 75\%}$  & ($x$,\,$y$) & $N_{\rm H}$ & $kT$ &
$L_{\rm X}$ & $\eta$\\
\#&                             &        & ks   &     & keV & & $10^{21}$\,cm$^{-2}$ & keV & $10^{28}$\,erg\,s$^{-1}$\\
\multicolumn{1}{c}{(1)} & \multicolumn{1}{c}{(2)} & (3) & (4) & \multicolumn{1}{c}{(5)} & (6) & (7) & (8) & (9) & (10) & (11)\\
\hline
2~  &   KPNO-Tau\,5  &  pn  &  17.2 &  25.3  & 0.71\,/\,0.82\,/\,1.02 & (-1.31,1.21) &  [0.1] & 0.2  & 0.8 & -4.1\\
4a  &   CFHT-Tau\,5  &  pn  &  26.1 & 118.4  & 1.08\,/\,1.45\,/\,1.89 & (-0.79,1.25) &  6.4 & 1.4  & 13.9 & -3.3\\
4b  &   CFHT-Tau\,5  &  MOS &  22.9 &  26.6  & 1.02\,/\,1.27\,/\,1.60 & (-0.90,1.42) & 11.1 & 0.5  & 20.1 & -3.2\\
5~  & CFHT-BD-Tau\,1 &  pn  &  20.6 &  64.7  & 1.11\,/\,1.36\,/\,1.98 & (-0.84,1.24) &  4.5 & 1.5  & 12.4 & -2.7\\
6~  & CFHT-BD-Tau\,3 &  pn  &  23.9 &  11.6  & 0.79\,/\,0.98\,/\,1.19 & (-1.12,1.27) & [0.1]& [1]  &  0.6 & -3.7\\
7~  &   CFHT-Tau\,6  &  pn  &  14.2 &  31.8  & 0.92\,/\,0.98\,/\,1.26 & (-1.13,1.65) & [0.7]& 0.1  &  2.8 & -3.5\\
9~  &   2MASS\,J0455 &  MOS & 123.9 &   5.0  & 0.75\,/\,0.95\,/\,1.15 & (-1.15,1.14) & [0.1]& [1]  &  0.2 & -4.5\\
\hline
\end{tabular}
\end{table*}

\begin{table}[!t]
\caption{Upper limits to the BD X-ray luminosities. The hydrogen column
density in unit of $10^{21}$\,cm$^{-2}$ given in Col.~(3) is obtained
from the optical extinction using the relation $N_{\rm H,21} =  1.6
\times A_{\rm V}$~cm$^{-2}$\,mag$^{-1}$ \citep{vuong03,cardelli89}. Exposure times in
Col.~(4) are for summed EPIC (pn+M1+M2) data, in units equivalent for
a pn-on-axis observation. Col.~(5) and (6) give the upper limits at
the 95\% confidence level for net counts in the 0.5--2\,keV energy
range and the X-ray luminosity in the 0.5--8\,keV energy range
corrected for absorption, respectively. The upper limit to the
X-ray fractional luminosity, $\eta=\log (L_{\rm X}/L_*)$, is given in
the last column.
}
\label{table:ul}
\centering
\begin{tabular}{@{}lcccccc@{}}
\hline\hline
\multicolumn{1}{c}{BD name} & XEST & $N_{\rm H,21}$ & Exp. & $N$ &
$L_{\rm X,27}$ & $\eta$ \\
& \# &  \,cm$^{-2}$ & ks & cnts & erg\,s$^{-1}$\\
\multicolumn{1}{c}{(1)}     & (2)      & (3) & (4) & (5) & (6) & (7)\\
\hline
2MASS\,J0414    & 20  &  1.1  &  44.7   &  115.5  &  15.8 & -3.6\\
2MASS\,J0421    & 11  &  4.8  &  38.4   &  19.5   &  6.1  & -3.3\\
KPNO-Tau\,2     &23+24&  0.6  &  22.1   &  30.2   &   6.0 & -3.6\\
KPNO-Tau\,4     & 02  &  3.9  &  51.8   &  25.8   &   5.0 & -3.5\\
KPNO-Tau\,6     & 14  &  0.1  &  26.8   &  31.1   &   6.4 & -3.3\\
KPNO-Tau\,7     & 14  &  0.1  &  25.6   &  16.9   &   2.5 & -3.8\\
KPNO-Tau\,9     &08+09&  0.1  &  51.5   &  37.9   &   2.8 & -3.1\\
CFHT-BD-Tau\,2  & 08  &  0.1  &  33.9   &  25.5   &   2.9 & -4.0\\
\hline
\end{tabular}
\end{table}

We noted that the emission lines of the spectral model produce folds
in the parameter grid, introducing some degeneracy of the X-ray
colours; i.e.\ several different values of (\nh, $kT$) have the same
$E_{\rm 25\%}$, $E_{\rm 50\%}$, $E_{\rm 75\%}$, and hence the same
X-ray colours. The degeneracy of the quantile diagram is studied in details in
online Appendix~\ref{degeneracy}. The result of this analysis is the
degeneracy map plotted in grey levels in Fig.~\ref{fig:QDx}, which 
indicates the number of different (\nh, $kT$) values at each position in the
quantile diagram.

We calculated for all \xmm~sources pn (or MOS1+MOS2) X-ray colours with background
subtraction (Table~\ref{table:quantiles}) and errors following the
prescriptions of \citet{hong04}. The MOS1+MOS2 quantile diagram is
used only for X-ray sources without available pn data. 
We checked for sources \#1, \#4a and \#5
that the plasma parameters derived from the pn X-ray colours are
consistent with the plasma parameters obtained from spectral fitting
marked with asterisks and diamond in Fig.~\ref{fig:QDx}. For comparison
purposes, we indicate in Fig.~\ref{fig:QDx} the X-ray colour locus
corresponding to the parameter uncertainties found in the fitting of
the \xmm/EPIC spectra of source \#4a, the brightest X-ray source
(Table~\ref{table:spectral_fit}).
The spectrum of source~\#3 requires to be fitted by a two-temperature
plasma model, and has X-ray colours which are outside the grid
computed with a single-temperature plasma model.  
Optical extinctions were used to disentangle temperature double
solutions (source \#2 on pn; see online Appendix~\ref{degeneracy}), and to
estimate the hydrogen column density for sources with low constrained
X-ray colours (sources \#2, \#6 and \#7 on pn, and \#9 on MOS) using
the relation $N_{\rm H} =  1.6 \times 10^{21} A_{\rm
  V}$~cm$^{-2}$\,mag$^{-1}$ \citep{vuong03,cardelli89}. Where negligible
optical extinction was measured, we adopted $N_{\rm H} = 0.1 \times
10^{21}$~cm$^{-2}$. When there was no constraint on the plasma
temperature, we adopted $kT=$1\,keV (Table~\ref{table:quantiles}). 

Then, in {\tt XSPEC} we
computed from these plasma parameters and the source's ARF and RMF the
X-ray luminosity, corrected for absorption in the 0.5--8\,keV energy
range needed to reproduce the observed net count rate in the
0.5--7.3\,keV energy range. For comparison, the X-ray luminosities
obtained by spectral fitting of sources \#5a and \#6 and the ones
obtained by quantile analysis show only a difference of $\sim$0.1\,dex.

In the following, for BDs observed and detected twice, we use the
logarithmic average of their X-ray luminosity. For CFHT-BD-Tau\,1, 
we use the quiescent X-ray luminosity.

        \subsection{Upper limit estimate of the X-ray luminosities for
the undetected brown dwarfs}

We calculated an upper limit to the X-ray luminosity of each
undetected  BD in the 0.5--8~keV band (see Table~\ref{table:ul}). 
Counts within 10\arcsec~of the  optical/near infrared position were
extracted from the summed EPIC  soft-band (0.5--2\,keV) image. The
expected number of counts in the absence of emission from the BD was
determined from the identical region of the corresponding
``reconstructed image'', which is composed of background and detected
sources output by the {\it XMM-Newton Science Analysing System} ({\tt
SAS}) source detection  algorithm, {\tt EMLDETECT}
\citep{guedel06b}. An upper limit, at the 95\% confidence level, to
the number of counts in the region from the BD was computed using the
prescription of \citet{kraft91}, which accounts for Poissonian
fluctuations in the counts from the background, and those from the
BD\footnote{The BD 2MASS\,J0414 is located on the highly structured
PSF wings of the bright X-ray source V773\,Tau, which are not
well-modeled, and consequently the contribution of V773\,Tau in the
extraction region of this BD is not well estimated. Therefore, we
assumed that the number of counts observed in the BD extraction region
($6\arcsec$-radius here instead of $10\arcsec$-radius) contains a
negligible number of counts from the BD, and so the number of observed
counts is identical to the number expected from background alone.}. 
This results in an upper limit to the sum of counts collected by three different
detectors. The expected  number of counts collected by
detector $i$, $N_{\rm i}$, from a source of flux  $f_{\rm X}$ is dependent on the
fraction of the total source counts that  were collected within the
extraction region, $\epsilon_{\rm i}$, the exposure time, $t_{\rm
exp,i}$, and the source energy flux required to produce a  count rate
of 1~count\,s$^{-1}$, $K_{\rm i}$, such that $N_{\rm i} =
\epsilon_{\rm i} \times t_{\rm exp,i} \times f_{\rm X}/ K_{\rm i}$. 
Therefore, the upper limit to the source  energy flux is
computed from the upper limit to the sum of source counts in all the
detectors, $N = \sum_{\rm i} N_{\rm i}$, as $f_{\rm X} = N / \sum_{\rm i}
\epsilon_{\rm i} \times t_{\rm exp,i} / K_{\rm i}  $. A model source,
constructed at the position of the BD from the calibration point
spread function used by {\tt EMLDETECT}, was used to calculate
$\epsilon_{\rm i}$, and convolved with exposure images generated by the {\tt
SAS} command {\tt EEXPMAP} to compute the effective on-axis exposure
time at the BD position, $ t_{\rm exp,i}$, on each detector. The
on-axis response of each detector was generated using the {\tt SAS}
command {\tt ARFGEN} and the appropriate canned response matrix. 
$K_{\rm i}$ was calculated in {\tt XSPEC} assuming an isothermal
coronal plasma of temperature 1\,keV (11.6\,MK) with metallicity 0.3 times
that of the solar corona \citep{anders89}, and a hydrogen column
density obtained from the optical extinction using the relation
$N_{\rm H} =  1.6 \times 10^{21} A_{\rm V}$~cm$^{-2}$\,mag$^{-1}$
\citep{vuong03,cardelli89}. Upper limits to X-ray luminosities in the 0.5--8\,keV
energy range were calculated assuming a distance of 140\,pc to the
TMC.

The median of the upper limits of our sample is about
$5\times10^{27}$\,erg\,s$^{-1}$. However, the deep archival observation
of SU\,Aur helped to detect 2MASS\,J0455 down to half this 
luminosity level.

{
\begin{figure}[!ht]
\centering
\includegraphics[height=0.98\columnwidth,angle=90]{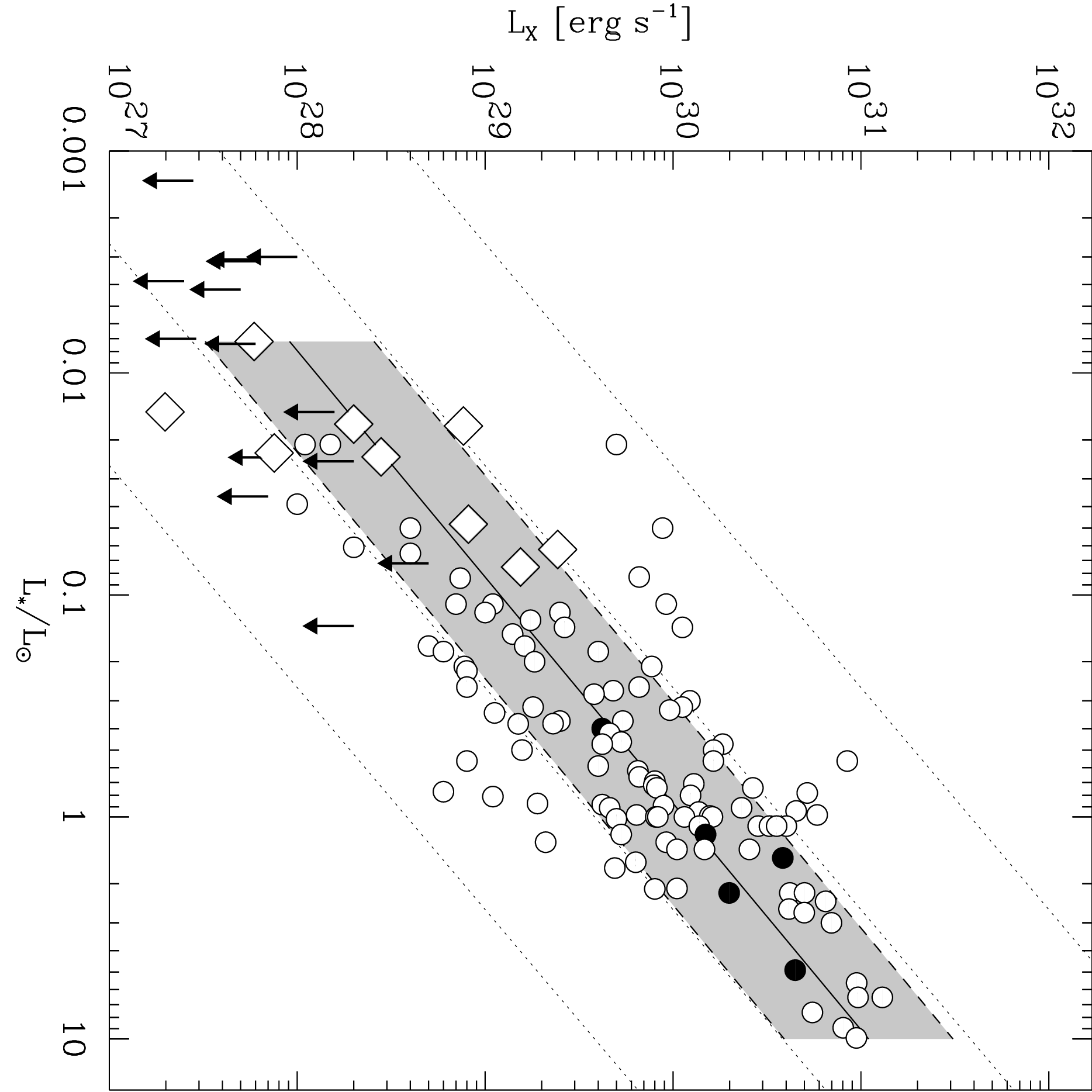}
 \caption{X-ray luminosity vs.\ bolometric luminosity for the TMC
   BDs (diamonds; and arrows for upper limits) and
   the TMC members detected in the XEST \citep[white and black dots
represent low-mass stars and protostars, respectively;][]{guedel06b}. The
dotted lines indicate from bottom to top an X-ray fractional
   luminosity, $\eta=\log (L_{\rm X}/L_*)$, of $-5$, $-4$, $-3$,
   $-2$. The grey stripe shows
a linear regression fit (continuous line) and standard deviation
(dashed lines) for detected objects with bolometric luminosities lower
than 10\,L$_\odot$.
}
\label{fig:lx_lbol}
\vspace{0.5cm}
\centering
\includegraphics[height=0.98\columnwidth,angle=90]{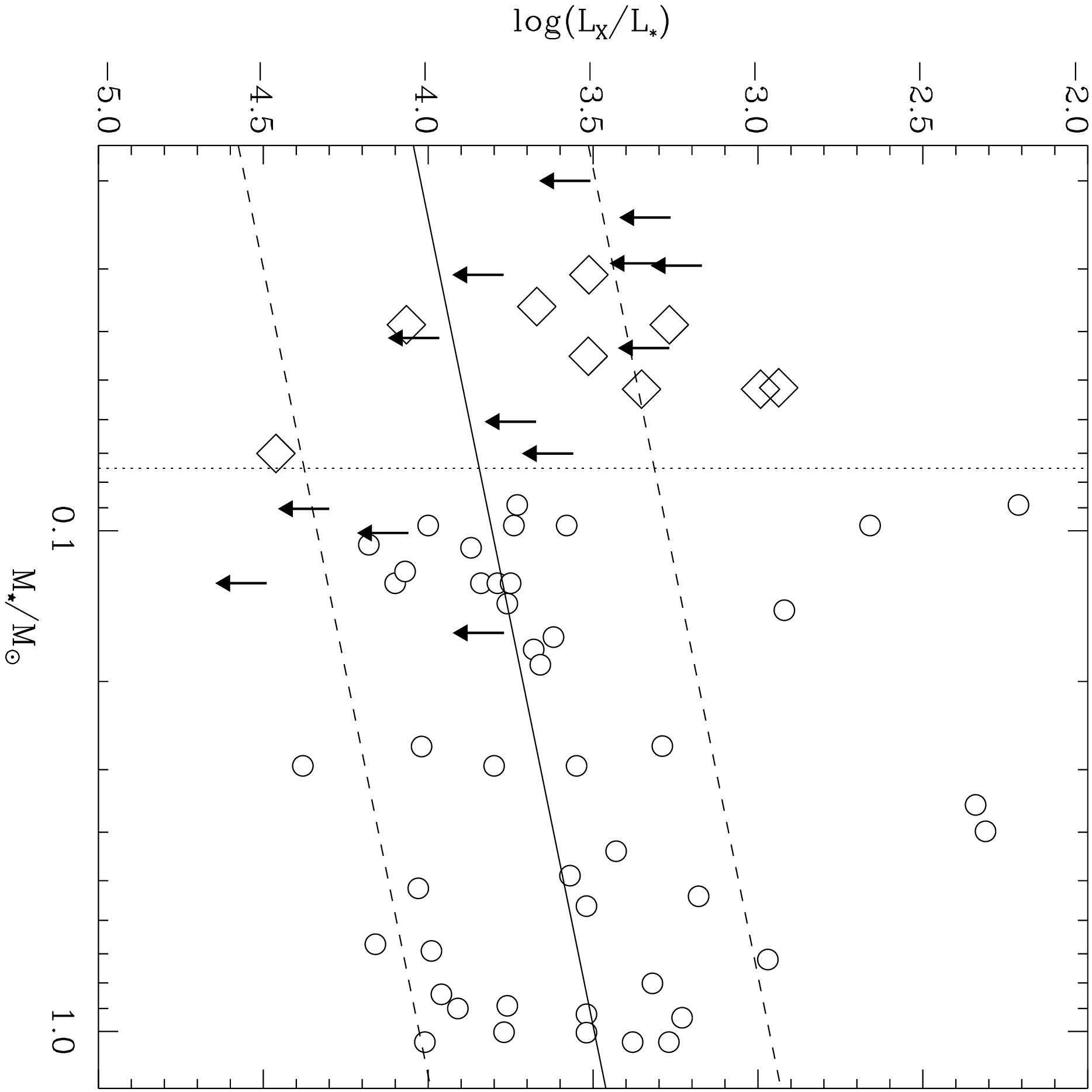}
 \caption{X-ray fractional luminosity vs.\ mass for the young
   BDs of the TMC and single TMC members of the XEST. The
symbols are as in Fig.~\ref{fig:lx_lbol}. The vertical dotted lines
indicates the stellar/substellar boundary. The solid and dashed lines show
a linear regression fit and standard deviation, respectively.  
}
\label{fig:eta_mass}
\end{figure}

\begin{figure}[!ht]
\centering
\includegraphics[height=0.98\columnwidth,angle=90]{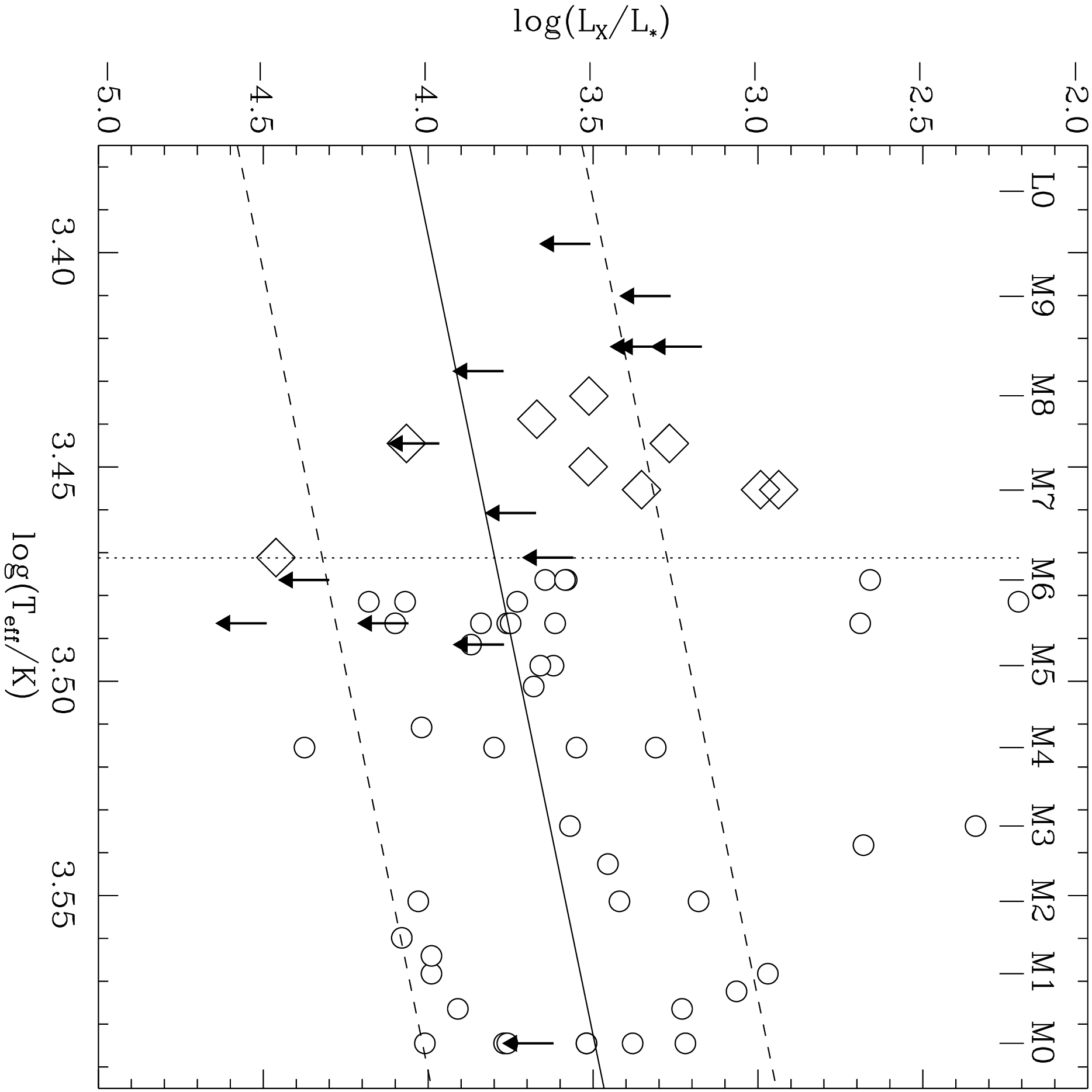}
 \caption{X-ray fractional luminosity vs.\ effective temperature for the young
   BDs of the TMC and single TMC members of the XEST. The
symbols are as in Fig.~\ref{fig:lx_lbol}. The vertical dotted lines
indicates the stellar/substellar boundary. The solid and dashed lines show
a linear regression fit and standard deviation, respectively.  
}
\label{fig:eta_teff}
\centering
\vspace{1.6cm}
\centering
\includegraphics[height=0.98\columnwidth,angle=90]{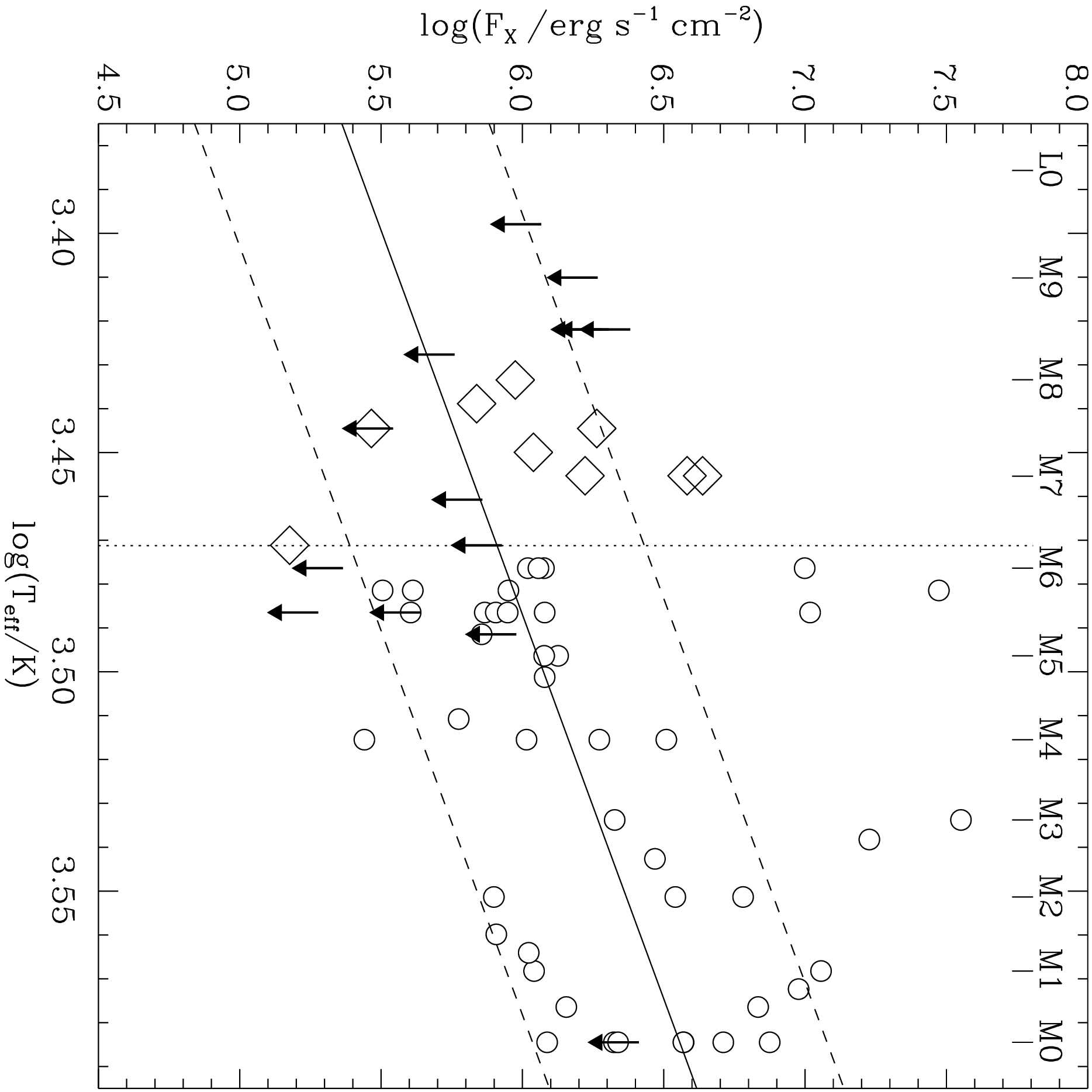}
 \caption{X-ray surface flux vs.\ effective temperature for the young
   BDs of the TMC and single TMC members of the XEST with spectral
type M0 or later. The symbols are as in Fig.~\ref{fig:lx_lbol}. The
solid and dashed lines show a linear regression fit and standard
deviation, respectively.
}
\label{fig:fx_teff}
\end{figure}
}
\section{X-ray luminosities of the XEST brown dwarfs compared to the XEST
low-mass stars}
\label{BD_vs_stars}

To put the X-ray luminosity properties of the XEST BDs into context,
we compare them to the X-ray luminosity of low-mass stars in the XEST
\citep[see][]{guedel06b}. 
For objects observed and detected twice, we use the
logarithmic average of their X-ray luminosity.
For TMC members that are unresolved multiple stars
in X-rays, we considered the total bolometric
luminosities. Fig.~\ref{fig:lx_lbol} shows, for the XEST low-mass
stars \citep[$L_* \le 10$\,L$_\odot$, which corresponds at
3\,Myr to $M \le 2$\,M$_\odot$;][]{siess00}, protostars (Class~I
sources) and BDs, the X-ray luminosities\footnote{For the low-mass
(proto)stars, we use the X-ray luminosities in the 0.3--10\,keV energy
band derived from the DEM method in \citet{guedel06b}. For a BD plasma with 0.3 times
the solar elemental abundances and a typical temperature of 1\,keV, the difference
of X-ray luminosity in the 0.3--10\,keV energy band and in the
0.5--8\,keV energy band is only 0.06 dex, and hence can be neglected.}
compared to the bolometric luminosities. Considering only the X-ray
detected BDs and low-mass (proto)stars, we determined with the parametric
EM (Expectation-Maximization) algorithm implemented in {\tt ASURV}
\citep{feigelson85} the following linear regression fit: $\log(L_{\rm
X}/{\rm erg\,s^{-1}})= (30.06\pm0.05)+(0.98\pm0.06) \times
\log(L_*/L_\odot)$, with a standard deviation of 0.4\,dex in
$\log(L_{\rm X})$ for the low-luminosity objects
($L_*$$\le$10\,L$_\odot$). This relation is very similar to the
relations found for low-mass stars in other young clusters
\citep{feigelson99,preibisch05b}. It is consistent with a linear
relation between X-ray and bolometric luminosity characterized by
$<\!\log(L_{\rm X}/L_*)\!>\,=-3.5 \pm 0.4$ which is valid from the
low-mass stars to the substellar regime. However, the bulk of the upper
limits of undetected BDs are below this average X-ray fractional 
luminosity, which suggests a lower X-ray fractional 
luminosity for the BDs with about $L_*$$\le$0.02\,L$_\odot$.
Taking into account upper limits of undetected BDs, the median of
$\log(L_{\rm X}/L_*)$ for the XEST BDs is $-4.0$ (see
\S\ref{XEST_COUP}). The X-ray fractional luminosity of XEST BDs is hence
lower than the one of XEST low-mass stars.

To investigate the relation between X-ray fractional luminosity
and physical parameters when one moves from low-mass stars to the
substellar regime, we focus on objects with spectral type M0 or later, which
corresponds for an age of 3\,Myr to masses and luminosities lower than
about 1\,M$_\odot$ and 0.7\,L$_\odot$, respectively. We used the temperature scale given in
Table~\ref{table:bd} to convert M spectral types to effective
temperatures.
We determined source masses from effective temperature
and luminosities, using the pre-main-sequence tracks of
\citet{baraffe98}. Masses were interpolated linearly along the
isochrones. For sources located above the 1\,Myr isochrone, we
prolonged the convective tracks vertically. We attributed
0.02\,M$_\odot$ to KPNO-Tau\,4, i.e.\ the only BD above the
0.02\,M$_\odot$ track. The source IRAS\,S04301+261 (spectral type
M0), which is located below the main-sequence was discarded.
We restricted our sample to stars which are not
multiple in order not to introduce extra assumptions when distributing the
X-ray flux to unresolved components. Fig.~\ref{fig:eta_mass} shows
the X-ray fractional luminosity versus mass for this
  sample. Spearman's rank correlation coefficient computed with {\tt
  ASURV} indicates a correlation with the probability of the null
hypothesis (i.e.\ no correlation) $P(0)\le 0.06$. The EM
algorithm yields the shallow linear regression fit: $\log(L_{\rm X}/L_{\rm
bol})=(-3.5\pm0.1)+(0.3\pm0.2)\times \log(M/{\rm M_\odot})$, with a
standard deviation of 0.5\,dex in $\log(L_{\rm X}/L_*)$.
This relation is consistent with the relation found by
\citet{preibisch05b} for the T~Tauri stars ($M \le 2$\,M$_\odot$) of
the ONC. This relation implies that the X-ray fractional luminosity
decreases by a factor of about 3 from 1\,M$_\odot$ stars to
0.03\,M$_\odot$ BDs.

Fig.~\ref{fig:fx_teff} shows the
X-ray fractional luminosity versus the effective temperature for our
sample ranging from $\sim$3840\,K to $\sim$2500\,K. Spearman's rank
correlation coefficient computed with {\tt ASURV} indicates a
correlation with the probability of the null hypothesis (i.e.\ no
correlation) $P(0)\le 0.11$. The EM algorithm yields the linear
regression fit: $\log(L_{\rm X}/L_{\rm
  bol})=(-13.1\pm5.4)+(2.7\pm1.5)\times \log(T_{\rm eff}/{\rm K})$,
with a standard deviation of 0.5\,dex in $\log(L_{\rm X}/L_*)$. This
relation implies that the X-ray fractional luminosity decreases by a
factor of about 3 from hot coronae of solar-mass stars to cooler
atmospheres of M9V BDs.

We computed the X-ray surface flux, i.e.\ the X-ray
luminosity divided by the surface area of the source, which is
computed from its bolometric luminosity and effective
temperature: $F_{\rm X}=\sigma T_{\rm eff}^4 \, L_{\rm X}/L_*$, where
$\sigma$ is the Stefan-Boltzmann constant. 
Combined with the decrease of the $L_{\rm X}/L_*$ ratio with the
effective temperature, this formula implies that the X-ray surface
flux decreases with the effective temperature with a power-law slope
greater than 4. Fig.~\ref{fig:fx_teff} shows the X-ray surface flux
versus the effective temperature for our sample. The X-ray surface
fluxes range from $\sim$$10^5$ to
$\sim$$3\times10^7$\,erg\,s$^{-1}$\,cm$^{-2}$. For comparison, the
X-ray surface flux of the solar corona during the solar cycle varies
from $\sim$$2\times 10^3$ to $\sim$$8\times 10^4$\,erg\,s$^{-1}$\,cm$^{-2}$
\citep{peres04}. Spearman's rank correlation coefficient indicates a
log-log correlation with a high confidence level ($P(0)=10^{-4}$). The
EM algorithm yields:
$\log(F_{\rm X}/{\rm
erg\,s^{-1}\,cm^{-2}})=(-13.9\pm4.8)+(5.7\pm1.4)\times\log(T_{\rm
eff}/{\rm K})$, with a standard deviation of 0.5\,dex in $\log(F_{\rm
X})$. The slope of this correlation is well steeper than 4 due to the
decline of the $L_{\rm X}/L_*$ ratio with the effective
temperature. This correlation predicts that a M9V BD should have an
X-ray surface flux in its corona $\sim$10 times weaker than in an M0
star. However, this is still 4 times higher than the X-ray surface
flux of the solar corona at the solar cycle maximum \citep{peres04}. A
similar trend was found by \citet{preibisch05} for the BDs of ONC (see
\S\ref{XEST_COUP} for further comparison between the XEST and the COUP
BD sample).

\section{X-ray fractional luminosity and H$\alpha$ emission of the XEST brown dwarfs}
\label{halpha}

\subsection{Correlation between X-rays and H$\alpha$ ?}

Studying a BD sample including young BDs (in the ONC, $\rho$
Ophiuchi, and IC\,348) plus one intermediate BD (TWA\,5B)
and one old (LP\,944-20) field BD, \citet{tsuboi03} found a log-log correlation
between the X-ray fractional luminosity (value independent of the
distance assumption), and $EW({\rm H}\alpha)$ ($\log{L_{\rm
X}/L_*}=-5.3+1.5\times\log{EW({\rm H}\alpha)}$), and noted that BDs
with $EW({\rm H}\alpha)$ greater than 100\,$\AA$ were not detected in
X-rays. We note that including the corresponding (six) X-ray upper
limits, this correlation vanishes. Moreover, this proposed correlation
is strongly biased by the low X-ray activity of LP\,944-20 -- detected
only during an X-ray flare \citep{rutledge00} -- which is likely
explained by the cooling of this old (500\,Myr) field BD
\citep{stelzer06}.

\begin{figure}[!t]
\centering
\includegraphics[width=\columnwidth]{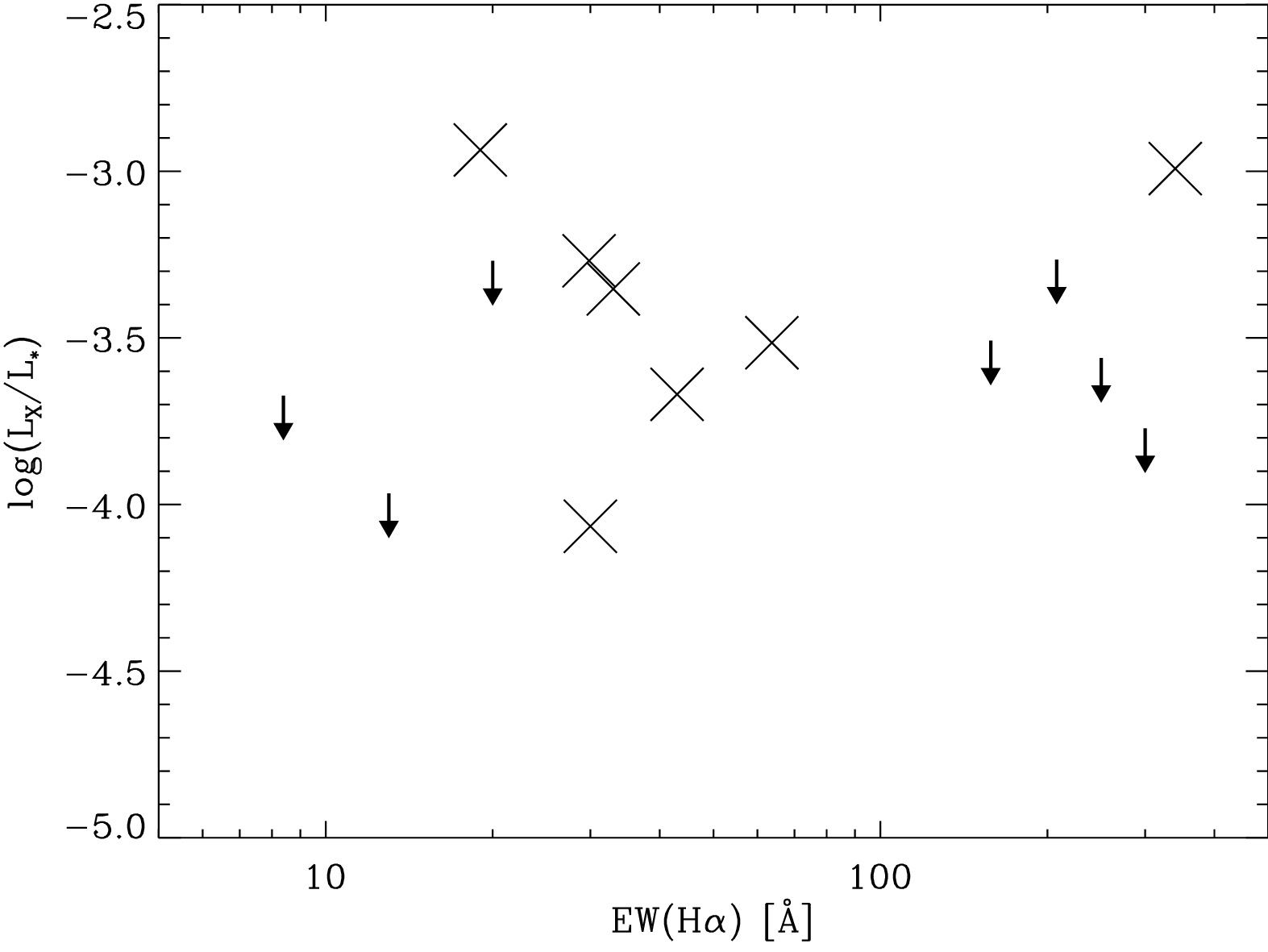}
 \caption{X-ray fractional luminosity of TMC BDs vs.\ ${\rm H}\alpha$. 
TMC BDs detected in X-rays are marked with `X'.}
\label{fig:eta_ha}
\end{figure}

$EW({\rm H}\alpha)$
measurements are available for all the XEST BDs but 2MASS\,J0455
(Table~\ref{table:bd}), allowing us to investigate the relationship
between X-ray fractional luminosity and H$\alpha$ emission from a
well-defined sample of young BDs. Fig.~\ref{fig:eta_ha} shows the
X-ray fractional luminosity versus $EW({\rm H}\alpha)$. We detected in
X-rays BDs both with low and high $EW({\rm H}\alpha)$. We compute Spearman's rank
correlation coefficient, and find no significant log-log correlation
between the X-ray fractional luminosity and $EW({\rm H}\alpha)$.

\subsection{X-ray activity and accretion}
\label{accretion}

\begin{figure}[!t]
\centering
\includegraphics[width=\columnwidth]{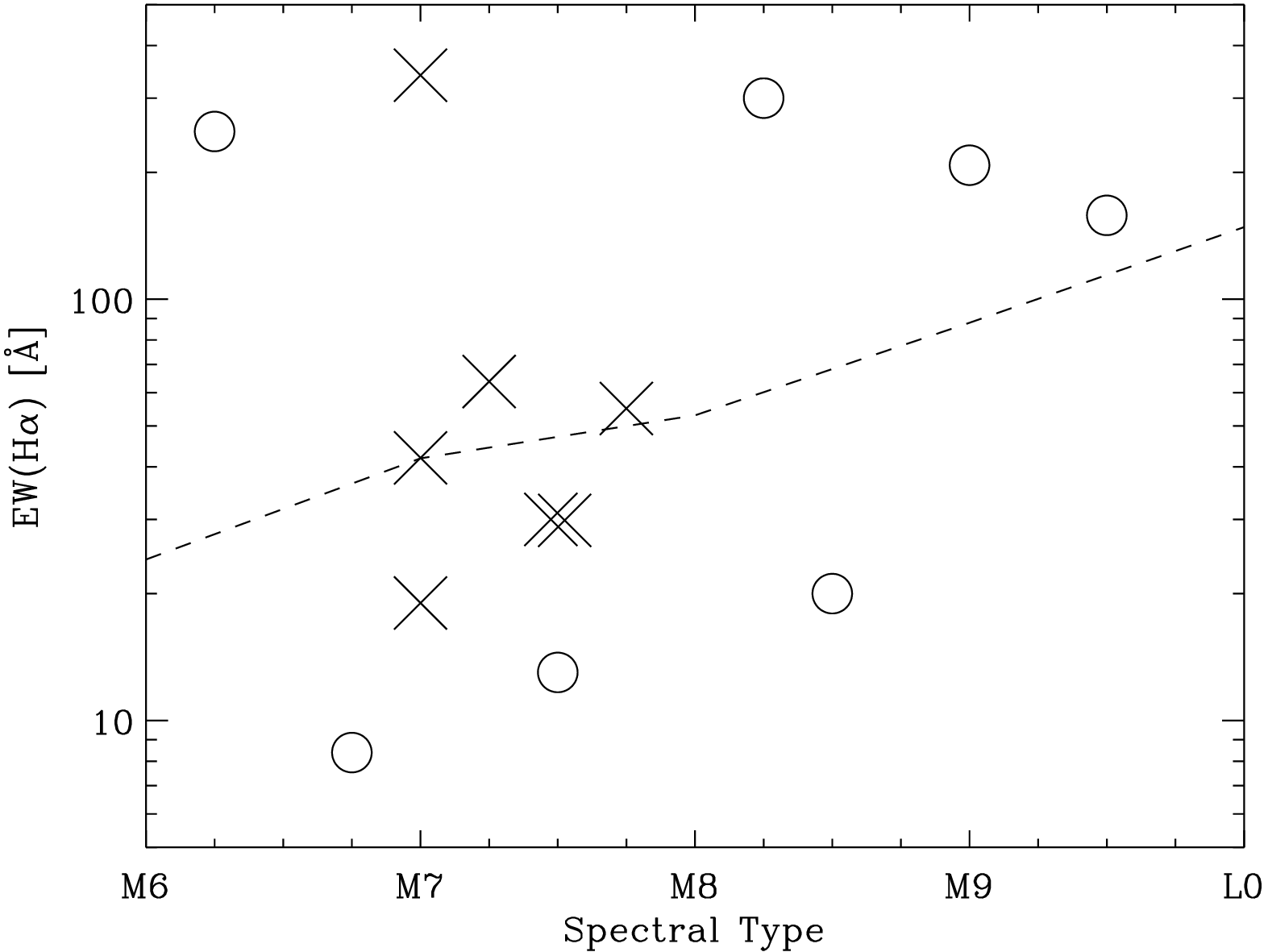}
 \caption{Equivalent width of H$\alpha$ emission lines vs.\ spectral
type for TMC BDs. TMC BDs detected in X-rays are marked with `X'. The dashed line
shows the saturation limit of chromospheric activity at $\log[L({\rm
  H\alpha})/L_*]=-3.3$ \citep[determined in the open
clusters;][]{barrado03}.
}
\label{fig:ha_spt}
\end{figure}

The H$\alpha$ emission line cannot be used as a tracer of accretion in
low-mass stars and BDs without a priori knowledge on the limit of pure
chromospheric H$\alpha$ emission in these objects. The canonical limit
of 10\,$\AA$ for $EW({\rm H}\alpha)$ was first used to disentangle
pure chromospheric activity in Weak-line T~Tauri Stars (WTTSs) and
H$\alpha$ emission line excess in Classical T~Tauri stars (CTTSs)
produced by accretion. \citet{martin98} proposed $EW({\rm H}\alpha)$
limits depending on the spectral type. 
\citet{barrado03} improved this criterion and extended it into
the substellar domain. This empirical criterion is the saturation at
$\log[L({\rm H\alpha})/L_*]=-3.3$, based on observations of
nonaccreting stars in young open clusters, which corresponds
physically to the maximum amount of energy that can be released in
nonthermal processes by the chromosphere, i.e.\ $\sim$$5 \times
10^{-4}$ of the total emitted energy. Low-mass stars or BDs
exceeding this limit are accreting. Fig.~\ref{fig:ha_spt} shows
$EW({\rm H}\alpha)$ versus spectral type for the BDs of XEST. The
dashed line shows the saturation at  $\log[L({\rm H\alpha})/L_*]=-3.3$
which increases from $EW({\rm H}\alpha)$$\sim$24\,$\AA$ at M6 to
$\sim$148\,$\AA$ at L0 \citep{barrado03}. 
MHO\,4 and CFHT-BD-Tau\,3 are just above this limit, but
high-resolution spectrum showed no indication of accretion
\citep{mohanty05}. Therefore, we classify these BD as
nonaccreting. For the other BDs, our classification based on $EW({\rm
H}\alpha)$ is in agreement with detailed high-resolution accretion
identification \citep{muzerolle05,mohanty05}. We find 6 accreting and
8 nonaccreting BDs.
The X-ray detection rates of
accreting and nonaccreting BDs are 33\%$\pm$27\% (2/6) and
63\%$\pm$36\% (5/8), respectively\footnote{To compute ratio errors we
considered that the number of X-ray detected BDs and the number of
(non)accreting BDs are not exactly known, and we combined both
Poissonian errors using Gaussian propagation, i.e.\ $\Delta(a/b)=a/b
\times \sqrt{(\Delta{a}/a)^2+(\Delta{b}/b)^2}$ with
$\Delta{a}=\sqrt{a}$ and $\Delta{b}=\sqrt{b}$.}.

\begin{figure}[!t]
\centering
\includegraphics[width=\columnwidth]{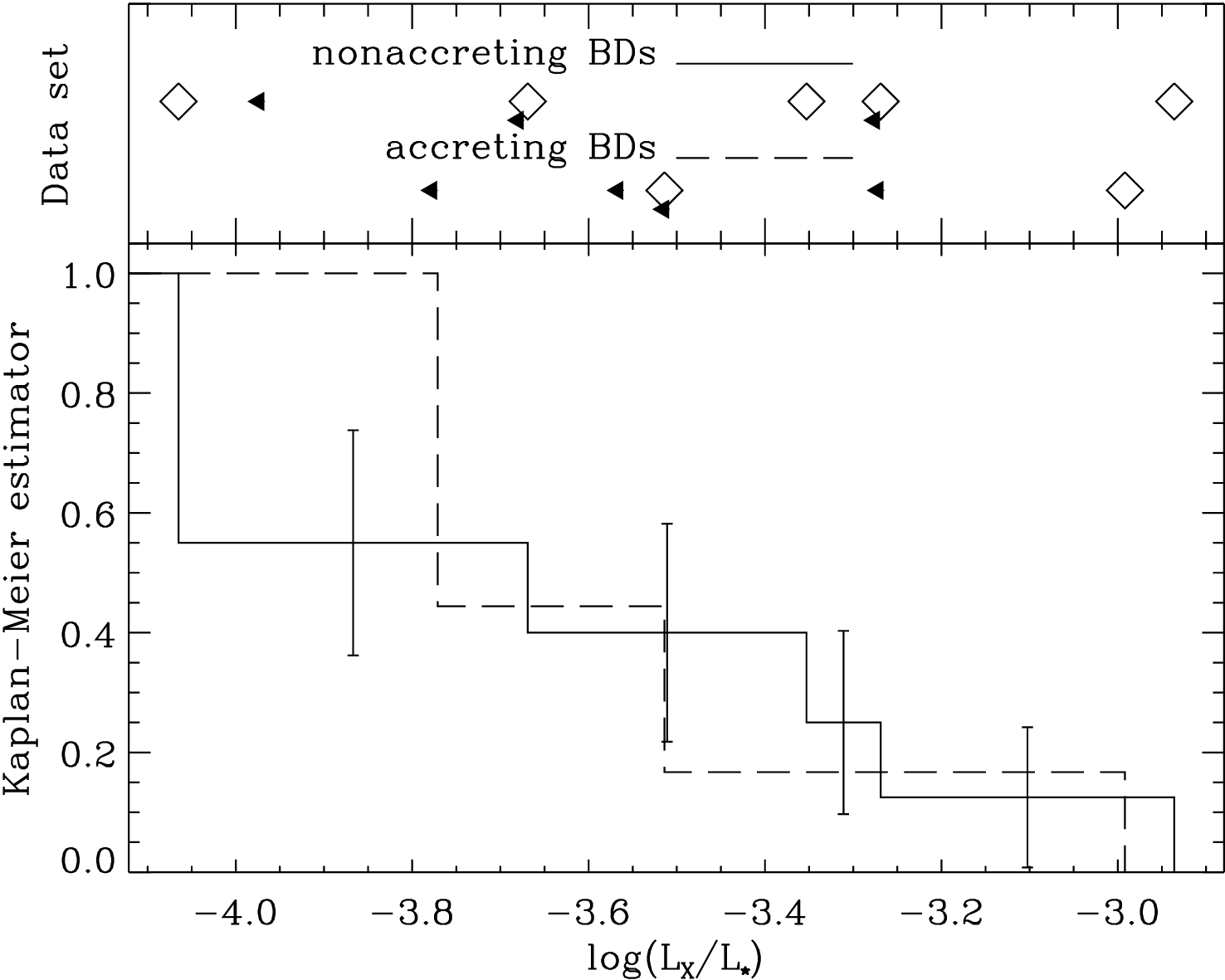}
 \caption{Cumulative distributions of the X-ray fractional luminosities
for nonaccreting (continuous line) and accreting (dashed line) BDs. In
the top panel, diamonds and arrows indicate detections and upper
limits, respectively of the two samples. Only Kaplan-Meier estimator
error bars of the nonaccreting sample are shown to clarify the
plot. Two-population statistical methods show that both samples
are drawn from the same underlying distribution.
}
\label{fig:eta_km}
\vspace{0.5cm}
\centering
\includegraphics[width=\columnwidth]{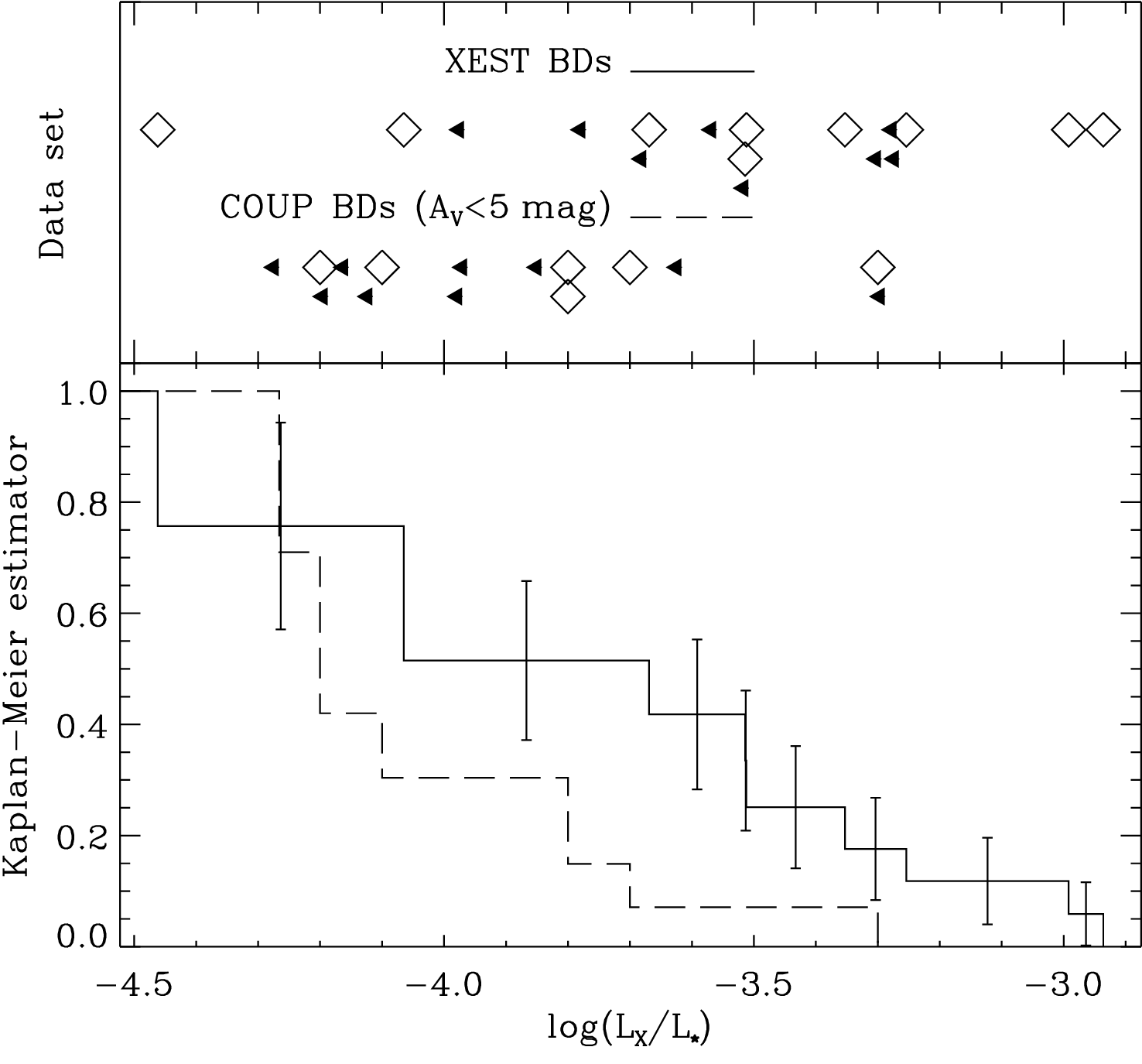}
 \caption{Cumulative distributions of the X-ray fractional luminosities
of XEST (continuous line) and COUP (dashed line) BDs. In the top
panel, diamonds and arrows indicate detections and upper limits,
respectively of the two samples. Only Kaplan-Meier estimator error
bars of the nonaccreting sample are plotted to clarify the
plot. Two-population statistical methods show that XEST BDs are
more active in X-rays than COUP BDs.
}
\label{fig:xest_coup_km}
\end{figure}

We compute with {\tt ASURV} the cumulative
distributions of the X-ray fractional luminosities -- including upper
limits -- with the Kaplan-Meier estimator for accreting and nonaccreting
BDs (Fig.~\ref{fig:eta_km}).
Two-population statistical methods provided by {\tt
ASURV}\footnote{Namely the Gehan and Peto-Peto generalized Wilcoxon
tests, and the Logrank test, which are standard methods of univariate
survival analysis as described by \citet{feigelson85}.} show that the
probability for the null hypothesis that both samples are drawn from
the same underlying distribution is 0.75. Therefore, the accreting and
nonaccreting BDs have similar X-ray fractional luminosities. The
median value of the X-ray fractional luminosities is $-3.9$ for the
nonaccreting BDs.

A large difference is found between accreting and nonaccreting
low-mass stars. In the TMC, the CTTSs are 2.2 times less luminous
than the WTTS \citep{guedel06a}. A similar result was found in the ONC
where the CTTSs (selected with 8542\,$\AA$ Ca\,{\small II} line) are 2.1
times less luminous than the WTTSs
\citep{flaccomio03,preibisch05b}. \citet{preibisch05b} proposed that
in accreting objects magnetic reconnection cannot heat the dense
plasma in mass-loaded accreting field lines to X-ray
temperatures. If true, this implies that coronal activity is less
affected by the accretion in BDs than in low-mass stars.

The median X-ray fractional luminosity of nonaccreting BDs in the XEST
is $\sim$4 times lower than the mean saturation value for rapidly rotating
low-mass ($0.22\le M_\star/M_\odot \le 0.60$) field stars
\citep[$\log{(L_{\rm X}/L_*)}=-3.3$;][]{pizzolato03}, whereas in
ONC the median X-ray fractional luminosity of nonaccreting low-mass
stars is consistent with this saturation level \citep{preibisch05b}.

\begin{figure*}[!t]
\centering
\includegraphics[height=1.3\columnwidth,angle=90]{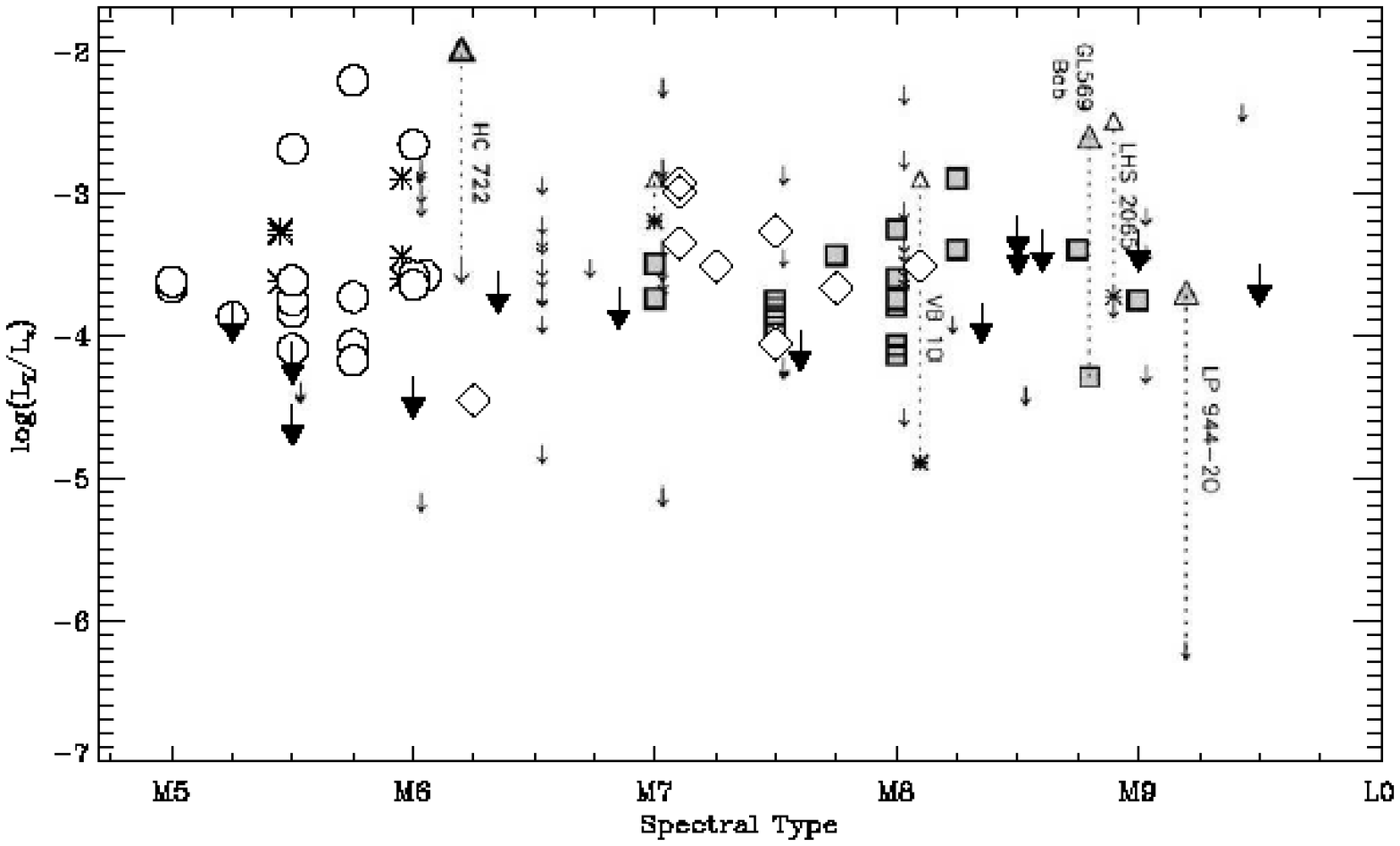}
 \caption{X-ray fractional luminosity vs.\ spectral type for objects of type M5 and
   later. Detections of late M field
   stars from \citet{fleming93} are shown as asterisks. The circles show
   low-mass stars of the TMC detected in X-rays \citep{guedel06b}. Diamonds and thick
   arrows show BDs in the TMC.  The other X-ray detected BDs
\citep[see][ and references therein]{preibisch05} are shown by gray
filled squares. For very cool objects with strong flares, the values
at flare peak are shown by triangles, connected by dotted lines to the
quiescent emission. Some symbols have been slightly moved in spectral
type to avoid overlaps.
}
\label{fig:eta_spectyp}
\end{figure*}

\section{Comparison of brown dwarf X-ray fractional luminosities in
the XEST and COUP}
\label{XEST_COUP}

\citet{preibisch05} in COUP studied the X-ray properties of young BDs
spectroscopically identified in the near-IR \citep{slesnick04} having
spectral types between M6V and M9V. Eight out of the 33
BDs\footnote{We suppressed from the young BD sample of
\citet{preibisch05} COUP\,344=HC\,722, which was identified as a
foreground (old) dwarf object in \citet{slesnick05}'s erratum. This
object, detected in COUP only during an X-ray flare, was also discussed
in \citet{preibisch05} but kept in the reference sample.}
were clearly detected as X-ray sources down to a detection limit of $\log(L_{\rm
X}/{\rm erg\,s^{-1}})= 27.3$. The near-IR selection helped to find extincted
BDs \citep{slesnick04}. Consequently, the apparently low detection
rate of BDs in COUP is in many cases related to the substantial
extinction of these BDs. Considering only the ONC BDs with $A_{\rm V}
\le 5$\,mag reduces the median visual extinction from 5.6 to 2.2\,mag,
and leads to an X-ray detection rate of 40\% (6 out of 15 BDs),
similar to the X-ray detection rate observed for TMC BDs.

We compute the cumulative distributions of the X-ray fractional luminosities
of the whole sample of XEST BDs, and of the COUP BDs with $A_{\rm V}
\le 5$\,mag (Fig.~\ref{fig:xest_coup_km}). Two-population statistical
methods show that the probability for the null hypothesis that both
samples are drawn from the same underlying distribution is lower than
0.05. The former and the latter samples have median X-ray
fractional luminosity equal to $-4.0$ and $-4.2$, respectively. 
Therefore, the XEST BDs are 1.6 times more active in X-rays than the COUP
BDs. The origin of this difference is not yet understood.

\section{Discussion}
\label{discussion}

We discuss now the origin of the BD X-ray emission in the broader
context of the X-ray emission of cool stars. 
Following \citet{preibisch05}, we compare the X-ray fractional
luminosities of TMC BDs with other cool objects. 

Fig.~\ref{fig:eta_spectyp} shows the X-ray fractional
luminosities of TMC objects with spectral type equal or later than M5,
X-ray detected BDs \citep[][ and references
therein; Stelzer et al.\ 2006]{preibisch05}, the evolved BD Gl\,569
Ba,b \citep{stelzer04}, and the field BD LP\,944-20
\citep{rutledge00,martin02}. Cool field stars with spectral type M5
or later are also indicated: M field stars from
\citet{fleming93}; very-low mass field stars VB\,10 \citep{fleming03},
LHS\,2065 \citep{schmitt02}, HC\,722
\citep{slesnick04,slesnick05,preibisch05}. All these cool objects show
a similar level of X-ray fractional luminosities ranging from
$\sim$$10^{-3}$ to $\sim$$10^{-4}$ (note that, compared to
Fig.~\ref{fig:eta_teff}, the spectral type range has been reduced
by a factor of two in Fig.~\ref{fig:eta_spectyp}).

Based on \citet{baraffe98} evolutionary tracks, objects of spectral
type M7 with an age of 1\,Gyr are not BDs, but low-mass stars twice
as massive as a typical TMC BD having an M7 spectral type and an age of 3\,Myr.  
Moreover, such very cool stars also have surface gravities about 40
times higher than in a typical TMC BD.
This shows, as found by \citet{preibisch05} from the COUP BDs, that
the X-ray activity of BD coronae is not strongly dependent of the BD
mass and the BD surface gravity. Therefore, this implies that the 
relation that we have obtained, for the young objects of TMC with
spectral type M, between X-ray fractional luminosity and mass
(Fig.~\ref{fig:eta_mass}) is in fact the consequence of the more
fundamental relation between X-ray fractional luminosity and effective
temperature (Fig.~\ref{fig:eta_teff}). This latter relation agrees
with the overall result (field dwarfs and young BDs) in
Fig.~\ref{fig:eta_spectyp}: of the 15 sources shown with spectral
types M8.5V or later, only 4 have any detected quiescent emission; the
rest are either not detected at all or (in 3 cases) detected only
during strong flares.
By considering a subsample of
evolved BDs (from $\sim$1\,Myr to $\sim$1\,Gyr) with a small
range of masses (0.05--0.07\,M$_\odot$) but with effective
temperatures ranging from 3000\,K down to 1000\,K, \citet{stelzer06} found
evidence for a similar decline of the fractional X-ray luminosity with the
effective temperature; showing that the atmospheric temperature plays
a crucial role in determining the level of X-ray activity. 

The growing evidence for a decline of the coronal activity with the
effective temperature is analogous to the decline in chromospheric
(H$\alpha$) activity seen in field dwarfs at similar spectral types,
which show a slow decline from mid- to late M in $L_{\rm
  H\alpha}/L_{\rm *}$ and a sharp drop-off around M9V \citep{mohanty03}.
Although a definitive turn-over of the fractional X-ray luminosity
 toward L-type BDs is not proven for the present sample
 due to sparse statistics, we note that the fractional luminosity
 declines from low-mass stars to M-type BDs, and as a sample, the
 BDs are less efficient X-ray emitters than low-mass stars.
 We thus conclude that while the BD atmospheres observed here are
 mostly warm enough to sustain coronal activity, a trend is seen
 that may indicate its gradual decline due to the drop in
 photospheric ionization degree \citep{mohanty03}.

\section{Summary}
\label{summary}

With the XEST we detected 9 (out of 17) young BDs with spectral
type ranging from M6.25V to M8V; 7 BDs are detected here for the
first time in X-rays. This BD sample surveyed in X-rays allowed us to
investigate the magnetic activity in the substellar regime. We recovered a
well-known relation between X-ray and bolometric luminosity for stars (with
$L_*$$\le$10\,L$_\odot$) and BDs detected in X-rays, which is
consistent with a mean X-ray fractional luminosity $<\!\log(L_{\rm
X}/L_*)\!>\,=-3.5 \pm 0.4$. For the XEST BDs, the median of
$\log(L_{\rm X}/L_*)$ (including upper limits) is $-4.0$. The X-ray
fractional luminosity of XEST BDs is hence lower than the one of XEST
low-mass stars.

A shallow relation is found between X-ray fractional luminosity and
mass. We show that the X-ray fractional luminosity decline by a
factor of about 3 from hot coronae of solar-mass stars to cooler
atmospheres of M9V BDs. Consequently, a relation is found between
X-ray surface flux and effective temperature, which implies a decrease
of about one magnitude in the X-ray surface flux from a M0V star to a
M9V BD.

No significant relation is found between the X-ray
fractional luminosity and $EW({\rm H}\alpha)$. We used $EW({\rm
H}\alpha)$ to identify accreting and nonaccreting BDs. 
Accreting and nonaccreting BDs have a similar X-ray fractional
luminosities. The median X-ray fractional luminosity of
nonaccreting BDs are about 4 times lower than the mean
saturation value for rapidly rotating low-mass field stars. BDs have higher X-ray
fractional luminosities in the TMC than in Orion.

We confirm, as previously observed in the Cha\,I \citep{stelzer04b}
and in the ONC \citep{preibisch05}, that there is no dramatic change
of the magnetic activity at the stellar/substellar boundary. Young BDs
of spectral type M are sufficiently warm to sustain an active
corona. The young BDs in the TMC, with a median spectral type of M7.5,
have on average an X-ray surface flux which is 7 times higher
than the one observed in the solar corona at the solar cycle maximum.

Deeper X-ray observations of the coolest M-type BDs in the TMC
are needed to investigate a possible turn-over of the fractional X-ray
luminosity of TMC BDs around spectral type M9V.

\begin{acknowledgements}
We thank the anonymous referee for his constructive comments that helped to
improve this paper; Beate Stelzer for a careful reading of the
manuscript; and the International Space Science Institute (ISSI) in Bern for
significant financial support of the project team. This research is
based on observations obtained with \xmm, an ESA science mission with
instruments and contributions directly funded by ESA Member States and
NASA. X-ray astronomy research at PSI has been supported by the Swiss
National Science Foundation (grants 20-66875.01 and
20-109255/1). M.A.\ acknowledges support from NASA grant NNG05GF92G.
\end{acknowledgements}

\bibliographystyle{aa}
\bibliography{biblio}

\vspace{0.5cm}
\begin{minipage}[h]{2\columnwidth}
{\it Note added in proofs.---} 2MASS\,J04335245+2612548, a TMC BD with
spectral type M8.5V \citep{luhman06b} which was reported after the
acceptance of this paper, is neither detected in X-rays by \xmm/EPIC
or in the $U$-band by the \xmm~optical/UV monitor; it is located on
the X-ray PSF wings and the $U$-band ``smoke ring'' of IT\,Tau
(XEST-18-030). Therefore, adding this newly identified BD to our TMC BD sample
wouldn't change our conclusions.
\end{minipage}


\Online
\appendix

\section{Log and data reduction of \cxo~observations}
\label{log}

\begin{table*}[!ht]
\caption{Archival \cxo~observations which surveyed
  serendipitously the TMC BDs. Col.~(1) gives the name of the
  instrument used: `I' ($17\arcmin\times17\arcmin$ field of view) and `S'
  ($8.5\arcmin\times25.5\arcmin$ field of view) for imaging and spectroscopy ACIS
  CCD \citep{garmire03}, respectively. Col.~(8) give the name of the
TMC BD whitin the \cxo~field of view. Col.~(9) indicates whether we use this archival
  data to supplement the XEST or not (see following notes). Notes:
  in the ACIS-S observation \#3364, the TMC BD KPNO-Tau\,2 was located
  13\arcmin~off-axis on ACIS-S2, and it is not detected; we obtained here a better
  constraint on the X-ray luminosity of this BD using the sum of the
two \xmm~exposures, rather than this ACIS-S observation. In the ACIS-S
observation \#4488, the TMC BD 2MASS\,J0421 was located
5.7\arcmin~off-axis on ACIS-S3, the pipeline detection algorithm found
no source at this location.
}
\label{table:log}
\centering
\begin{tabular}{@{}cclcccccc@{}}
\hline\hline
ObsID & ACIS- & \multicolumn{1}{c}{Target} & \multicolumn{2}{c}{Nominal
  pointing} & Start/End obs. & Exposure & TMC BD & Used\\
 & & &  $\alpha_{\rm J2000}$ & $\delta_{\rm J2000}$         &      &     ks \\  
(1) & (2) & \multicolumn{1}{c}{(3)} & (4) & (5) & (6) & (7) & (8) & (9)\\
\hline
1866 & I & L1551            & 04$^{\rm h}$31$^{\rm m}$32\fs7  & 18\degr08\arcmin08\arcsec  & 2001-07-23T05:10:11\,/\,24T03:52:12 &   79 & MHO\,4 & yes\\
2563 & I & L1527            & 04$^{\rm h}$39$^{\rm m}$52\fs7  &  26\degr03\arcmin05\arcsec  & 2002-12-06T08:30:12\,/\,06T14:21:15 &  20 & CFHT-BD-Tau\,4 & yes\\
3364 & S & V410\,Tau & 04$^{\rm h}$18$^{\rm m}$34\fs6 & 28\degr22\arcmin47\arcsec &2002-03-07T06:16:32\,/\,07T11:45:24 & 18 & KPNO-Tau\,2 & no\\
4488 & S & FS\,Tau &  04$^{\rm h}$22$^{\rm m}$00\fs1  &  26\degr58\arcmin07\arcsec &2003-11-08T12:57:58\,/\,08T21:56:34 &30 & 2MASS\,J0421 &no\\
\hline
\end{tabular}
\end{table*}

Table~\ref{table:log} gives the log of \cxo~observations which
surveyed serendipitously the TMC BDs, and used to supplement the XEST.

Data reduction of archival \cxo~data were performed with {\tt CIAO}
software package version 3.2.1, {\tt XSPEC} version 11.3, the Penn State charge transfer
inefficiency (CTI) corrector
version 1.45, and the {\tt acis\_extract} package version
3.67\footnote{Descriptions and codes for CTI correction and {\tt
acis\_extract} can be found at {\tt
http://www.astro.psu.edu/users/townsley/cti} and {\tt
http://www.astro.psu.edu/xray/docs/TARA},
respectively.}. 
We followed Penn State University (PSU) procedures
\citep[e.g.,][]{getman05b} starting from data reduction from Level 1
event files provided by the {\it Chandra X-Ray Center} (CXC). We used
the PSU CTI Corrector \citep{townsley02} to correct partially the data
for CCD CTI caused mainly by radiation damage at the beginning of the
\cxo~mission. Source detections were performed with the {\tt
CIAO}'s task {\tt wavdetect}. 2MASS counterparts were used to
correct small boresight errors. The {\tt acis\_extract} package was
then used to extract source photons, estimate local background,
construct source and background spectra, compute redistribution matrix
files (RMFs) and auxiliary response files (ARFs), construct light
curves and time-energy diagrams, and perform automated spectral
grouping and fitting.


\section{Quiescent X-ray light curves of the TMC BDs}
\label{quiescent}

\begin{figure*}[!t]
\centering
\begin{tabular}{@{}c@{}c@{}c@{}}
\includegraphics[height=0.675\columnwidth,angle=-90]{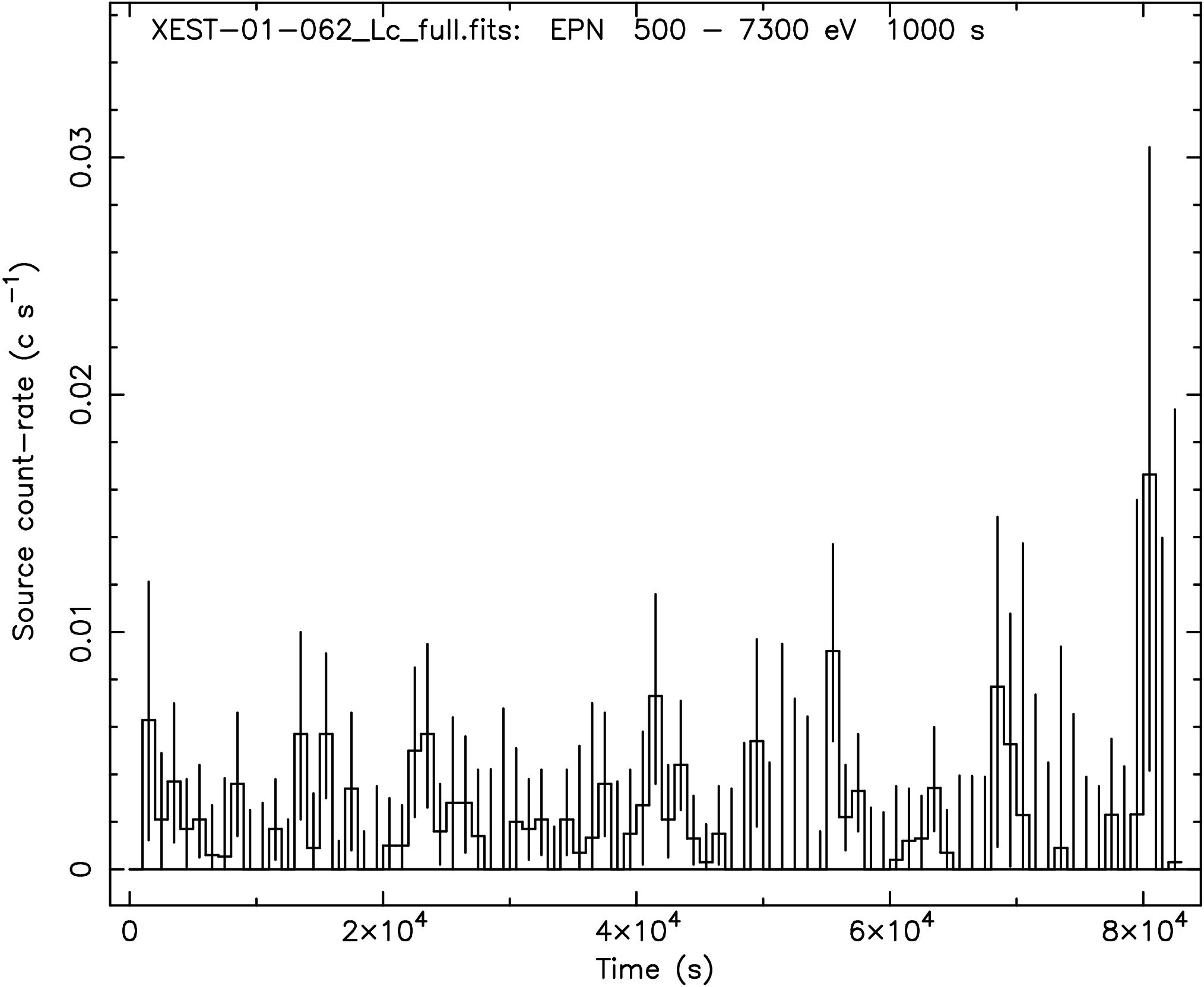}
&\includegraphics[height=0.675\columnwidth,angle=-90]{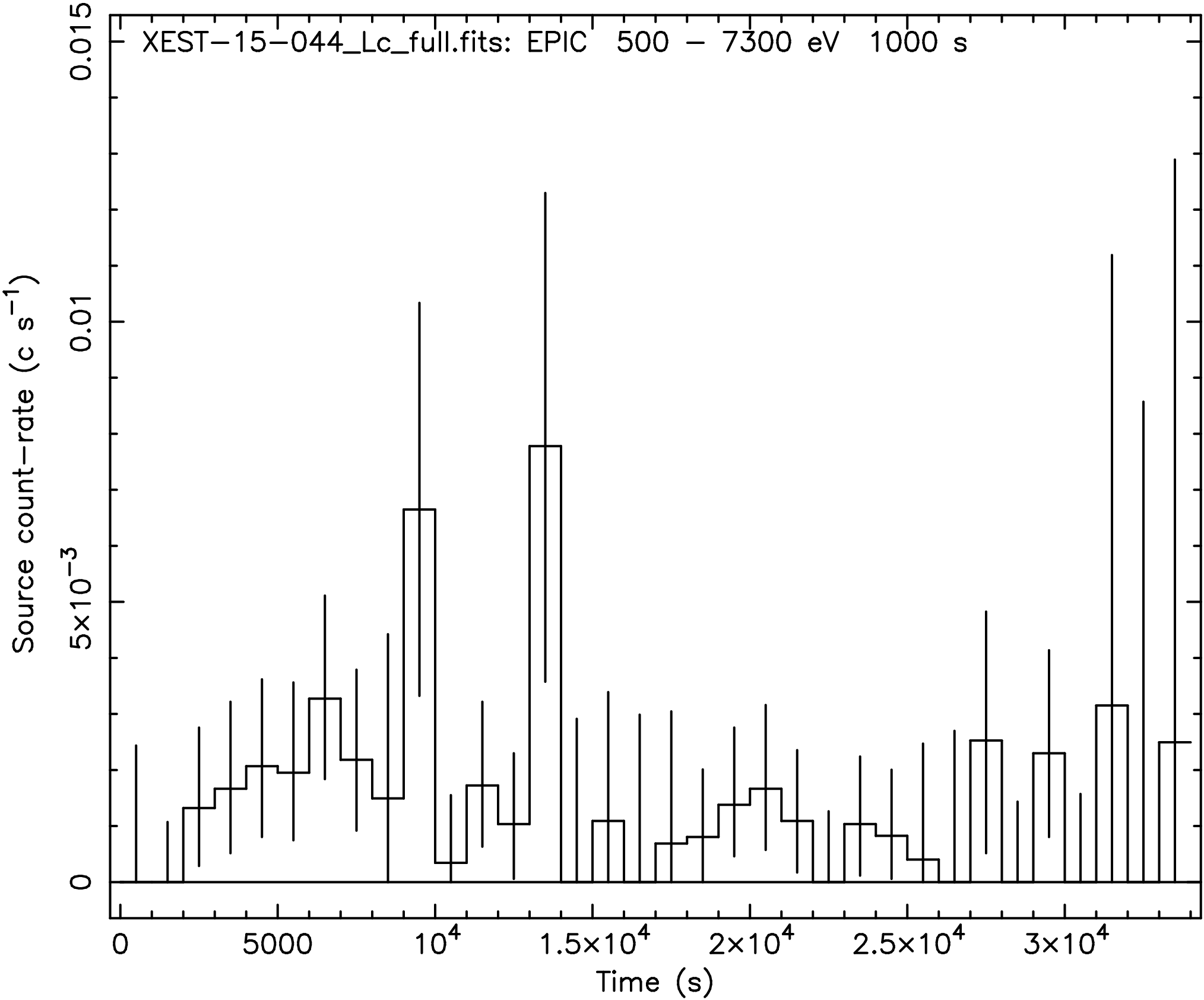}
&\includegraphics[height=0.675\columnwidth,angle=-90]{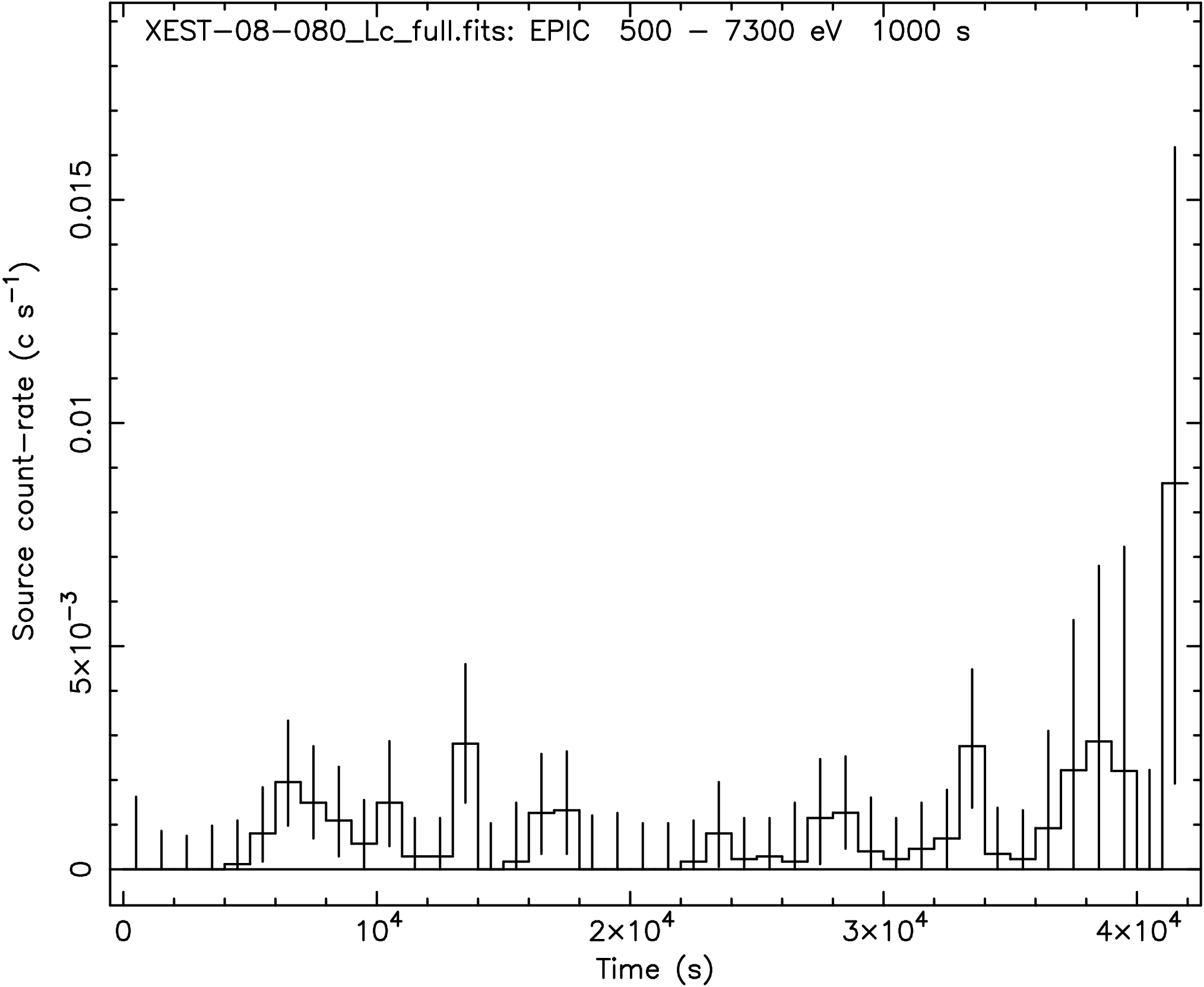}\\
\includegraphics[height=0.675\columnwidth,angle=-90]{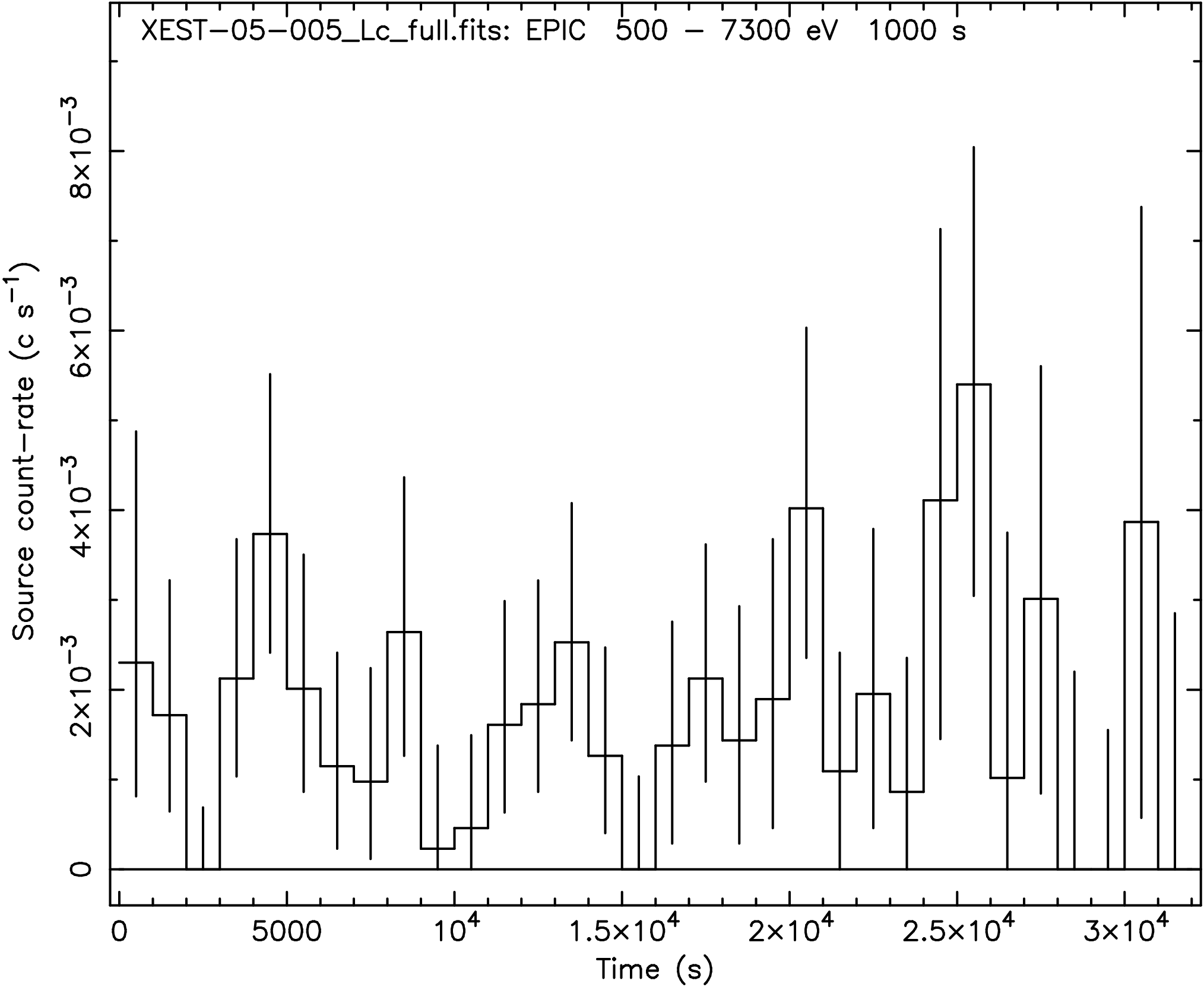}
&\includegraphics[height=0.675\columnwidth,angle=-90]{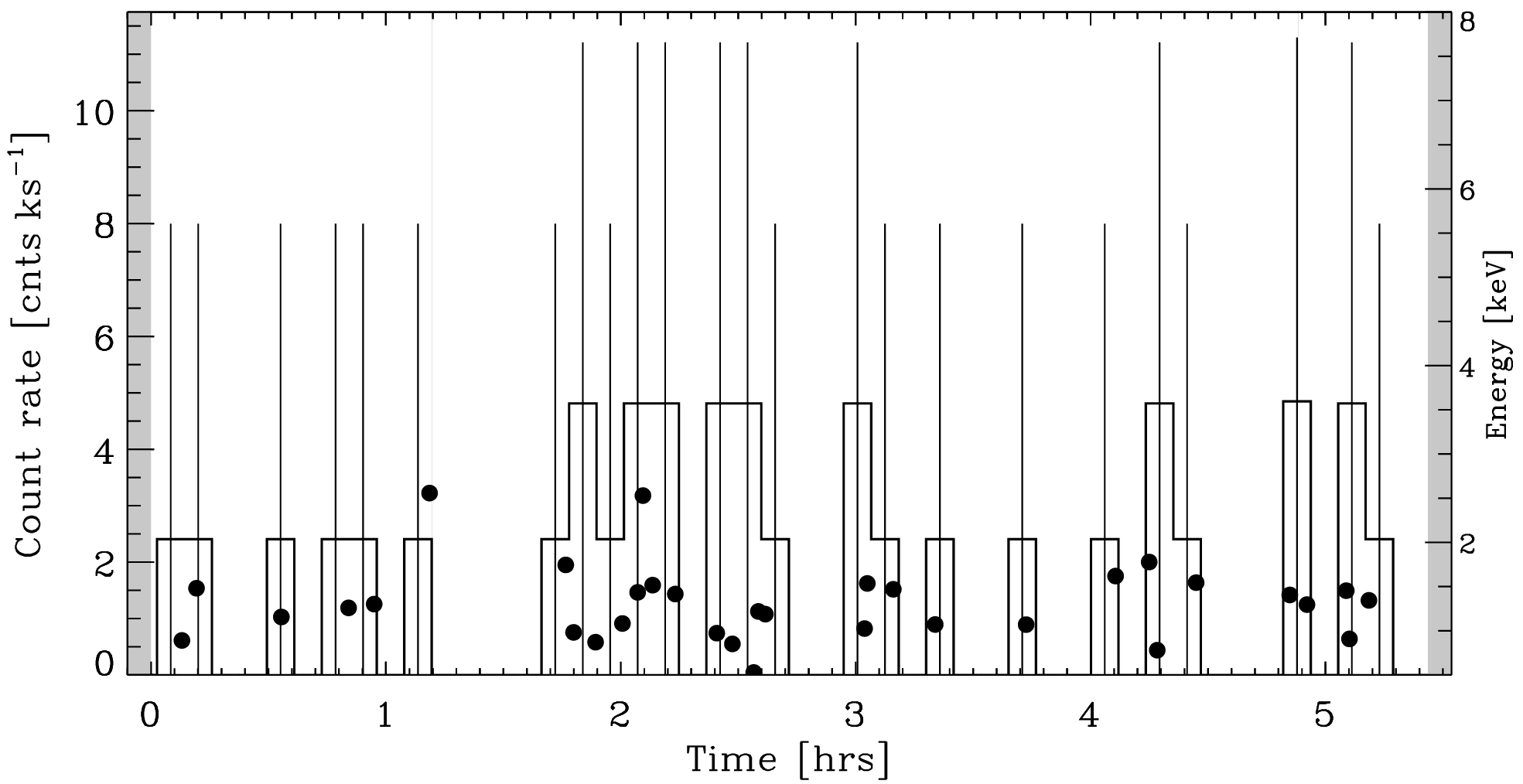}
&\includegraphics[height=0.675\columnwidth,angle=-90]{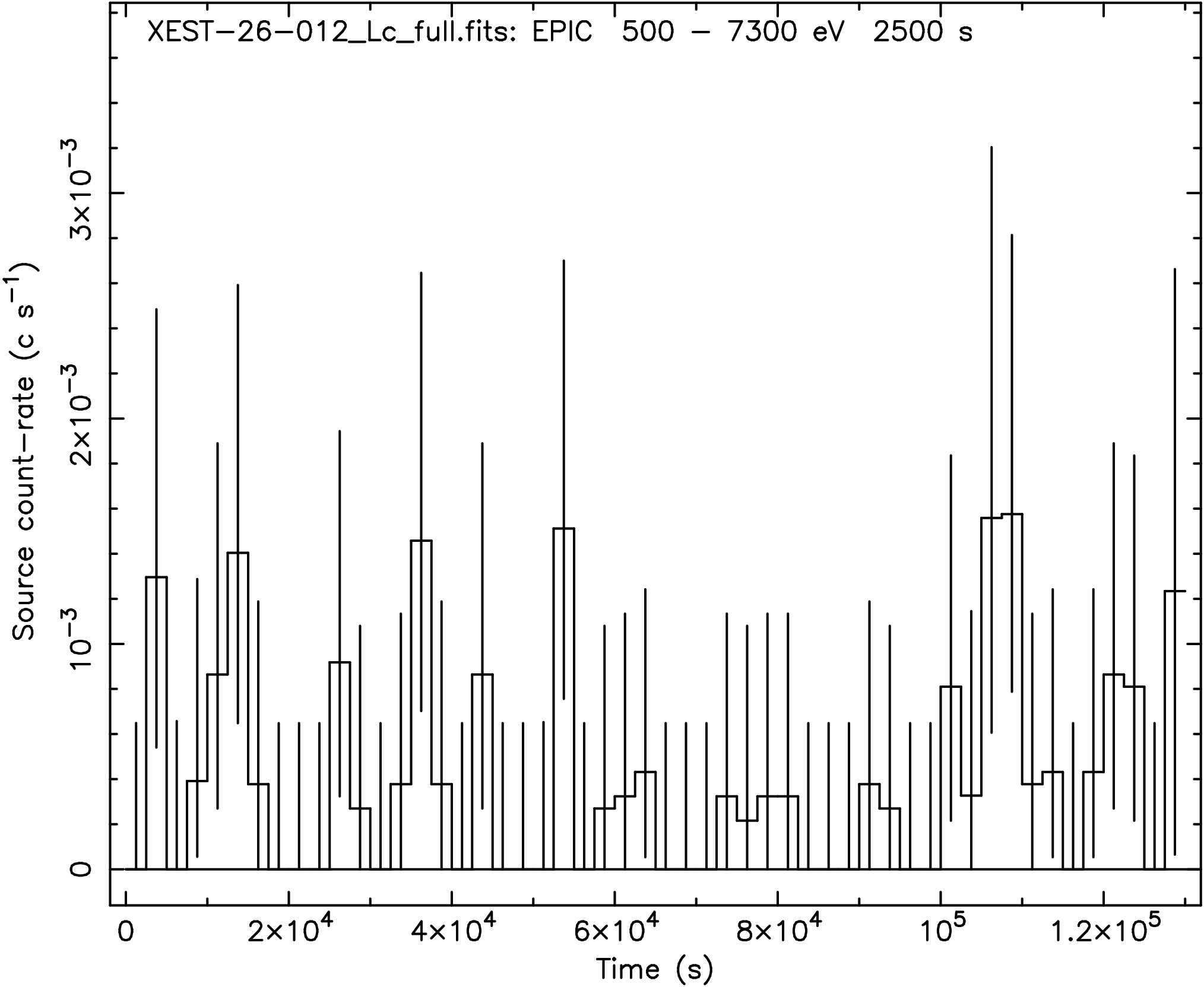}
\end{tabular}
 \caption{X-ray light curves of BDs in the TMC. From top to bottom and left to
right: 2MASS\,J0422 (only pn data), KPNO-Tau\,5, CFHT-BD-Tau\,3,
CFHT-Tau\,6 (EPIC), CFHT-BD-Tau\,4 (ACIS-I), and 2MASS\,J0455 (only
MOS data). On the \cxo~light curve, black dots indicate the arrival
time and energy of individual X-ray photons. These light curves are consistent with quiescent emission.
}
\label{fig:lc}
\vspace{0.5cm}
\centering
\begin{tabular}{@{}cc@{}}
\includegraphics[height=0.675\columnwidth,angle=-90]{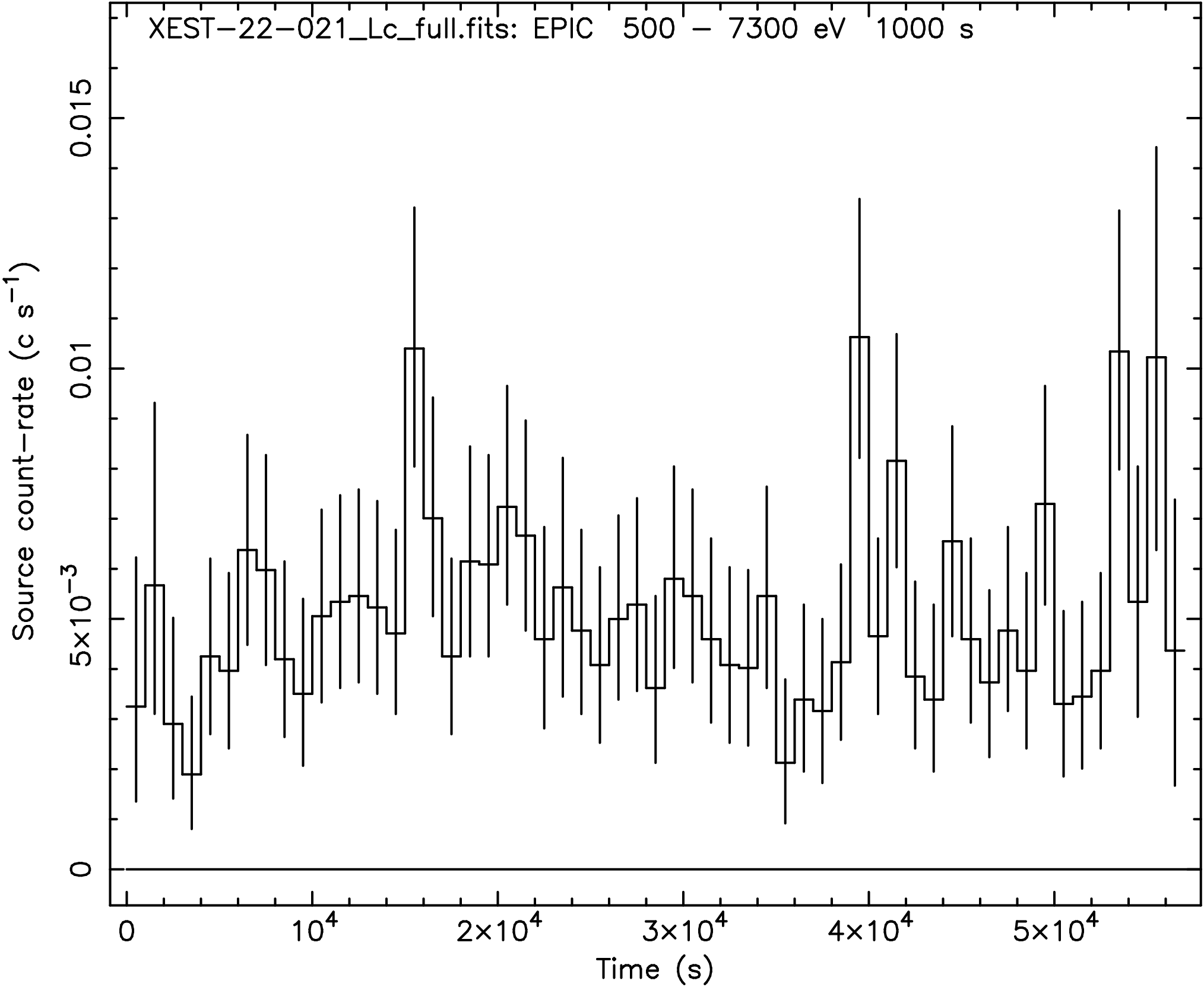} & \includegraphics[height=0.675\columnwidth,angle=-90]{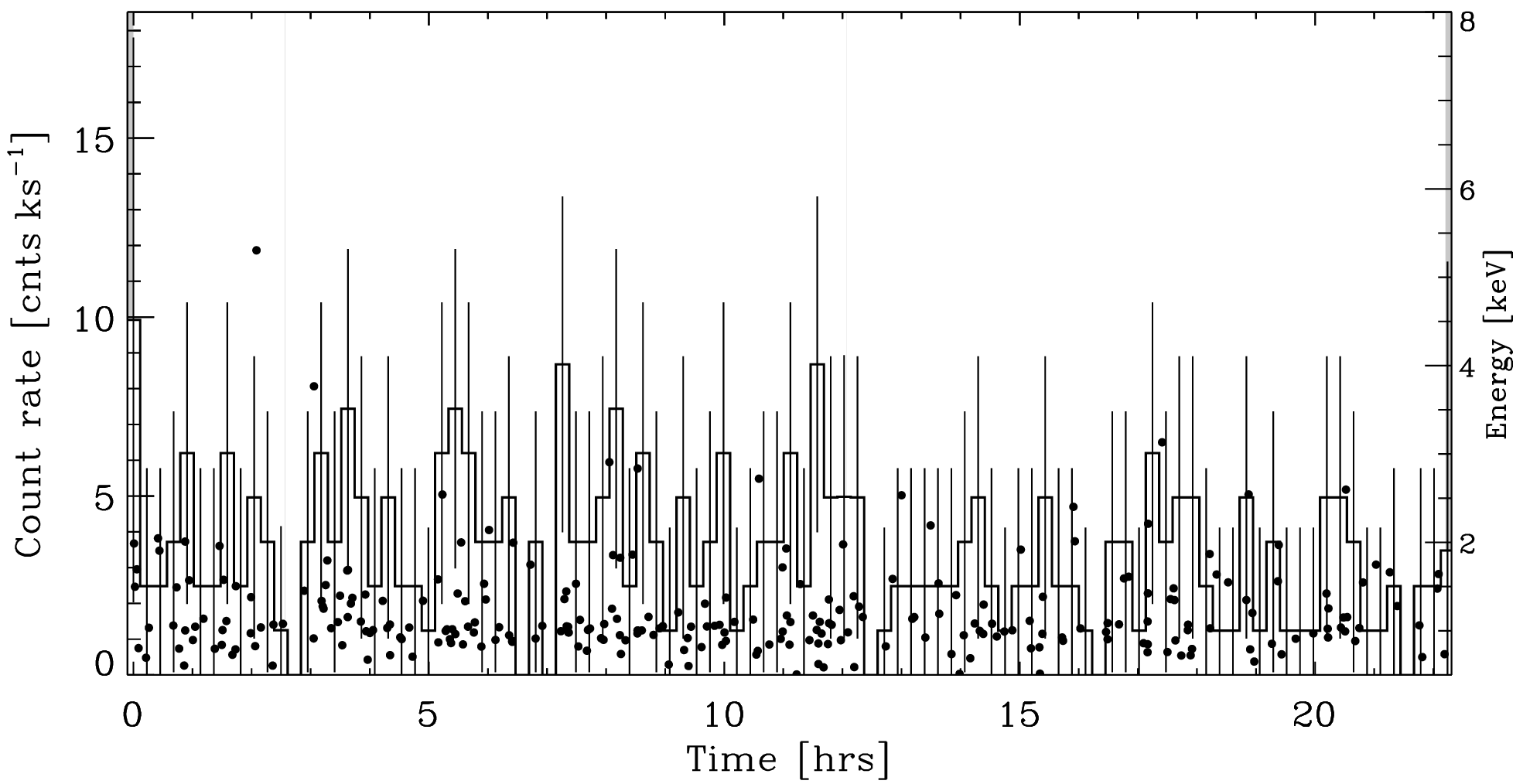}\\
\includegraphics[height=0.675\columnwidth,angle=-90]{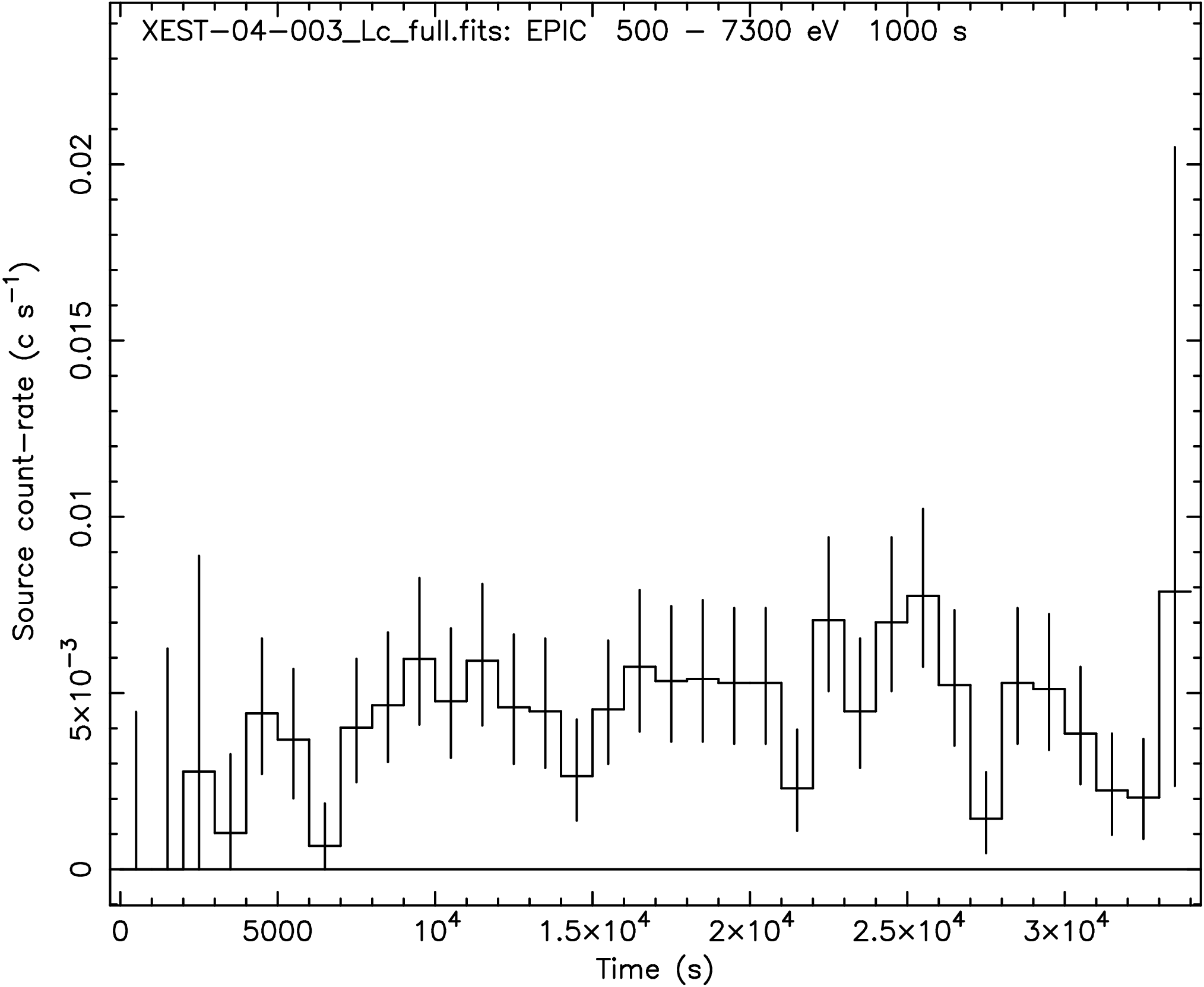} & \includegraphics[height=0.675\columnwidth,angle=-90]{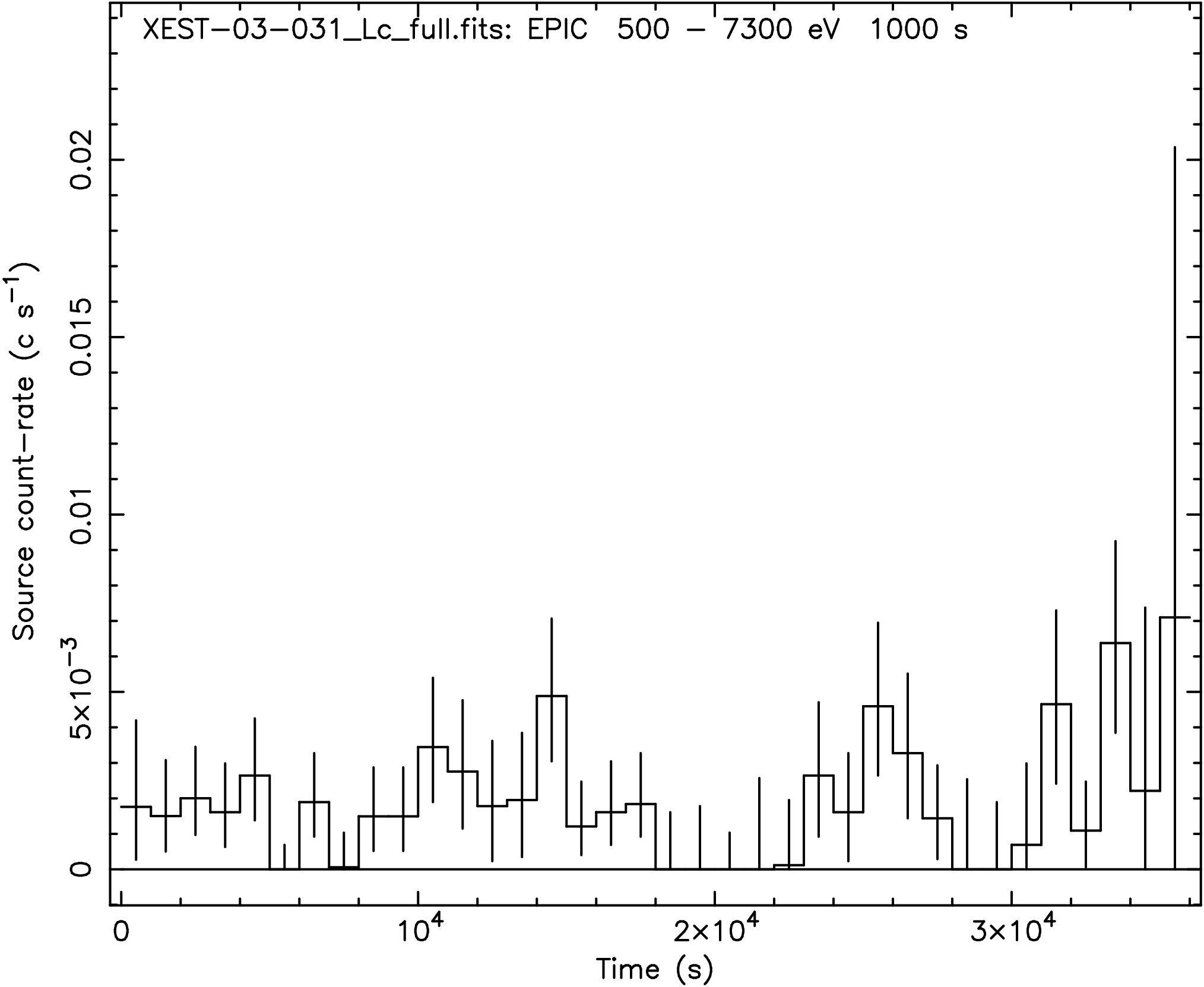}\\
\end{tabular}
 \caption{X-ray light curves of BDs in the TMC observed at two different
epochs. Top and bottom row show MHO\,4 and CFHT-Tau\,5 light curves,
respectively. During the second \xmm~observation of CFHT-Tau\,5, the
source is located on an EPIC pn gap, which explains the count rate
difference observed between the two epochs.
}
\label{fig:lc2}
\end{figure*}

Fig.~\ref{fig:lc} shows the X-ray light curves of the brightest TMC BDs,
which are consistent with quiescent activity. Fig.~\ref{fig:lc} shows
X-ray light curves of BDs observed at two different epochs.

\section{Degeneracy of the quantile diagram}
\label{degeneracy}

In contrast to the continuum models used as illustration of quantile
diagram by \citet{hong04}, which connect one single couple of physical
parameters to unique X-ray colours, we found that the spectra of
optically thin plasma (i.e.\ a continuum bremsstrahlung plus emission
lines) may have several couple of physical parameters producing the
same X-ray colours. To use the quantile analysis safely, we need to
reckon for each physical parameter the number of possible values per
X-ray colour.
 
We divided first the quantile diagram in a regular grid with a
$0.03\times0.03$ resolution. Then, the uneven grid of X-ray colours of
each physical parameter was projected to this regular grid using
spline interpolation. Each physical parameter therefore generates a
3-dimensional surface regularly sampled in the X-ray colour
plane. Finally, we found numerically the number of intersections
between this surface and a line of constant colour, i.e.\ the
degeneracy level.

\begin{figure*}[!t]
\centering
\begin{tabular}{@{}cc@{}}
\includegraphics[width=\columnwidth]{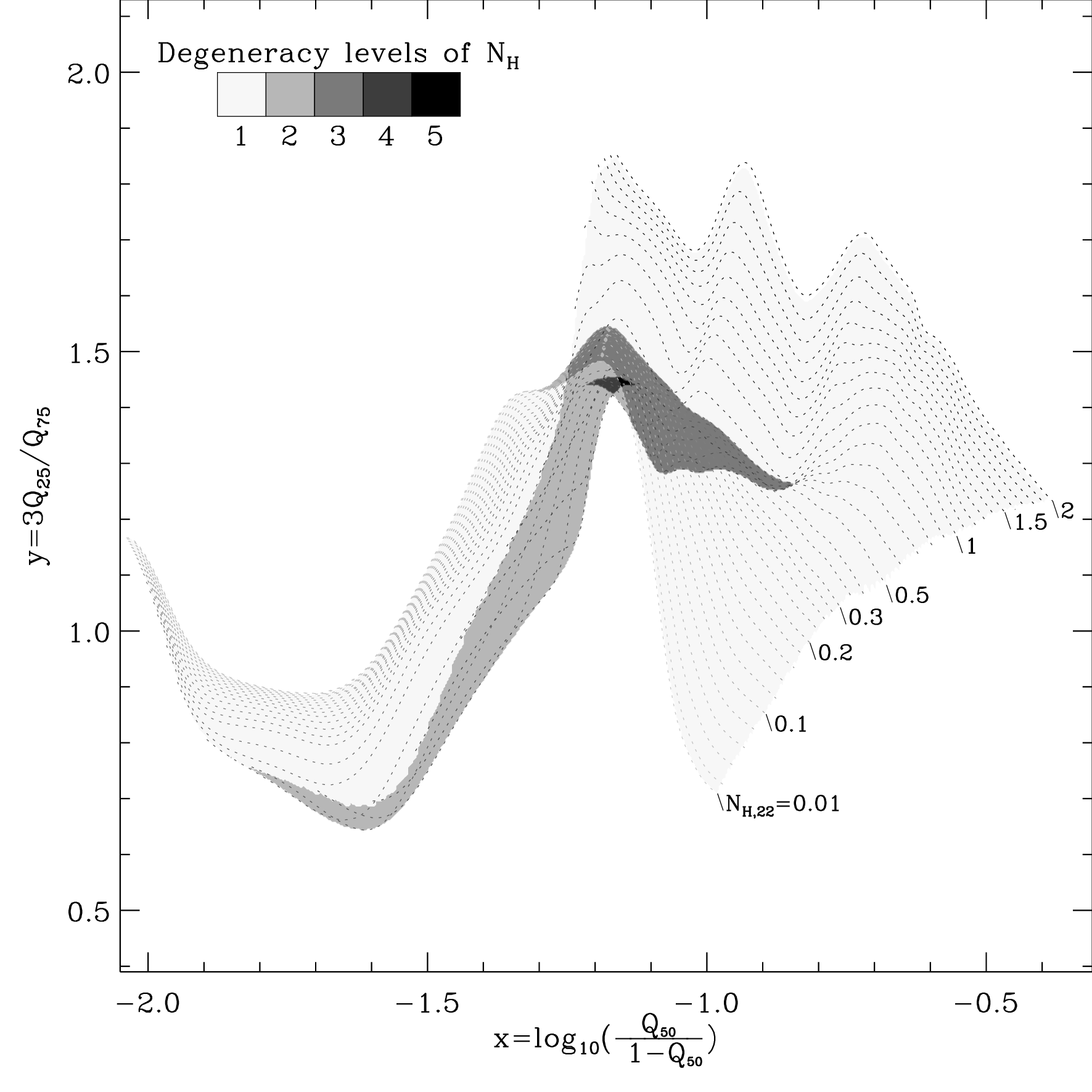}
& \includegraphics[width=\columnwidth]{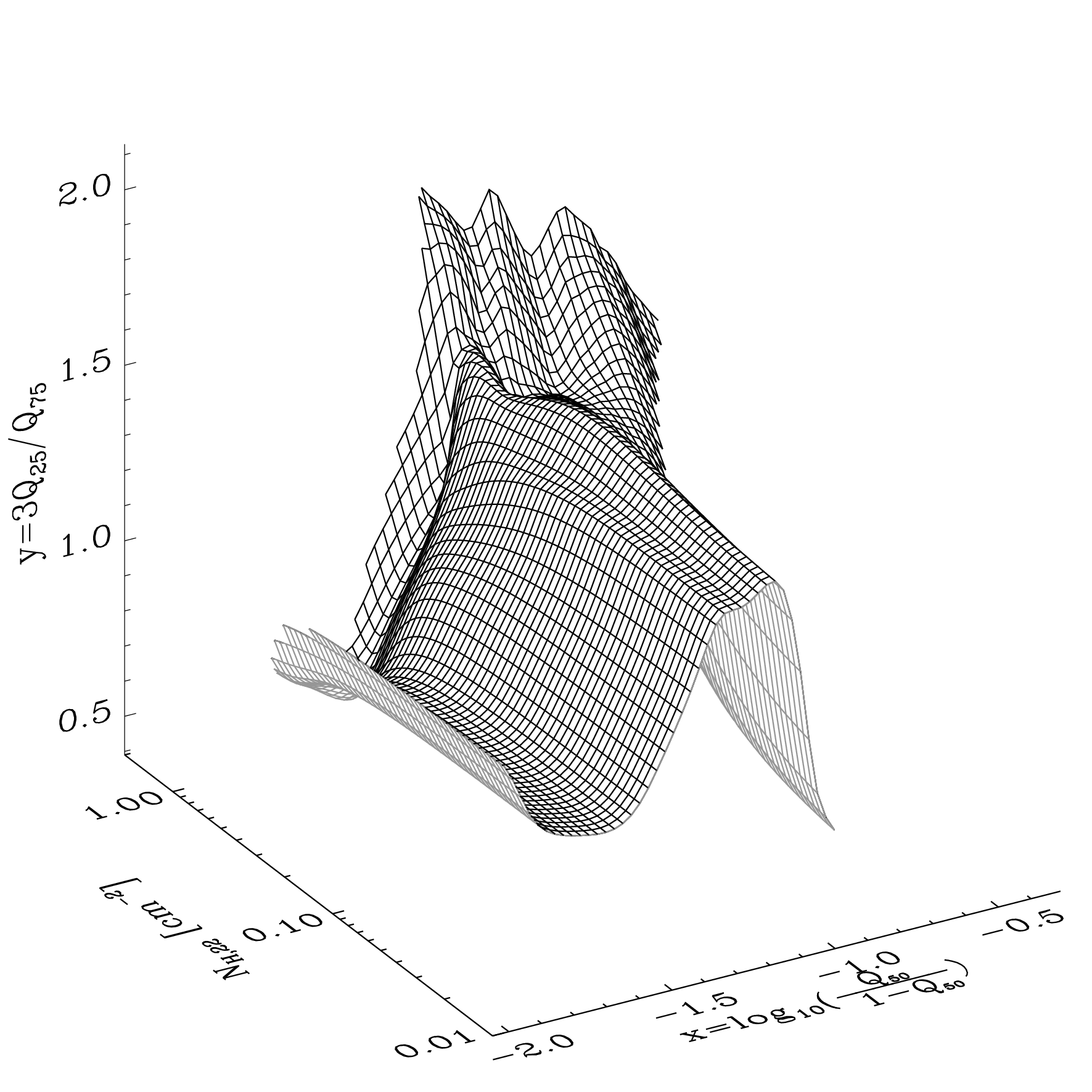}
\end{tabular}
\vspace{-0.4cm}\\
\caption{Degeneracy map of the absorption column density parameter (\nh) in
the quantile diagram. Left: quantile diagram showing in the ($x$,\,$y$)
space colours the loci of constant \nh~values (dotted lines) for a
{\tt WABS $\times$ MEKAL} plasma model with 0.3 times the solar elemental
abundances using \xmm~EPIC pn RMF and ARF. Grey levels indicate for
each ($x$,\,$y$) value the number of corresponding
absorption column density values. Regions where a ($x$,\,$y$) value
correspond to a unique value of \nh~are coloured in light grey.
Right: 3-dimensional shape of the surface generated by \nh~from the colour
space of the quantile diagram ($x$,\,$y$). Folds produce the
degeneracy observed in the quantile diagram.
}
\label{fig:nh_degeneracy}
\medskip
\medskip
\begin{tabular}{@{}cc@{}}
\includegraphics[width=\columnwidth]{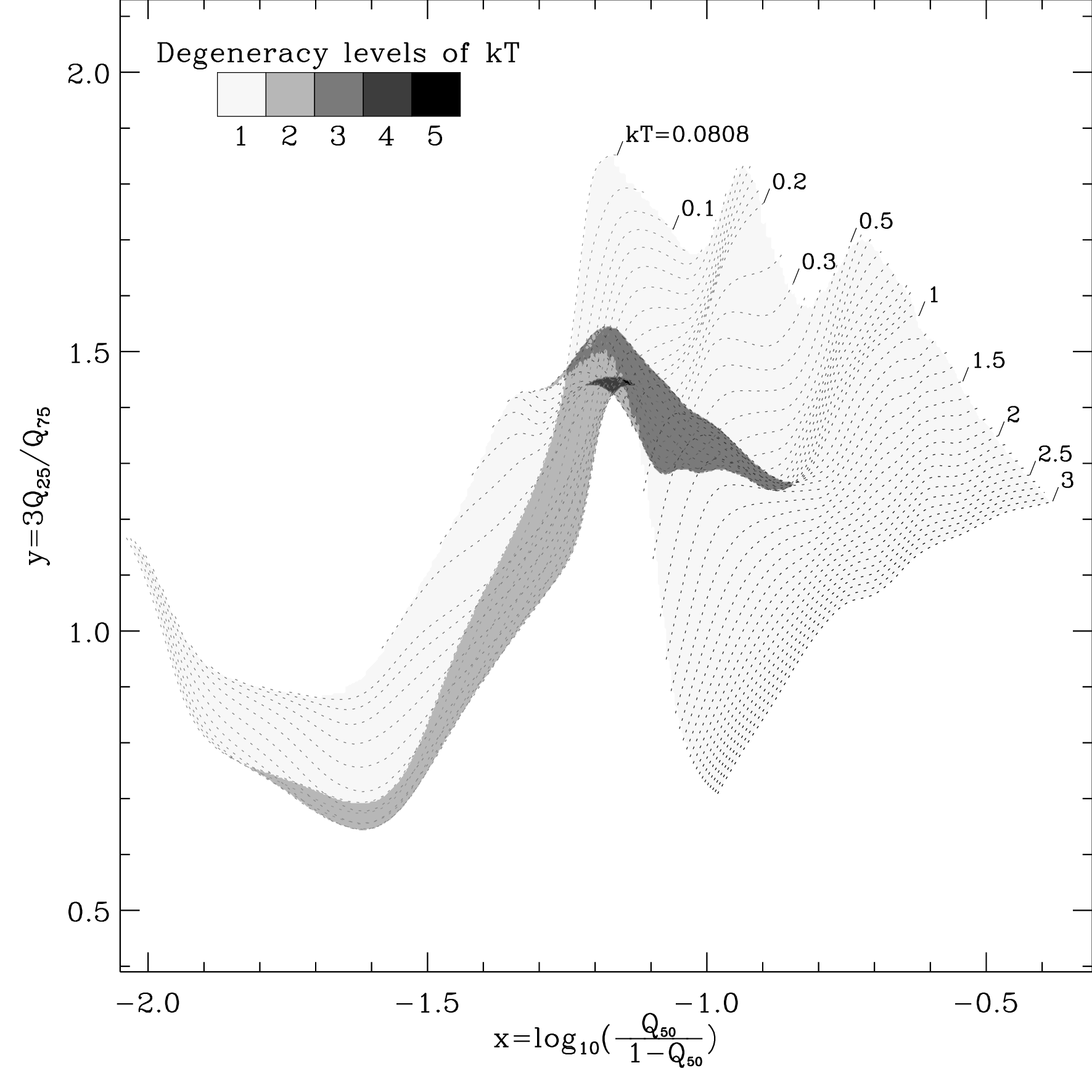}
& \includegraphics[width=\columnwidth]{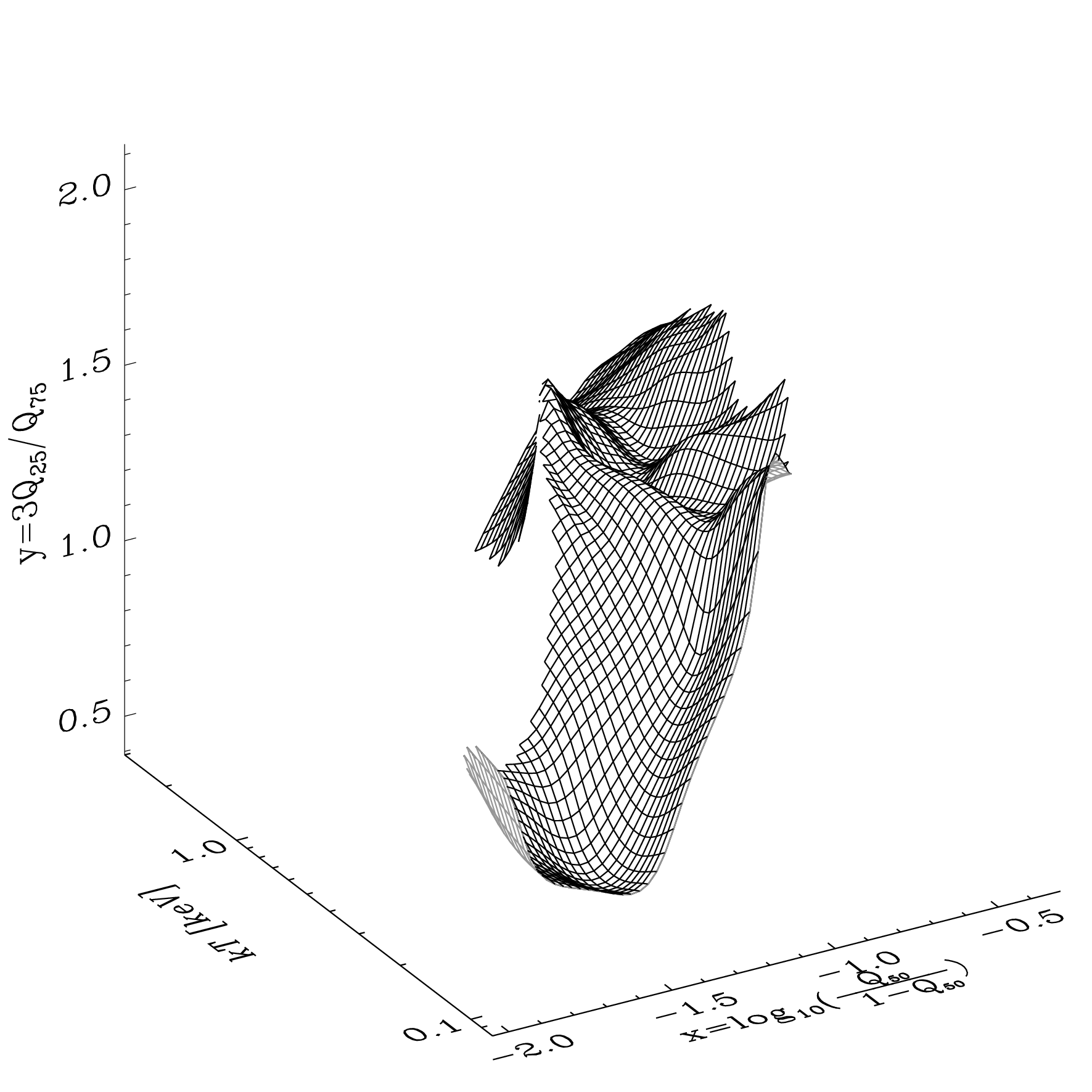}
\end{tabular}
\vspace{-0.4cm}\\
 \caption{Degeneracy map of the plasma temperature parameter ($kT$) in
the quantile diagram. Left: quantile diagram showing in the ($x$,\,$y$)
space colours the loci of constant $kT$ values (dotted lines) for a
{\tt WABS $\times$ MEKAL} plasma model with 0.3 solar elemental
abundance using \xmm~EPIC pn RMF and ARF. Grey levels indicate for
each ($x$,\,$y$) value the number of corresponding
plasma temperature values. Regions where a ($x$,\,$y$) value correspond
to a unique value of $kT$ are coloured in light grey.  Right: 3D
shape of the surface generated by $kT$ from the parameter space of the
quantile diagram ($x$,\,$y$). Folds produce the
degeneracy observed in the quantile diagram.
}
\label{fig:kt_degeneracy}
\end{figure*}

Fig.~\ref{fig:nh_degeneracy} and ~\ref{fig:kt_degeneracy} show the
result for the hydrogen column density and the plasma temperature grids,
respectively, computed with {\tt WABS $\times$ MEKAL} plasma model
with 0.3 times the solar elemental abundances and \xmm~EPIC pn RMF and ARF. The left
pannels show the degeneracy maps, which are nearly identical when one
considers the uneven sampling of the model parameter; in
Fig.~\ref{fig:QDx} we plotted conservatively the maximum of these two
degeneracy maps. The right pannels show for each parameter the
corresponding 3-dimensional surface in the X-ray colour plane with
folds producing multiple solutions. The quantile diagram is mainly composed
of areas where X-ray colours correspond to a unique couple of
parameters, but areas with multiple solutions -- from 2 to 5 -- are also
present. However, the area surface decreases with the number of
multiple solutions.  

\begin{figure*}[!ht]
\centering
\begin{tabular}{@{}cc@{}}
\includegraphics[width=\columnwidth]{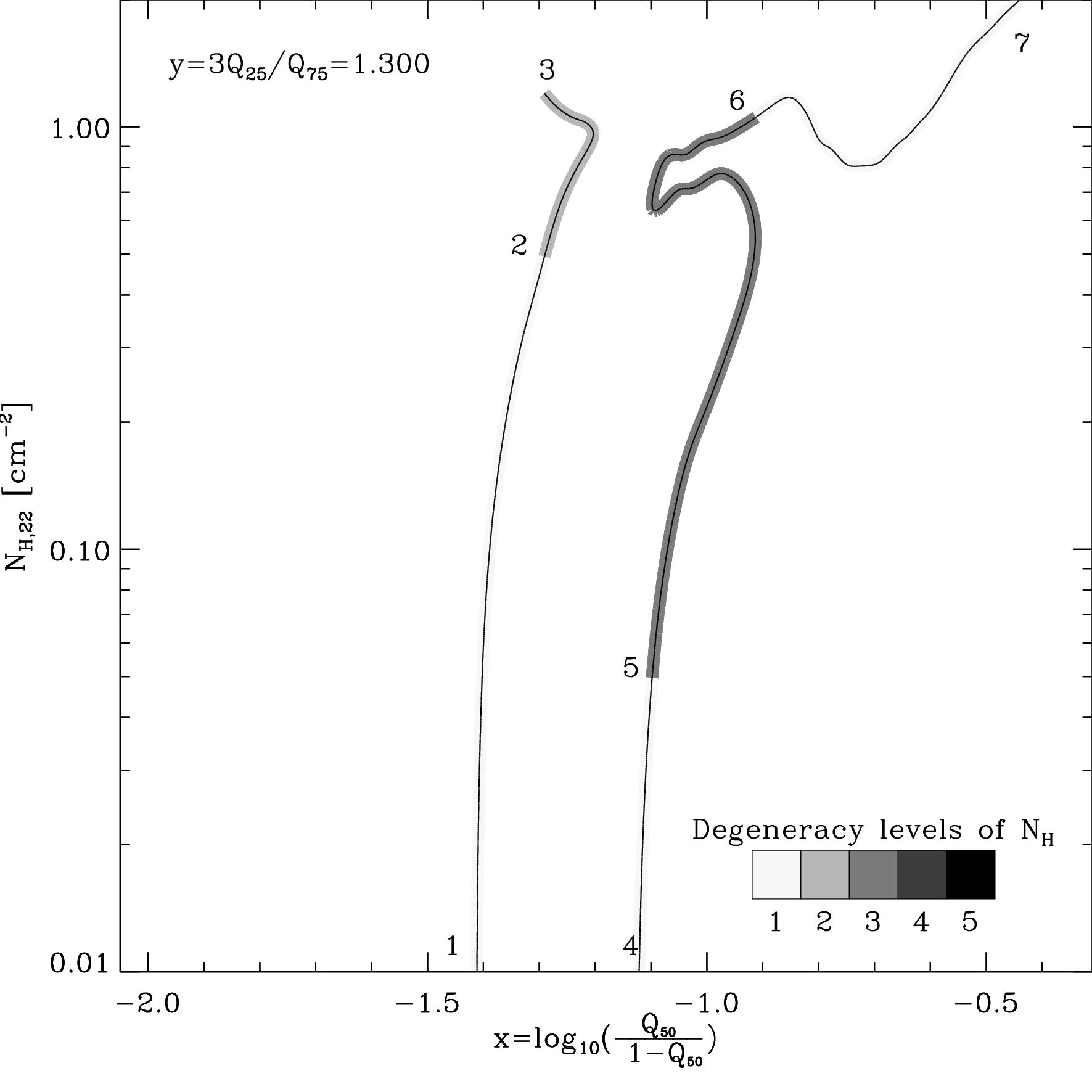} & \includegraphics[width=\columnwidth]{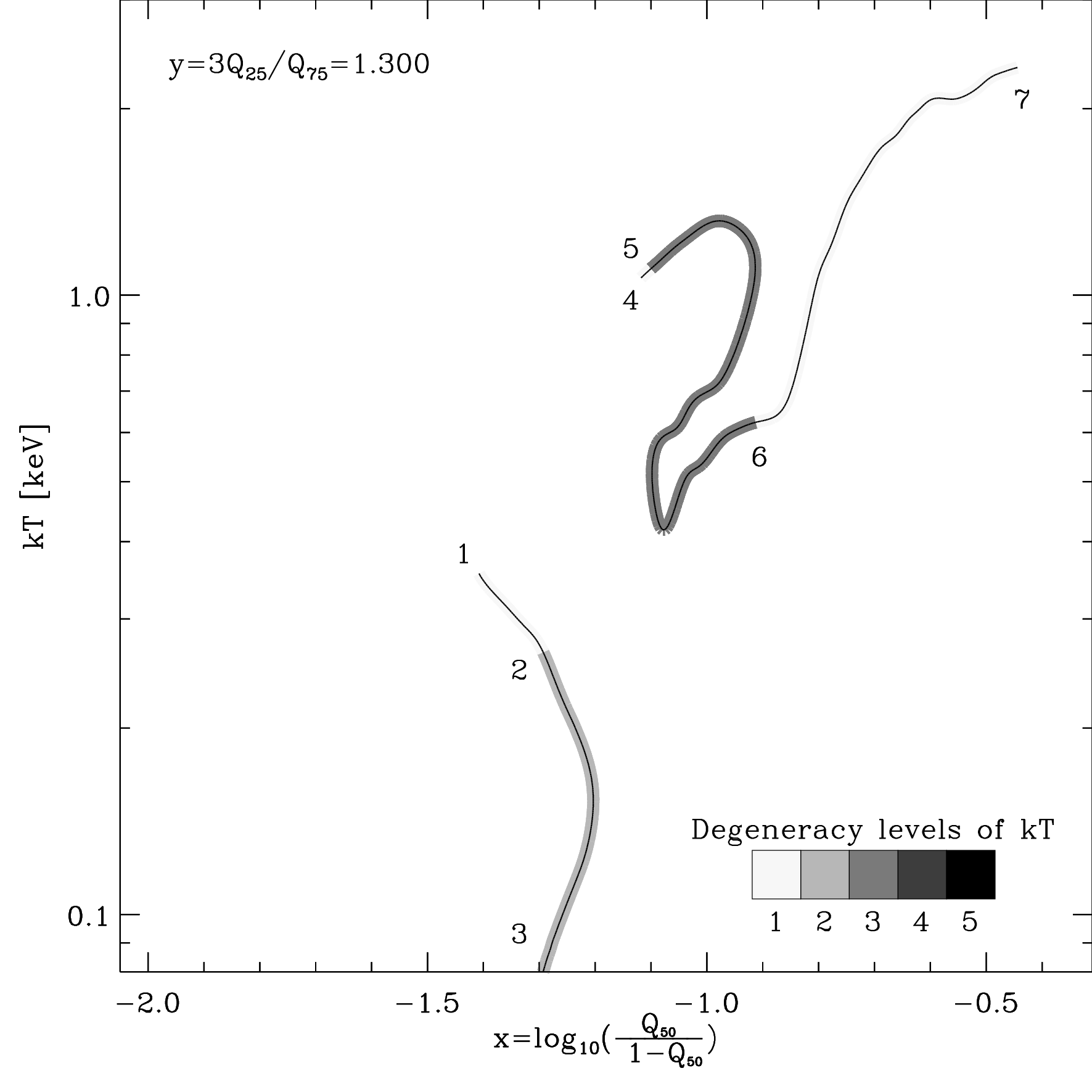}
\end{tabular}
\vspace{-0.4cm}\\ 
\caption{Illustration of degenerated solutions. Left: horizontal cut of
the \nh~surface (Fig.~\ref{fig:nh_degeneracy} right) at
$y=1.3$. Right: horizontal cut of the $kT$ surface
(Fig.~\ref{fig:kt_degeneracy} right) at the same elevation. The line
colour indicates the degeneracy level of the parameter. Numbers label 
the boundary/changing points when one moved on these surfaces along the $x$
direction, and give the correspondance between the two parameters.
}
\label{fig:degeneracy_cut}
\end{figure*}

Fig.~\ref{fig:degeneracy_cut} shows an illustration of degenerated
solutions if $y=1.3$. For example if $x_1 \le x \le x_2$ there is
a unique solution for $N_{\rm H}$ and $kT$. If $x_2 \le x \le x_3$
there is a double solution: a first solution with low $N_{\rm H}$ and high
$kT$, and a second solution with high $N_{\rm H}$ and low
$kT$. Generally speaking, if there are $N$ solutions of $N_{\rm H}$
and $kT$ ordered by increasing values, the solution number $n$
(with $n$ ranging from 1 to $N$) is defined as the $n^{\rm th}$-value
of $N_{\rm H}$ and the $N+1-n^{\rm th}$-value of $kT$. Degeneracy on
both parameter can then be fixed using extra knowledge on only one
parameter.

We used the knowledge of the optical extinction to derive the value
of the hydrogen column density using
the relation $N_{\rm H} = 1.6 \times 10^{21} A_{\rm
V}$~cm$^{-2}$\,mag$^{-1}$ \citep{vuong03,cardelli89}. The best estimate of the
temperature is then the temperature solution corresponding to the
hydrogen column density solution which is the closest to this value.


\end{document}